\documentclass[11pt]{article}
\usepackage{jheppubmodnew}
\pdfoutput=1
\usepackage{hyperref}
\usepackage{psfrag}
\usepackage{array}
\usepackage{amssymb}
\usepackage{amsmath}
\usepackage{amsthm}
\usepackage{graphicx}
\usepackage{etoolbox}
\usepackage[labelsep=quad]{subcaption}
	
\usepackage[punctsep]{collref}
\collectsep[]{;}
\usepackage{bm}
\usepackage{epsfig}
\usepackage{wrapfig}
\usepackage{tikz}
\usetikzlibrary{arrows,plotmarks,positioning,calc}

\newtheorem*{result}{Result}

\def\al{A}
\def\op{{\cal O}}
\def\obdry{{\cal O}}

\def\pb[#1,#2]{\{#1, #2\}}
\def\deb[#1,#2]{[#1,#2]_{\text{D.B.}}}

\def\l{\lambda}
\def\la{\lambda}

\def\w{\omega}

\def\e{\eta}

\def\a{\alpha}

\def\Or[#1]{\text{O}\!\left(#1\right)}
\def\dotl[#1,#2]{\left\langle #1,\, #2 \right\rangle}
\def\dotlb[#1,#2]{\left\langle #1,\, #2 \right\rangle}
\def\dotlm[#1,#2]{\left[ #1,\, #2 \right]}
\def\dotp[#1,#2]{(\vect{#1} \cdot\vect{#2})}
\def\aff[#1,#2]{\hat{#1}(#2)}

\def\n4sym{{\cal N}=4 SYM}
\def\>{\rangle}
\def\<{\langle}

\def\projsho[#1]{{\cal P}^{\text{sho}}_{#1}}
\def\transsho[#1,#2]{{\cal T}^{\text{sho}}_{#1,#2}}
\def\weight[#1,#2,#3]{\{(#1),#2,#3\}}
\def\ads[#1]{$\text{AdS}_{#1}$}

\newcommand{\be}{\begin{equation}}
\newcommand{\ee}{\end{equation}}
\newcommand{\ba}{\begin{align}}
\newcommand{\ea}{\end{align}}
\newcommand{\bs}{\begin{split}}
\def\sess\end{split}
\newcommand{\vect}[1]{{\boldsymbol{#1}}}

\newcommand{\bea}{\begin{eqnarray}}
\newcommand{\eea}{\end{eqnarray}}

\def \bes {\begin{equation*}}
\def \ees {\end{equation*}}

\def \rt  {\right }

\def \pa  {\partial}

\def \scrip{{\cal I}^{+}}
\def \scrim{{\cal I}^{-}}

\def\alcut[#1]{{\cal A}_{#1, \epsilon}}
\def\alseg[#1,#2]{{\cal B}_{#1, #2}}

\def\supcharge[#1]{\{#1\}}

\def\projsupeig[#1]{{\cal P}_{{\ell, m}}[{#1}]}
\def\transop[#1, #2]{T_{\{#1\}, \{#2\}}}
\def\supket[#1]{|\{#1\} \rangle}
\def\supbra[#1]{\langle \{#1\} | }

\def\rsop[#1]{X_{#1}}

\def\projlow[#1,#2]{P_{{#1}<{#2}}}
\def\rd[#1]{^{(#1)}R}

\newcommand{\htt}[1][]{\ifstrempty{#1}{h^{\text{TT}}}{h^{\text{TT},#1}}}
\newcommand{\htr}{h^{\text{T}}}
\newcommand{\hlo}{h^{\text{L}}}
\newcommand{\ztt}{\zeta^{\text{TT}}}
\newcommand{\ztr}{\zeta^{\text{T}}}
\newcommand{\zlo}{\zeta^{\text{L}}}

\newcommand{\pitt}[1][]{\def\ArgI{{#1}} \pittrelay}
\newcommand{\pitr}[1][]{\def\ArgI{{#1}} \pitrrelay}
\newcommand{\pilo}[1][]{\def\ArgI{{#1}} \pilorelay}

\newcommand\pittrelay[1][]{\ifstrempty{#1}{\Pi_{\text{TT}}^{\ArgI}}{\Pi_{\text{TT}, #1}^{\ArgI}}}
\newcommand\pitrrelay[1][]{\ifstrempty{#1}{\Pi_{\text{T}}^{\ArgI}}{\Pi_{\text{T}, #1}^{\ArgI}}}
\newcommand\pilorelay[1][]{\ifstrempty{#1}{\Pi_{\text{L}}^{\ArgI}}{\Pi_{\text{L}, #1}^{\ArgI}}}

\newcommand{\whtr}{\widetilde{h}^{\text{T}}}
\newcommand{\wpitr}[1][]{\def\ArgI{{#1}} \wpitrrelay}
\newcommand\wpitrrelay[1][]{\ifstrempty{#1}{{\widetilde{\Pi}}_{\text{T}}^{\ArgI}}{\Pi_{\text{T}, #1}^{\ArgI}}}

\newcommand{\psifull}{\Psi[h, \phi]}

\newcommand{\psifullg}{\Psi[g, \phi]}
\newcommand{\psifullexp}{\Psi[\htt, \htr, \hlo, \phi]}

\newcommand{\psifullfour}[1][]{\Psi_{#1}[\htt, \pitr, \hlo, \phi]}

\newcommand{\psifullfourE}{\Psi[\htt, \pitr, \hlo, \phi]}
\newcommand{\psifullfourEtilde}{\Psi[\htt, \wpitr, \hlo, \phi]}
\newcommand{\psigrav}{\psi_{\rm g}[\htt_{}]}
\newcommand{\psigravnoarg}{\psi_{\rm g}}
\newcommand{\psimat}{\psi_{\rm m}[\phi]}
\newcommand{\psimatnoarg}{\psi_{\rm m}}
\newcommand{\psivacg}{\psi_{0}[\htt_{}]}
\newcommand{\psifullE}{\Psi^{E,\{a\}}[h_{}, \phi]}

\newcommand{\psifullEnoarg}{\Psi^{E,\{a\}}}
\newcommand{\psifullEpnoarg}{\Psi^{E',\{a'\}}}

\newcommand{\psifulltildeE}{\Psi^{E',\{a'\}}[\tilde{h}_{}, \tilde{\phi}]}
\newcommand{\rhofullE}{\rho^{E,E', \{a\}, \{a'\}}[h_{}, \phi, \tilde{h}_{}, \tilde{\phi}]}
\newcommand{\rhofull}{\rho[h_{}, \phi, \tilde{h}_{}, \tilde{\phi}]}
\newcommand{\rhofullEnoargs}{\rho^{E,E', \{a\}, \{a'\}}}

\newcommand{\psivacmat}{\psi_{0}[\phi]}
\newcommand{\indices}{{E,E',\{a\}, \{a'\}}}
\newcommand{\indicesrev}{{E',E,\{a'\}, \{a\}}}
\newcommand{\indicesdiag}{{E,E,\{a\}, \{a\}}}

\newcommand{\psifockE}[1][]{\psi_{\text{F}}^{E_{#1},\{a_{#1}\}}[\htt, \phi]}

\newcommand{\psifockEnoarg}[1][]{\psi_{\text{F}}^{E_{#1},\{a_{#1}\}}}
\newcommand{\psifockEpnoarg}{\psi_{\text{F}}^{E',\{a'\}}}

\newcommand{\intcur}{\Hbdy}
\newcommand{\dbdy}{d^{d-1} \Omega}
\newcommand{\psiint}{\psi_{\text{I}}^{E,\{a\}}[\intcur, \htt, \phi]}
\def\innerp[#1,#2]{\left({#1, #2} \right)}
\def\hbulk{\cH_\text{bulk}} 

\def\hG{\hat{\G}}

\def\hn{\hat{\n}}
\def\hga{\hat{\g}}
\def\hh{\hat{h}}

\def\bt{\mathbf{t}}

\def\Hbdy{H_\partial}

\def\ss{\subsection}

\def\pg{\paragraph}

\def\ie{\emph{i.e.} }

\def\nt{\notag}

\def\wt{\widetilde}

\def\R{\mathbb{R}}
\def\Z{\mathbb{Z}}

\def\cD{\mathcal{D}}

\def\cH{\mathcal{H}}

\def\cL{\mathcal{L}}

\def\cN{\mathcal{N}}

\def\cP {\mathcal{P}}

\def\cR{\mathcal{R}}
\def\cS{\mathcal{S}}

\def\bZ{\mathbb{Z}}

\def\dg {\dagger}
\def\p{\partial}

\def\/{\over}
\def\ov{\over}

\def\t{\theta}

\def\e{\epsilon}
\def\ve{\varepsilon}

\def\a{\alpha}

\def\d{\delta}
\def\k{\kappa}
\def\g {\gamma}
\def\la {\lambda}
\def\w {\omega}

\def\l{\ell}
\def\mn{{\mu\nu}}

\def\n {\nabla}
\def\L{\Lambda}
\def\D{\Delta}
\def\G {\Gamma}
\def\Om {\Omega}
\def\S{\Sigma}

\def\ra{\rightarrow}

\def\r{\mathrm}

\def\_{\hspace{2cm}}
\def\-{\\\notag}
\def\={&=&}

\newcommand{\bpm}{\begin{pmatrix}}
\newcommand{\epm}{\end{pmatrix}}

\newcommand{\bit}{\begin{itemize}}
\newcommand{\eit}{\end{itemize}}

\newcommand{\ben}{\begin{enumerate}}
\newcommand{\een}{\end{enumerate}}

\newcommand\bsp{\begin{split}}
\newcommand\esp{\end{split}}

\def\le{\left}
\def\ri{\right}

\def\l{\ell}

\def\qq{\qquad}

\title{Holography from the Wheeler-DeWitt equation}

\author[a]{Chandramouli Chowdhury,}
\author[a]{Victor Godet,}
\author[b]{Olga Papadoulaki}
\author[a]{and Suvrat Raju}
\affiliation[a]{International Centre for Theoretical Sciences, Tata Institute of Fundamental Research, Shivakote, Bengaluru 560089, India.}
\affiliation[b]{International Centre for Theoretical Physics,
Strada Costiera 11, 34151 Trieste, Italy.}
\emailAdd{chandramouli.chowdhury@icts.res.in}
\emailAdd{victor.godet@icts.res.in}
\emailAdd{papadoulaki@ictp.it}
\emailAdd{suvrat@icts.res.in}
\date{}

\abstract{In a theory of quantum gravity, states can be represented as wavefunctionals that assign an amplitude to a given configuration of matter fields and the metric on a spatial slice. These wavefunctionals must obey a set of constraints as a consequence of the diffeomorphism invariance of the theory, the most important of which is known as the Wheeler-DeWitt equation. We study these constraints perturbatively by expanding them to leading nontrivial order in Newton's constant about a background AdS spacetime. We show that, even within perturbation theory, any wavefunctional that solves these constraints must have specific correlations between a component of the metric at infinity and energetic excitations of matter fields or transverse-traceless gravitons. These correlations disallow strictly localized excitations. We prove perturbatively that two  states  or two density matrices that coincide at the boundary for an infinitesimal interval of time must coincide everywhere in the bulk. This analysis establishes a perturbative version of holography for theories of gravity coupled to matter in AdS.}

\begin{document}
\maketitle

\section{Introduction}
It has recently been argued that theories of gravity localize quantum information very differently from local quantum field theories \cite{Laddha:2020kvp}. This argument can be encapsulated in a principle of holography of information: in a theory of quantum gravity, information that is available in the bulk of a Cauchy slice is also available near its boundary \cite{Raju:2020smc}. This principle can be made precise and proved in asymptotically AdS spacetimes and in four-dimensional asymptotically flat spacetimes. In \cite{Chowdhury:2020hse}, a physical protocol was presented that exploited this effect to allow observers near the boundary of AdS to extract information about low-energy states in the bulk without directly exploring the bulk.

In the presence of a negative cosmological constant, these effects may be expected from  the  AdS/CFT conjecture \cite{Maldacena:1997re,Gubser:1998bc,Witten:1998qj}. But a study of how quantum gravity localizes information sheds light on the {\em physical origin} of holography for gravitational theories. It also indicates how holography should be extended beyond asymptotically AdS spacetimes to asymptotically flat spacetimes.

In this paper, we present a direct perturbative analysis of the allowed wavefunctionals in a theory of gravity coupled to matter in an asymptotically AdS spacetime. We find that any two wavefunctionals that coincide at the boundary for an infinitesimal interval of time  must also coincide in the bulk.  This is a uniquely gravitational effect; wavefunctionals in a local quantum field theory do not have such a property.

In gravity, the metric is one of the dynamical degrees of freedom. In the Hamiltonian formalism, which we adopt in this paper, the degrees of freedom are divided into the metric on a spatial slice and its conjugate momentum, which is related to the extrinsic curvature of the slice. We consider theories that might have additional matter fields.  The values of these fields on a spatial slice provide another set of canonical variables whose conjugate momenta are related to the time derivatives of these fields. A wavefunctional assigns a complex number to any specification of the metric and other fields on a spatial slice. 

Not every wavefunctional is a valid state in a theory of gravity. A valid wavefunctional must take on the same value for configurations that can be related by a diffeomorphism that vanishes asymptotically. This leads to a set of  constraints on the wavefunctional, of which the most important constraint is called the Wheeler-DeWitt (WDW) equation \cite{DeWitt:1967yk}. 

In this paper, we present a direct perturbative analysis of the WDW equation. We build on an important old paper by Kuchar \cite{Kuchar:1970mu} who analyzed the solutions of the WDW equation about flat space in the free limit. We  extend this analysis by expanding the constraints to leading nontrivial order in perturbation theory in the gravitational interaction in the presence of a negative cosmological constant. This analysis is already sufficient to reveal the  remarkable property of these solutions alluded to above.

The structure of the constraints that we find can roughly  be described as follows. The metric degrees of freedom can be divided into a longitudinal component, a transverse-traceless component and, what we call, a ``T-component'' that keeps track of the trace \cite{Arnowitt:1962hi}. The transverse-traceless component can be freely specified, just like another dynamical field. Invariance of the wavefunctional under spatial diffeomorphisms fixes its dependence on the longitudinal component of the metric.  The so-called Hamiltonian constraint, which imposes invariance of the state under diffeomorphisms that mix space and time,  fixes the dependence of the wavefunctional on the T-component. We show that an  important role is played by a specific integral of the Hamiltonian constraint on the entire Cauchy slice which relates the asymptotic T-component of the metric to the total energy of the transverse-traceless gravitons and matter-fields on the Cauchy slice.

We prove that these constraints are sufficient to disallow any deformations of the wavefunctional which alters its form in the bulk without changing its boundary values. The reason can be understood as follows. A bulk deformation that changes the energy must necessarily also change the T-component of the metric near the boundary. So deformations that leave the asymptotic T-component unchanged can only ``move'' energy from one part of space to another and must have zero total energy. But  the Heisenberg uncertainty principle tells us that an operator that implements such a deformation must be completely delocalized. Therefore, while such an operator may commute with the asymptotic metric, it must fail to commute with some {\em other} dynamical operator near the boundary. The final result  is that correlators of the T-component of the metric and of other dynamical operators at the boundary of AdS for an infinitesimal amount of time completely fix the wavefunctional.

\begin{figure}[!ht]
\centering
\begin{subfigure}{0.4\textwidth}
\centering
\includegraphics[height=0.3\textheight]{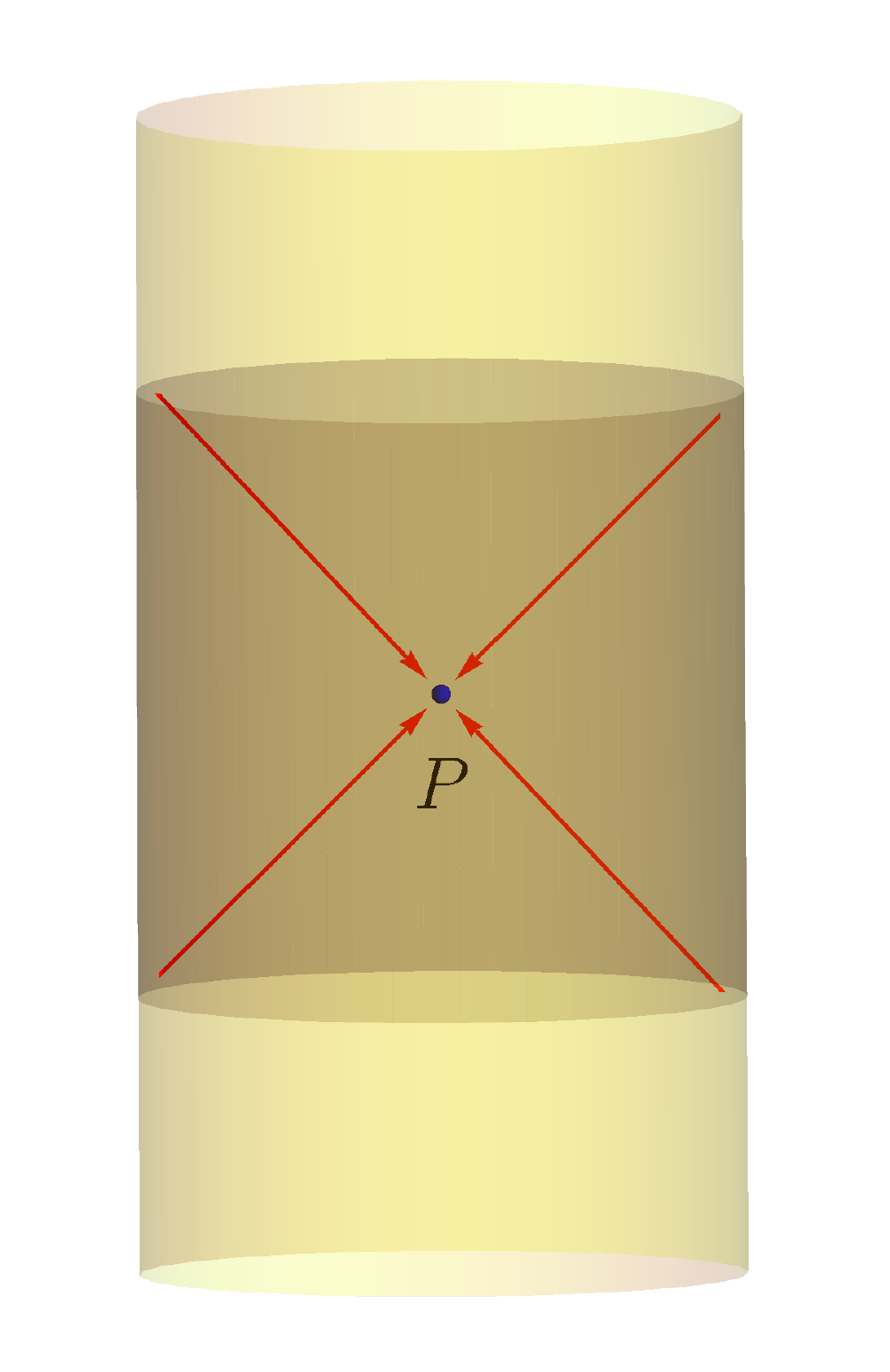}
\caption{\label{figmisunderstanding}}
\end{subfigure}
\hspace{0.15\textwidth}
\begin{subfigure}{0.4\textwidth}
\centering
\includegraphics[height=0.3\textheight]{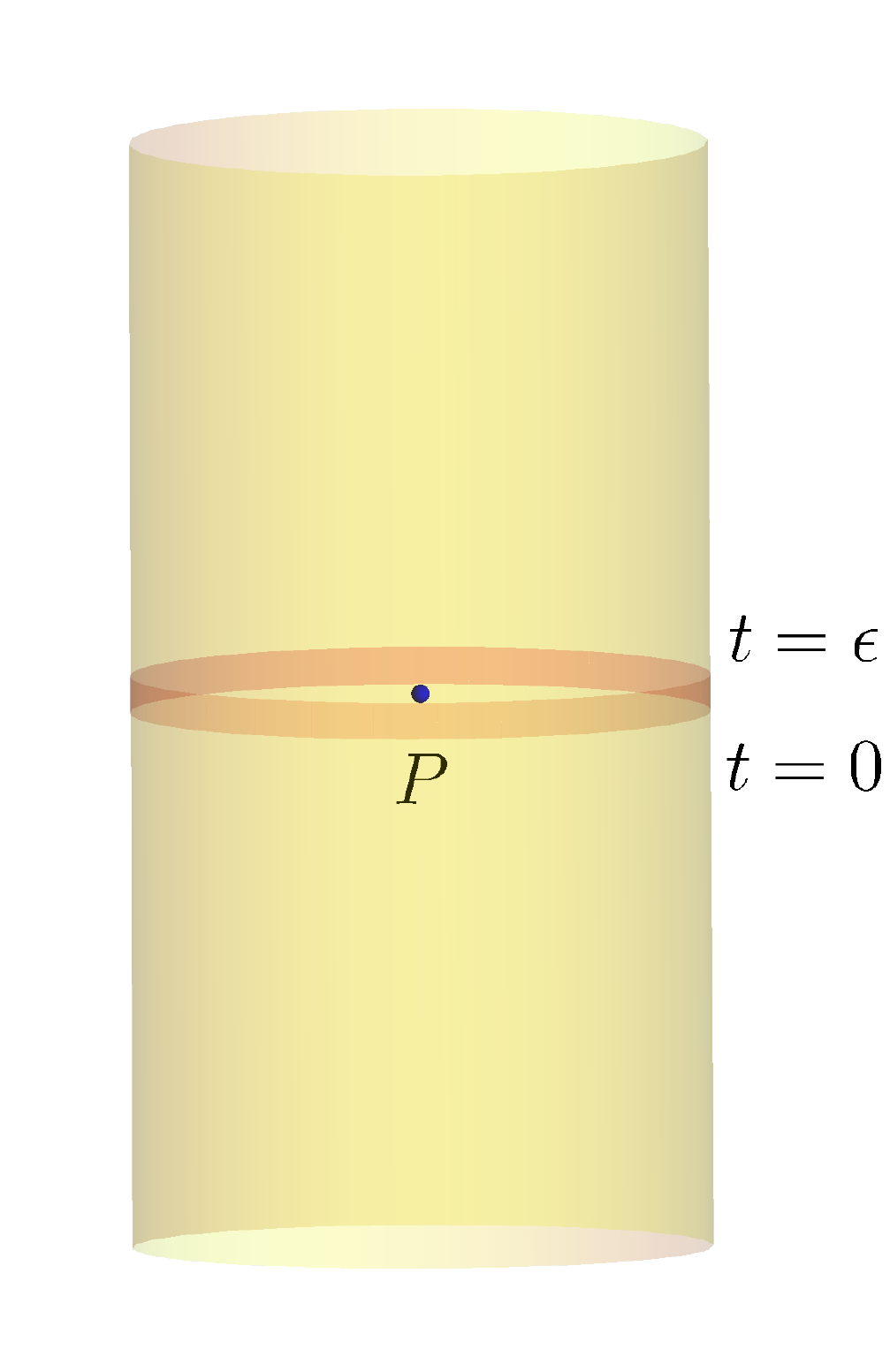}
\caption{\label{figunderstanding}}
\end{subfigure}
\caption{\em A common misunderstanding of ``holography'' is that it only tells us that data on the timelike boundary of AdS can be used to reconstruct physics at the bulk point P as shown in the left subfigure. But this statement follows from the causal structure of AdS and does not require holography. Gravitational theories are genuinely holographic. In this paper we show how, in gravity, data on an infinitesimal time band (right subfigure) can be used to reconstruct physics in the bulk.
}  \label{figholmisundund}
\end{figure}

This result establishes, in the perturbative approximation, that one of the central aspects of holography follows from the constraints of gravity. The significance of this result can be illustrated by studying the contrast between gravitational and non-gravitational quantum field theories in AdS. Even in a non-gravitational theory, the specification of data on the entire timelike boundary of AdS is sufficient to reconstruct physics in the bulk. See Figure \ref{figmisunderstanding}. This is just a property of the causal structure and is not indicative of holography. What our result shows is that, in a gravitational theory,  data on an infinitesimal time band on the boundary of AdS is already sufficient to reconstruct the state in the bulk. See Figure \ref{figunderstanding}.  

We emphasize that in a non-gravitational theory, our final result could not possibly be true. The action of a unitary operator in a bulk at the same time would commute with all observations on this infinitesimal time band on the boundary by microcausality. Therefore, in a non-gravitational theory,  it is impossible to distinguish a given state from the state obtained after the action of such a unitary. 

\paragraph{\bf Relationship to previous work.} As mentioned above, it has already been argued previously \cite{Laddha:2020kvp,Raju:2020smc} that gravitational theories localize information very differently from ordinary quantum field theories. These previous arguments \cite{Laddha:2020kvp,Raju:2020smc}, which  built on \cite{Banerjee:2016mhh,Raju:2019qjq}, relied on weak assumptions about the structure of the Hilbert space, and the nature of the gravitational Hamiltonian to arrive at nonperturbative results.

Although the analysis in this paper is perturbative,  it is more explicit. We make no prior assumptions either about the Hilbert space or about the gravitational Hamiltonian. Instead, we explicitly construct the low-energy Hilbert space by studying solutions to the gravitational constraints and we explicitly show that such solutions must have correlations between a component of the asymptotic metric and the energy of the state. This analysis also reveals how the unusual localization of quantum information in quantum gravity is visible at the level of wavefunctionals.

The analysis in this paper takes advantage of the infrared cutoff that is provided by global AdS boundary conditions. (See comment \ref{infraredcomment} in section \ref{secholproof}.) For this reason the analysis presented here reproduces  Result 5 of \cite{Laddha:2020kvp} --- which pertains  to asymptotically AdS spacetimes and was proved there using  operator-theoretic techniques --- but cannot immediately be used to make contact with Result 1 and Result 2 of \cite{Laddha:2020kvp}, which apply to asymptotically-flat spacetime. We expect that it should be possible to generalize the proof of the holography of information presented here to address the infrared subtleties present in flat space and hope to report on this in forthcoming work.

The relationship between the bulk constraints and holography was also explored previously in \cite{Marolf:2008mf,Marolf:2006bk} and more recently in \cite{Jacobson:2019gnm} (see also the prescient essay \cite{Jacobson:2012gh}) although the techniques used in this paper are quite different.  A radial version of the  WDW equation was studied in the context of AdS/CFT \cite{Freidel:2008sh}, which was analyzed further in \cite{Cianfrani:2013oja} and has proved to be useful in the context of the study of $T\overline{T}$ deformations \cite{McGough:2016lol,Caputa:2020lpa,Kruthoff:2020hsi,Coleman:2019dvf,Donnelly:2018bef,Tolley:2019nmm,Ireland:2019vvj,Hirano:2020nwq,Belin:2020oib} and bulk reconstruction \cite{Caputa:2019pam}. Here our analysis is different since we are considering the conventional WDW equation that governs wavefunctionals on a Cauchy slice. 

The Wheeler-DeWitt equation has been studied in the mini-superspace approximation --- where we found \cite{Kenmoku:1998he,Hajicek:1984mz,Fischler:1989ka,Fischler:1989se,Halliwell:1984eu} useful --- and also in the context of two-dimensional models \cite{Hori:1993zq,Louis-Martinez:1993bge,Gegenberg:1994pv,Maldacena:2019cbz,Betzios:2020nry} and in terms of the Ashtekar variables \cite{Ashtekar:1986yd}. See \cite{rovelli2004quantum,Halliwell:1989myn,coleman1991quantum} for a more detailed list of references. However, there has been relatively little work on a straightforward perturbative analysis of the equation in higher dimensions. As already mentioned, Kuchar \cite{Kuchar:1970mu} studied this problem at zeroth order in the gravitational constant, and here we will show that, even at leading order, the structure of the constraints is interesting and leads to surprising properties of the solutions.

The question of how the gravitational constraints affect the localization of quantum information was also studied, from another perspective, in \cite{Giddings:2019hjc,Donnelly:2018nbv}. (See also \cite{Donnelly:2016rvo,Donnelly:2017jcd}.)  However, these papers  reached the opposite conclusion from the one we will reach here: in \cite{Giddings:2019hjc,Donnelly:2018nbv} it was claimed that it should be possible to perturbatively construct states that differ inside a bounded region but are asymptotically identical. It appears to us that this  conclusion was reached because \cite{Giddings:2019hjc,Donnelly:2018nbv} focused on the asymptotic gravitational field but failed to consider quantum correlators of the metric and the dynamical scalar field that was included in the analysis there.  As we will see in section \ref{secholproof} this  latter class of correlators, involving both the metric and dynamical fields, plays an important role and cannot be neglected.

\paragraph{\bf Organization of the paper.}
This paper is organized as follows. We provide a technical summary of our results in section \ref{secsummary}. In section \ref{secsetting}, we provide a quick review of the constraints on valid wavefunctionals in any theory of quantum gravity. In section \ref{secpert}, we expand these constraints in perturbation theory and explain how they can be conveniently organized by dividing the degrees of freedom in the metric into a transverse-traceless, longitudinal and T-component. In section \ref{sec:solving}, we show that focusing on the integral of the Hamiltonian constraint leads to a significant simplification. We solve this integrated Hamiltonian constraint revealing a structure where the T-component of the metric at infinity is correlated with eigenfunctionals of the bulk matter energy. We also propose  a procedure to solve the pointwise Hamiltonian and momentum constraints and we give explicit expressions for the leading order solutions. In section \ref{secholproof} we prove that correlators of the T-component of the metric and boundary operators, in an infinitesimal time band, are sufficient to completely fix the structure of the bulk wavefunctional.  

As already stated, the results we derive here are valid for theories of gravity and do {\em not} have an analogue in non-gravitational gauge theories. To illustrate this difference, in Appendix \ref{appem}, we analyze the constraints in electromagnetism. We show that they are significantly weaker than the constraints in a theory of gravity. Consequently, QED and other nongravitational gauge theories localize information much like ordinary local quantum field theories and does not share the unusual constrained properties of gravitational wavefunctionals. Appendices \ref{App:GravitonModes}, \ref{appEnergy} and \ref{appleadingorder} provide additional technical details.

\section{Summary of our results \label{secsummary}}

We now provide a concise summary of our results. This section is meant to provide a guide for the rest of the paper, and some of the notation used here is defined carefully in later sections. The equations in this section are all linked to corresponding equations in later sections, which provide a more detailed discussion of the physics.

When gravity is quantized using the canonical formalism, the physical states of the theory are given by wavefunctionals of the metric $g_{i j}$ on a spatial slice, and the matter fields $\phi$ that obey the so-called Hamiltonian and momentum constraints, 
\be
\label{constraints}
\cH(f) \psifullg =0,\qq \cH_i (f)  \psifullg =0~.
\ee
These are the constraints displayed in equation \eqref{HamConstraint} --- where the conjugate momenta for the metric and the matter fields are denoted $\pi^{ij}= -i{\d\/\d g_{ij}}$ and $\pi= - i{\d\/\d\phi}$ --- which have been smeared with a function $f$ that vanishes at the boundary. The momentum constraint is linear in momenta while the Hamiltonian constraint is quadratic. 

To study these constraints, we first expand the metric about a background AdS metric as $g_{i j} = \g_{i j} + \kappa h_{i j}$ where $\g_{i j}$ is the AdS metric and  $\kappa=\sqrt{8 \pi G}$. We also introduce a corresponding momentum operator $\Pi^{i j} = -{i \over \sqrt{\gamma}} {\d\/\d h_{ij}}$ that is more appropriate for understanding perturbation theory. We further decompose this metric fluctuation as 
\be
h_{ij} = \htt_{ij}+\hlo_{ij}+\htr_{i j},
\ee
in terms of the transverse-traceless component, the longitudinal component, and what we term the ``T-component''. This decomposition was introduced about flat space in \cite{Arnowitt:1962hi}, and we generalize it to AdS. The precise definition of the three components is given in equation \eqref{admdecomp}. Similarly, the conjugate momentum can be decomposed as  $\Pi^{ij} = \pitr[ij] + \pitt[ij] + \pilo[ij]$ and we show below equation \eqref{pidecomposition} that each component is the canonical momentum associated with the corresponding metric component. 

We then expand the constraints in perturbation theory.  It is convenient to set the AdS scale, $\ell=1$ and treat $\k$ as a small dimensionless parameter that allows us to organize the perturbative expansion. The validity of perturbation theory then requires that any numbers that emerge from the action of derivative operators on the wavefunctional should not scale with ${1 \over \kappa}$ and we ensure this below.

At leading  order in $\k$, the momentum constraint implies that the wavefunctional $\psifull$ is independent of $\hlo$. This is simply the statement that the wavefunctional should be invariant under linearized spatial diffeomorphisms. At next order, it gives
\be
\label{momsummary}
\le( -2 \n_j \pilo[ij]+\k \, Q^i\ri) \psifull =0~,
\ee
where $Q^i$ is quadratic in the canonical variables and is given in \eqref{defQi}. We have $\pilo[ij]=-{i\/\sqrt{\g}}{\d\/\d \hlo_{ij}}$ so the second order momentum constraint determines the dependence of $\Psi$ in $\hlo_{ij}$. 

At leading order in $\kappa$, the Hamiltonian constraint implies that the T-component of the metric vanishes: $\htr = 0 + \Or[\kappa]$. At next order, the Hamiltonian constraint fixes the T-component of the metric via
\be
\label{hamsummary}
\le(-\cD^{ij} \htr_{ij}+ \k\, Q\ri) \psifull= 0~,
\ee
where  $\cD^{ij}$ is given in \eqref{dijexplicit} and $Q$ is given in \eqref{defQ}. This sets $\htr_{ij}$ to a non-trivial $\Or[\k]$ value. 

To analyze these constraints, we first integrate the Hamiltonian constraint over a Cauchy slice $\S$ to obtain a simpler constraint, which takes the form 
\be
\label{simpham}
\le(-\Hbdy +\int_\S d^d x\sqrt{\g}\,N \hbulk  \rt)\psifull = 0~,
\ee
where
\be
\label{summaryhbdy}
\Hbdy \equiv {1\/2\k}\int_{\p \S}  \dbdy \,J^i n_i ~.
\ee
Here the ADM current $J^i$, which  is integrated over the boundary $\p \S$ after contracting with the normal $n_i$,  is linear in the metric fluctuation and defined in \eqref{defJi}. It depends only on the T-component of the metric as shown in \eqref{honlytdepend} and gives the ADM energy $H_\p$.  In \eqref{simpham}, $N$ is the lapse function; $\hbulk$ is quadratic in the canonical variables and its precise definition is given in equation \eqref{defhbulk}. It can be viewed as the ``bulk energy density'' involving the transverse-traceless gravitons and the matter. Thus, the integrated Hamiltonian constraint gives a quantum version of the familiar statement that the energy is a boundary term in canonical gravity.

Since the integrated Hamiltonian constraint is so simple, we can {\em explicitly} obtain wavefunctionals that solve it. The solutions take the form of a ``dressed'' Fock space that we construct as follows. First, we obtain wavefunctionals of $\htt$ and $\phi$ that form an ordinary free-field Fock space in AdS and are eigenstates of the free-field Hamiltonian. We choose a basis for these wavefunctionals that we denote by
\[
\psifockE~.
\]
The superscript $E$ indicates the energy of the state in the Fock space, and  the superscript $\{a\}$ is an additional label for degenerate energy eigenstates.
 
These Fock space wavefunctionals can be promoted to a solution of the integrated constraint by additionally specifying that they are eigenstates of the integral of the boundary metric that appears on the left of equation \eqref{simpham}:
\be
\label{intsol}
\psiint = \psifockE \otimes |\Hbdy= E \rangle~.
\ee

The constraints \eqref{hamsummary} and \eqref{momsummary} constitute an infinite number of constraints --- one at each point of the Cauchy slice. So the solution to the integrated Hamiltonian constraint obtained above needs to be improved further to obtain a solution to these constraints. We present an explicit leading order solution to the pointwise constraints in section \ref{sec:pointwise} and Appendix \ref{appleadingorder}. In addition, we give a simple discussion of a procedure that makes it clear that each solution of the integrated constraint \eqref{intsol} can be uniquely uplifted to a solution of the pointwise constraint \eqref{constraints}:
\be
\psiint \rightarrow \psifullE~.
\ee
This argument is enough to ensure that once the dependence of the wavefunctional on $\htt$ and $\phi$ in the auxiliary Fock space is chosen, there is no further freedom to specify its dependence on $\htr$ and $\hlo$. The integrated constraint fixes the detailed form of $\htr$ at the boundary and, although the solution to the pointwise constraints that we find is both new and interesting, we do not require the explicit form of the dependence of the wavefunctional on $\htr$ and $\hlo$ in the bulk for obtaining our main result.

We then define a natural inner product on the space of solutions (see section \ref{sec:innerproduct}) and show that it is compatible with the structure of the constraints. This allows us to meaningfully compute correlation functions of observables using these wavefunctionals.

The above analysis of the constraints allows us to obtain a striking result. We show that any two pure or mixed states in a theory of gravity that agree on the boundary of AdS for an infinitesimal interval of time must agree everywhere in the bulk. To demonstrate this result we consider a general density matrix that depends on two metric perturbations, $h_{i j}$ and $\tilde{h}_{i j}$ and two matter perturbations, $\phi$ and $\tilde{\phi}$. We write it in the form 
\be
\label{summrhostate}
\rhofull =\sum_{\indices} c(\indices) \rhofullE~, 
\ee
where  $c(\indices)$ is a list of coefficients and a basis of density matrices
\be
\rhofullE \equiv \psifulltildeE \psifullE^\ast~,
\ee
is obtained by combining the solutions to the constraints obtained above.

We consider a simple class of gauge invariant operators that are supported only on the boundary, and therefore automatically commute with the constraints \eqref{constraints}.  One such operator is $\Hbdy$ displayed in \eqref{summaryhbdy}, whereas other operators --- which we denote by $\op(t, \Omega)$ --- correspond to the boundary limit of fluctuations of the dynamical fields, including the transverse-traceless graviton and matter fields.  We first show that if two density matrices $\rho_1$ and $\rho_2$ yield the same correlators of the following combination of such operators 
\be
\label{summsumzerocond}
\langle \Hbdy^n\, \op(t_1, \Omega_1) \ldots \op(t_q, \Omega_q) \Hbdy^m \rangle_{\rho_1} = \langle \Hbdy^n\, \op(t_1, \Omega_1) \ldots \op(t_q, \Omega_q) \Hbdy^m \rangle_{\rho_2}~,
\ee
then the respective coefficients $c_1(\indices)$ and $c_2(\indices)$ must satisfy the following identity at each individual value of $E$ and $E'$ 
\be
\label{summindivzerocond}
\sum_{\{a\}, \{a'\}}  \Big[c_1(\indices) - c_2(\indices) \Big]   \langle \op(t_1, \Omega_1) \ldots \op(t_q, \Omega_q) \rangle_{\rhofullEnoargs} =0~.
\ee
We only demand that the equations above hold at $\Or[1]$ and not at $\Or[\kappa]$ so that we can study them reliably within our perturbative setup. In particular, this means that $n,m,q$ are limited to $\Or[1]$ integers as well and cannot scale with an inverse power of $\kappa$ and the passage from \eqref{summsumzerocond} to \eqref{summindivzerocond} can be performed reliably provided that the energy of the state in \eqref{summrhostate} does not scale with $\log{1 \over \kappa}$ in AdS units. We show that there is no non-trivial solution to these equations if the $t_i$ above are allowed to range in the infinitesimal interval $[0, \epsilon]$. Therefore if two pure or mixed states agree on the boundary for even an infinitesimal time interval then they must be the same.

This last result that we obtain is central to the notion of holography since it tells us that, in a theory of gravity, the state in the bulk is completely determined by boundary data in an infinitesimal time interval. Here we see that this surprising aspect of gravity follows directly from the constraints of the theory.

\section{Preliminaries  \label{secsetting}}

In this section, we set the stage for our analysis,  establish some notation, and review the constraints that must be satisfied by physical states in any theory of gravity.

\subsection{Action and boundary conditions}

We will study gravity  with a negative cosmological constant  in $d+1$ dimensions, as described by the action 
\be
\label{actionInitial}
S = {1 \over 2 \kappa^2} \int dt d^{d} x\,\sqrt{-\hat{g}}\,  (\hat{R} - 2 \Lambda)  +S_{\text{GHY}}+ S_{\text{matter}},
\ee
where $\kappa = \sqrt{8 \pi G}$, $\hat{R}$ is the $d+1$-dimensional Ricci scalar,  $S_{\text{GHY}}$ is the Gibbons-Hawking-York  boundary term and $\Lambda$ is a cosmological constant. We will use hats to differentiate spacetime quantities with Cauchy slice quantities. The specific details of the matter sector will not be important in the subsequent analysis although we will use scalar fields as an example for illustration. 

We are interested in spacetimes that are asymptotically AdS. Note that in both the classical and the quantum theory it is necessary to fix asymptotic boundary conditions on the metric. The metric is then allowed to fluctuate in the bulk. We introduce a coordinate $r$ so that the conformal boundary is attained as $r \rightarrow \infty$. We then demand that near this boundary
\be\label{Adsmetric}
ds^2 \underset{r \rightarrow \infty}{\longrightarrow} \ell^2 \left(- (1+r^2) dt^2 + {d r^2 \over 1+r^2} + r^2 d \Omega_{d-1}^2 \right)
\ee
where the AdS length $\ell$ 
will be set to one for the rest of the paper.  Note that, in these units, $\k$ is a dimensionless number and we will assume that $\k \ll 1$ which is simply the assumption that the Planck length is much smaller than the cosmological length.

This means that we allow for the standard {\em normalizable} boundary conditions for fluctuations of the metric and matter fields  following \cite{cmp/1103942446, Ashtekar:1999jx, deHaro:2000vlm}, demanding that the metric and matter fluctuations have appropriate falloffs near the boundary.\footnote{It is of interest to consider other kinds of boundary conditions \cite{Freidel:2020xyx,Freidel:2020svx}. However, if the boundary conditions allow energy to escape from AdS, then one-loop effects generically generate a mass for the graviton in the bulk \cite{Aharony:2003qf} leading to a theory that might have qualitatively different properties from standard theories of gravity.}

\subsection{Canonical formalism}

In the canonical formalism for gravity described by ADM \cite{Arnowitt:1962hi}, the line element is written using a $d+1$ split
\be
\label{spacetimesplit}
ds^2 = -N^2 dt^2 + g_{i j} (d x^i + N^i d t) (d x^j + N^j d t)~,
\ee
where $N$ is called the lapse function, and  $N^i$ is called the shift vector. The metric on a Cauchy slice $\S$,  at a fixed value of $t$, is $g_{ij}$ where $i,j,\dots$ run only over the spatial coordinates.

We can rewrite the action as
\be\label{actionRewrite}
S = {1\/2\k^2}\int dt d^{d}x\, N \sqrt{g} \left(K_{i j} K_{k l} g^{i k} g^{j l} - K^2 + R -2 \Lambda \right) + S_{\text{GHY}} + S_{\text{matter}}~,
\ee
using the extrinsic curvature of the slice of constant $t$, given by
\be
K_{i j} = {1 \over 2 N} \left(-\dot{g}_{i j} + D_{j} N_i + D_{i} N_j \right),
\label{extrinsiccurv}
\ee
where $D_i$ is the covariant derivative with respect to $g_{i j}$, $K = g^{i j} K_{i j}$ and $R$ is the Ricci scalar on the slice.

The canonical momentum is defined as
\be
\label{momextrinsic}
\pi^{i j} =  {\d S \over \d \dot{g}_{i j}} = -{1\/2\k^2} \sqrt{g} \left(g^{i l} g^{j k} K_{l k} - g^{i j} K \right)\,.
\ee
The conjugate momenta for the lapse and shift vanish identically leading to the primary constraints \cite{dirac2001quantum} 
\be
\pi_{N} ={\d S \over \d \dot{N}} =  0,\qquad \pi_{N_i} ={\d S \over \d \dot{N}_i}  = 0~.
\label{primaryconst}
\ee
The Hamiltonian can be written in the form 
\be\label{Hcan}
H = H_0 + \Hbdy~,
\ee
where
\be
H_0= \int_\S d^d x \sqrt{g}\,  ( N \cH + N^i \cH_i)~,
\ee
and $\cH$ and $\cH_i$ are given by
\bea\label{HamConstraint}
\cH \=2 \k^2 g^{-1}\le(  g_{i k} g_{j l} \pi^{kl }\pi^{ij} -{1\/d-1}(g_{ij}\pi^{ij})^2\ri)  - {1\/2\k^2}(R-2\L) + \cH^\r{matter} ~,\\\label{MomConstraint}
\cH_i\= -2 g_{ij} D_k {\pi^{jk} \over \sqrt{g}} +\cH_i^\r{matter},
\eea
where $\cH^\r{matter}$ is the matter Hamiltonian density, $\cH_i^\r{matter}$ is the matter momentum density and $\Hbdy$ is a boundary contribution \cite{Regge:1974zd} whose explicit form we give below in \eqref{defHbdy}.
 
 The matter Hamiltonian is obtained in a standard way using canonical quantization. Let us illustrate this in the example of a scalar field of mass $m$, described by the action
\be
S_\text{matter} = -{1\/2}\int  dt d^{d}x \sqrt{g}\, N \le((\p\phi)^2 +m^2\phi^2 \ri)~.
\ee
 the conjugate momentum is $\pi = \sqrt{g} N^{-1}( \p_t\phi- N^i \p_i \phi)$ and the Hamiltonian and momentum density are
\be\label{scalarConstraints}
\cH^\text{matter} =  {1\/2  }g^{-1}\pi^2 + {1\/2}\le( g^{ij} \p_i \phi \p_j\phi+m^2\phi^2\ri)~,\qq \cH_i^\text{matter}= {1\/\sqrt{g}}\,\pi \,\p_i \phi~.
\ee

We obtain secondary constraints by demanding that the primary constraints are preserved by time evolution.  These secondary constraints are nontrivial and are called the Hamiltonian and momentum constraints. They can be described as follows. Let $f$ be any function that dies off smoothly as $r \rightarrow \infty$ and let
\be
\label{smearing}
\cH(f) \equiv \int_{\S} d^d x\, \cH f, \qquad \cH_i(f) \equiv \int_{\S} d^d x \,\cH_i f~.
\ee
Then the Hamiltonian and momentum constraints are
\be
\label{secondconst}
\cH(f) = 0,\qq \cH_i(f)  =0~.
\ee
Note that \eqref{secondconst} are equivalent to imposing $\cH = 0$ and $\cH_i = 0$ at all points except for the conformal boundary. 

The exclusion of the boundary can be understood using a simple physical argument. The constraints \eqref{secondconst} express the diffeomorphism invariance of the theory. But, as is standard in gauge theories, only {\em small diffeomorphisms} --- those diffeomorphisms that vanish smoothly at the conformally boundary --- are redundancies in the description. Large diffeomorphisms --- those diffeomorphisms that act nontrivially at the conformal boundary --- generate physical transformations and should not be viewed as trivial.

\subsection{Quantum theory}
So far our description has been classical. In the quantum theory, the states are given by wavefunctionals
\[
\psifullg~.
\]
Note that, to lighten the notation,  we do not display the indices on $g$ and on other tensors when they appear in an argument of the wavefunctional.  Here, $\phi$ is used as a collective variable for the matter fields in the theory. The wavefunctional returns a complex number upon being given a configuration of the metric and matter fields on the entire Cauchy slice.

The conjugate momenta act on these wavefunctionals via 
\be
\label{momentumopaction}
\pi_{i j} \psifullg = -i {\d \over \d g_{i j}} \psifullg, \qquad \pi \psifullg = -i{\d \over \d \phi} \psifullg~.
\ee
In the quantum theory, we demand that all valid wavefunctionals are annihilated by the constraints. The primary constraints tell us that the wavefunctional is {\em independent} of $N$ and $N_i$ since they imply that
\be
\label{primconst}
{\d \over \d N} \psifullg = 0,\qq {\d \over \d N^i} \psifullg = 0~.
\ee
In the quantum theory, the information about how the $d$-geometries are glued together into a spacetime geometry must be extracted from the canonical momentum and not from the values of $N$ or $N^i$.  If one takes the classical limit in the quantum theory, then the expectation value of the momentum operator can be related to the classical extrinsic curvature via \eqref{momextrinsic}.

Finally, the wavefunctional must be annihilated by the Hamiltonian and momentum constraints
\be
\cH(f)  \psifullg =0,\qq \cH_i(f)  \psifullg =0~.
\ee
These constraints can be understood as imposing the gauge invariance of the wavefunctional in the quantum theory. As usual, we do not impose invariance under large gauge transformations which may act non-trivially on the state. For mixed states, the corresponding condition is that the density matrix must commute with the constraints.

A valid observable in the theory, denoted $\op$, is a Hermitian operator that commutes with the constraints
\be
\label{opconstraint}
[\op, \cH(f)] = 0 ,\qquad [\op, \cH_i(f)] = 0~.
\ee
A simple set of gauge-invariant observables are just given by the {\em boundary limits} of bulk operators. Such observables manifestly satisfy \eqref{opconstraint} because $\cH(f)$ and $\cH_i(f)$ vanish near the boundary. Such observables may depend on the boundary coordinates including the boundary time and, in the discussion below, we display this dependence as  $\obdry(t, \Omega)$. We discuss these observables further in section \ref{secholproof}.

\section{Perturbative expansion \label{secpert}}

In this section, we will expand the constraints in the perturbative regime about the AdS background. We start by introducing the perturbative variables and then proceed to the perturbative expansion. All the derivations described in this section are checked using xAct \cite{xAct} and xPert \cite{Brizuela:2008ra} in a Mathematica notebook associated with this paper \cite{xActgithub}.

\subsection{Perturbative setup}
\paragraph{\bf Metric fluctuation.}
In perturbation theory, we expand the metric as 
\be
\label{hijdefn}
g_{ij} = \g_{ij} + \k h_{ij}~,
\ee
where $\kappa = \sqrt{8 \pi G}$ and the background metric, $\g_{ij}$, corresponds to the metric on a constant time slice of global  AdS$_{d+1}$. 
\be
\label{adsbackground}
 \g_{ij}dx^i dx^j = {dr^2\/1+r^2} + r^2d\Om_{d-1}^2 ~,
\ee
Equation \eqref{hijdefn} should be taken as the definition of the perturbative variable $h_{i j}$.  Note that, for now, this equation is just an exact change of variables although below  we will perform a perturbative expansion in $\kappa$.   
We will find it convenient to represent states  as wavefunctionals of this new variable  using the notation
\[
\psifull~.
\]
\paragraph{\bf Momentum operator.}
It is also convenient, in perturbation theory, to work in terms of the momentum operator
\be
\Pi^{ij} = {\k\/\sqrt{\g}} \pi^{ij}~.
\ee 
In the wavefunctional representation, the action of this operator is just
\be
\label{pilinearized}
 \Pi^{ij} = - {i\/\sqrt{\g}} {\d\ov \d h_{ij}}~,
 \ee
and so this operator is canonically conjugate to $h_{i j}$ up to a factor of ${1 \over \sqrt{\g}}$ that is included so that $\Pi^{i j}$ transforms like a tensor field on the background.

\paragraph{\bf Derivatives and indices.} 
We will use $\n_i$ to denote the covariant derivative associated to the background metric $\g_{ij}$. This should be distinguished from $D_i$ which is the covariant derivative associated to the full metric $g_{ij}$. Furthermore, for the rest of this paper, we will raise and lower indices using only the background metric $\g_{i j}$.  We remind the reader that indices are summed only over the spatial coordinates and if time appears in a formula, it is displayed separately.

\paragraph{\bf Shift and lapse.} The primary constraints imply that neither $N$ nor  $N^i$ enter in any wavefunctional or observable. Nevertheless, in our analysis it will be convenient to fix the background value of $N$ to be
\be
\label{backgroundn}
N^2 = 1+r^2.
\ee
Since  $N$ is not an observable,  the reader can just take equation \eqref{backgroundn} to specify a certain function of the coordinates that will be useful in the analysis.

\paragraph{\bf Background properties.}
In our computations, it will be useful to use the following identities satisfied by the background quantities:
\begin{align}\nt
R_{ijk\l} & = \g_{i\l}\g_{jk} - \g_{ik}\g_{j\l}, & R_{ij} & = -(d-1) \g_{ij}~,\\\label{AdSbackgroundValues}
 R & = -d (d-1), & \n_i\n_j N  &= \g_{ij}N,
\end{align}
and the cosmological constant is $\L  = -d(d-1)/2$. We are using conventions where $\l_\r{AdS}=1$.

\subsubsection{\bf Boundary Hamiltonian.} In terms of the notation introduced above,  the boundary contribution to the Hamiltonian in \eqref{Hcan} takes on a simple form. This can be viewed as an AdS version of the ADM energy. It is given as 
\be\label{defHbdy}
\Hbdy = {1\/2\k} \int_{\p\S} d^{d-1}\Om  \,n_i  J^i~,
\ee
where 
\be\label{defJi}
J_i\equiv  N \n^j(h_{ij} - h \g_{ij}) -\n^j N (h_{ij}-  h\g_{ij}) 
\ee
will be called the ADM current.
We show in Appendix \ref{appEnergy} that this agrees with various prescriptions for the gravitational energy in AdS. Here $\dbdy$ denotes the appropriate measure for boundary integration and $n_i$ denotes the normal to the boundary.\footnote{For concreteness, we can take $d^{d-1}\Om$ to be the volume form of the unit sphere and $n_i = r^{d-1} {\bf n}_i$ where ${\bf n}_i$ is the unit normal to the boundary.}  Note that, in the coordinates \eqref{adsbackground} the area of a sphere at large $r$ grows like $\Or[r^{d-1}]$ which precisely compensates the large $r$ falloff of $J^i$. Also note that as a consequence of  \eqref{secondconst}, the bulk contribution to the energy of any state vanishes. The nonzero contribution to the energy comes only from the boundary term \eqref{defHbdy}.

\subsection{ADM decomposition \label{subsecadmdecomp}}

In order to better understand the Hamiltonian and momentum constraints given in \eqref{HamConstraint} and \eqref{MomConstraint}, it is convenient to use the ADM decomposition of symmetric tensors \cite{Arnowitt:1962hi}. ADM originally introduced this decomposition about flat space, and here we present the generalization to an AdS background. We refer the reader to \cite{york1973conformally} for related discussion.

We decompose the metric perturbation as 
\be
\label{admdecomp}
h_{i j} = \htt_{i j}  + \htr_{i j} + \hlo_{i j}~,
\ee
and the three terms in the sum are called the transverse-traceless component, the T-component and the longitudinal component respectively. We will perform precisely analogous decompositions for other tensor fields below and, in each case, the three components will be labeled by ``TT'', ``T'' and ``L''  as above.

The transverse-traceless component obeys
\be
\label{ttconds}
\nabla^i \htt_{i j} = 0, \qquad \gamma^{i j} \htt_{i j} = 0~. 
\ee
The T-component of the metric is also transverse 
\be
\label{trconds}
\nabla^i \htr_{i j} = 0~,
\ee
but only captures information about the trace of the transverse part of the decomposition. The longitudinal component is of the form
\be\label{hLdiffeo}
h_{ij}^\r{L} = \n_i\e_j + \n_j\e_i ~,
\ee
in terms of an arbitrary vector field $\e_i$ that vanishes at the conformal boundary.

Given any tensor field $h_{i j}$, the decomposition \eqref{admdecomp} is unique and can be obtained by solving a set of elliptic partial differential equations as we now describe. The transversality conditions \eqref{trconds} and \eqref{ttconds} imply that $\e_i$ is obtained as the solution to
\be
\nabla^i \nabla_i \epsilon_j + \nabla^i \nabla_j \epsilon_i = \nabla^i h_{i j}~,
\ee
which has a unique solution for $\e_i$ subject to our boundary conditions and thereby yields $\hlo_{i j}$.
Note that the Killing vectors of the background cannot be added to a solution of the equation above to obtain another solution since they do not vanish asymptotically.

We denote the trace of the transverse part of the metric by
\be
\htr = \gamma^{i j} \left(h_{i j} - \hlo_{i j} \right) = \gamma^{i j} \htr_{i j}~.
\ee
We want the T-component of the metric to depend linearly on the metric, correspond to a single degree of freedom, and vanish when $\htr$ vanishes. This is achieved by introducing an auxiliary scalar field $\chi$ and writing
\be
\htr_{i j} = {1 \over d}\,\htr \gamma_{i j}  -  {1 \over d-1}\left[\nabla_i \nabla_j - {1 \over d} \gamma_{i j} \D \right]\chi~,
\ee
where $\D \equiv \nabla^i \nabla_i$. The condition \eqref{trconds} implies that $\chi$ must obey
\be
\label{alphaeqn}
 (\D-d)\,\chi  = \htr~.
\ee
Once the longitudinal and T-component have been determined as above,  the $\htt$ component of the metric is what remains: $\htt_{i j} = h_{i j} - \hlo_{i j} - \htr_{i j}$. Note that, by construction, the conditions \eqref{ttconds} are met.

It is also clear that the degrees of freedom on both sides of equation \eqref{admdecomp} match. The propagating modes of the graviton are contained in $\htt_{ij}$ and represent $(d+1)(d-2)/2$ degrees of freedom. There are $d$ degrees of freedom in $h_{ij}^\r{L}$ corresponding to the components of $\e_i$ and $1$ degree of freedom in $\htr_{i j}$. This gives a total of $d(d+1)/2$ as appropriate for a symmetric tensor.

The terms in the decomposition \eqref{admdecomp} are orthogonal when contracted and integrated over the Cauchy slice. For instance,
\be
\int_{\S}  d^{d} x \sqrt{\gamma} \,\htt[i j] \hlo_{i j} = -2 \int_{\S} d^d x\sqrt{\gamma}\, \nabla_i \htt[i j]  \epsilon_j  = 0~,
\ee
where we have integrated by parts and utilized \eqref{ttconds}. A similar argument shows that the integral of a T-component with the longitudinal component vanishes. We also find that
\be
\int_{\S} d^{d} x \sqrt{\gamma} \,\htt[i j] \htr_{i j}  = {1 \over 1 - d} \int_{\S} d^{d} x  \sqrt{\gamma} \, \htt[i j] \nabla_i \nabla_j \chi = 0~,
\ee
where in the first step we  used the fact that $\htt[i j]$ is traceless and in the second step we integrated by parts and used the property \eqref{ttconds}.

We now turn to the canonical momenta. Note that \eqref{pilinearized} tells us that $\Pi^{i j}$ is an operator-valued field. Nevertheless we can perform a decomposition similar to \eqref{admdecomp}. We write
\be
\label{pidecomposition}
\Pi^{ij} = \pitt[ij] + \pitr[ij] + \pilo[ij] ~.
\ee
The canonical generator \cite{Arnowitt:1962hi} that induces an infinitesimal shift in the metric fluctuation, $h_{i j} \rightarrow h_{i j} + \zeta_{i j}$, is simply 
\be
G = i \int_\S d^d x\sqrt{\g} \,\Pi^{ij} \zeta_{ij}~.
\ee
Using the orthogonality of the components demonstrated above, it is clear that the canonical generator diagonalizes so that 
\be
G = i \int_\S  d^d x \sqrt{\gamma}\le( \pitt[ij] \ztt_{ij}+\pitr[ij] \ztr_{ij}+\pilo[ij] \zlo_{ij}\ri),
\ee
which implies that 
\be
\label{pidecomp}
\pitt[ij] = - {i \over \sqrt{\gamma}} {\d\ov \d \htt_{ij}}~,\qq 
\pitr[ij] = - {i \over \sqrt{\gamma}}  {\d\ov \d \htr_{ij}}~,\qq
\pilo[ij] = - {i \over \sqrt{\gamma}}  {\d\ov \d \hlo_{ij}}~.
\ee

\subsection{Expansion of the constraints}

In this section, we present the perturbative expansion of the constraints. A similar analysis was performed in \cite{Kuchar:1970mu} about Minkowski space.

\subsubsection{Momentum constraint}

Let us start by considering the momentum constraint \eqref{MomConstraint}. We consider successive approximations to the constraint which we write as
\be
\sqrt{g}\,  \cH_i  =\sqrt{\g} \,\cH^{(n)}_i + \Or[\k^{n-1}],\qq n=0,1,2,\dots
\ee
by which we mean that $\cH^{(n)}_i$ captures all terms in the expansion of the left hand side up to terms of order $\k^{n-2}$. 

The zeroth order term in the momentum constraint vanishes trivially,
\be
\cH_i^{(0)} = 0~.
\ee

\pg{First order.} At leading order, the momentum constraint takes the form
\be
\cH^{(1)}_i  =  -{2 \over \kappa} \g_{ij} \n_k{\Pi^{jk}}~.
\ee
The momentum constraint simply tells us that the wavefunctional is independent of $\hlo_{i j}$ to leading order. This can be seen as follows. Consider the infinitesimal gauge transformation $x^i \rightarrow x^i + \xi^i$. Then we see that 
\be
\begin{split}
\Psi[h_{i j} + \n_i \e_{j} + \n_j \e_i, \phi] &= \psifull  + \int_\S d^d x\, (\n_i \e_{j} + \n_j \e_i)(x) {\d \over \d h_{i j}(x)} \psifull \\ 
&= \psifull -2 i \int_\S d^dx\sqrt{\g}\, \e_j(x) \n_i  \Pi^{i j}(x) \psifull = \psifull
\end{split}
\ee
at leading order in $\e_i$, where we have used the leading-order momentum constraint in the last equality.

Alternately, this can also be seen from the decomposition \eqref{pidecomposition}. The leading order momentum constraint tells us that
\be
-2 \g_{i j}  \n_k \pilo[j k] \,\psifull = 0 + \Or[\kappa].
\ee
which is equivalent to $\pilo[j k] \psifull = 0 + \Or[\kappa].$

\pg{Second order.} At next order, we have
\be
\cH_i^{(2)} = ( \n_i h_{jk}  -2 \n_k h_{ij})\Pi^{jk} -2 h_{ij} \n_k\Pi^{jk} - {2\/\k} \g_{ij} \n_k\Pi^{jk}+ \cH_i^\r{matter}~.
\ee
The first order constraint implies that $ h_{ij} \n_k\Pi^{jk}  = \Or[\k]$ which is subleading. We can then rewrite the constraint as
\be\label{secondOrderMomentum}
{2\/\k} \g_{ij} \n_k\Pi^{jk}_\r{L} = Q_i
\ee
where we have defined
\be\label{defQi}
Q_i\equiv ( \n_i h_{jk}-2 \n_k h_{ij})\Pi^{jk} + \cH_i^\r{matter}~,
\ee
and $\cH_i^\r{matter}$ is the contribution of the matter to momentum constraint. For a free scalar field, we have from  \eqref{scalarConstraints}
\be
\cH_i^\r{matter}= {1\/\sqrt{\g}}\pi \,\p_i\phi~.
\ee
This shows that the second order momentum constraint determines the $\Or[\k]$ part of $\Pi_\r{L}^{jk}$  in terms of $\Or[1]$ quantities.

\subsubsection{Hamiltonian constraint}

We now consider the perturbative expansion of the Hamiltonian constraint. We consider successive approximations
\be\label{HconstraintExpansion}
\sqrt{g}\,   \cH  = \sqrt{\g} \, \cH^{(n)} + \Or[\k^{n-1}],
\ee
by which we mean that $\cH^{(n)}$ includes all the terms from the Hamiltonian constraint up to terms of order $\k^{n-2}$.

At zeroth order, we simply have
\be
\cH^{(0)} = -{1\/2\k^2}(R-2\L)~.
\ee
Plugging in the values from \eqref{AdSbackgroundValues}, we see that this term vanishes identically:  $\cH^{(0)} = 0$.

\pg{First order.}
At first order, we obtain
\be\label{Hpert1}
N \cH^{(1)} =-{1\/2\k} N\le( \n^i \n^j h_{ij} -\n_i\n^i h +(d-1) h \ri) = -{1\/2\k}\n^i J_i~,
\ee
which we have written as a total derivative in terms of the ADM current \eqref{defJi}. Note the factor of $N$ that we have inserted on the LHS of \eqref{Hpert1}. It is only with this factor that the expression turns into a total derivative, and this fact will play an important role in the analysis below.

In the decomposition \eqref{admdecomp}, it can be seen that this expression \eqref{Hpert1} involves only $\htr_{ij}$ and not $\htt_{ij}$ or $\hlo_{ij}$. It is clear that $\htt_{ij}$ disappears  because of the transverse-traceless condition. The longitudinal component also disappears from this expression.  This can be checked explicitly from \eqref{hLdiffeo} by evaluating $\cH^{(1)}$ on \eqref{hLdiffeo} and commuting the covariant derivatives and using the background identities \eqref{AdSbackgroundValues}.
This can also be understood from the fact that, at first order, the longitudinal component corresponds to an infinitesimal spatial diffeomorphism. Hence, it doesn't change the Ricci scalar which is constant according to \eqref{AdSbackgroundValues}. This implies that $\cH^{(1)}$ doesn't depend on $\hlo_{ij}$. The end result is that
\be
\label{honlytdepend}
N \cH^{(1)} ={1\/2\k} N\le( -\n_i\n^i h^\r{T} +(d-1) h^\r{T} \ri) .
\ee
where we denote $\htr = \g^{ij}\htr_{ij}$. Since $h^\r{T}_{ij}$ has only one degree of freedom, which can be taken to be $h^\r{T}$, the first order Hamiltonian constraint implies that
\be
\htr_{i j} = 0 + \Or[\kappa].
\ee
In the sections below, we will work out aspects of the $\Or[\kappa]$ correction to this equation, which will play a central role in our analysis.

\paragraph{Second order.}

At second order, we have contributions from the term quadratic in $\Pi_{ij}$ and the matter stress tensor:
\be
N \sqrt{\g}\, \cH^{(2)} = 2 N  \sqrt{\g}\le(  \Pi^{ij}\Pi_{ij} -{1\/d-1}\Pi^2\ri) - {1\/2} N \le[\sqrt{g} (R-2\L)\ri]^{(2)} +  N \sqrt{\g} \,\cH^\text{matter}~,
\ee
and we should expand the term in the brackets to second order in $\k$.  The expansion is performed in the accompanying Mathematica script \cite{xActgithub}. It leads to many terms which we can organize as 
\bea\label{expHamSecond}
{-}N\le[ \sqrt{g} (R-2\L) \ri]^{(2)} \= {1\/4}\sqrt{\g}N \le( -  h^{ij} (\D_N +2)h_{ij} + h (\D_N-d) h \ri) \-
&&+{1\/2}\sqrt{\g}N \le(  2 h^{ij}\n_i \n^k h_{jk} +  \n_i h^{ij} \n^k h_{jk}+ 2 \n_i h \n_j h^{ij} +  h \n_i\n_j h^{ij}\ri)\-
 &&+ {1\/2} \sqrt{\g} \n_i L^i- {1\/2\k} \n_i J^i ~,
\eea
where we have introduced a Laplace-type operator
\be\label{DeltaN}
\D_N h_{ij} = N^{-1} \n_k (N \n^k h_{ij})~.
\ee
The total derivative involves a current
\be
\label{lidef}
L_i \equiv -N \n^j L_{ij} + L_{ij}\n^j N + {1\/2}N \le( h_{jk} \n_i h^{jk} -  h\n_i h\ri)~,
\ee
where we have defined
\be
L_{ij} \equiv 2  h h_{ij} - h_{ik} h_j^{~k} +\g_{ij} h_{k\l}h^{k\l} -{1\/2}\g_{ij} h^2~.
\ee
There is also a contribution from the ADM current $J^i$  evaluated on the $\Or[\k]$ part of $h_{ij}$, the order one part being zero by the first order constraint. Finally, the second order Hamiltonian constraint takes the form
\be
\label{finalform}
N \cH^{(2)} = N Q  - {1\/2\k} \n_i J^i~,
\ee
where we have defined
\be\label{defQ}
\begin{split}
Q &\equiv 2  \le(  \Pi^{ij}\Pi_{ij} -{1\/d-1}\Pi^2\ri)  +{1\/8} \le( -  h^{ij} (\D_N +2)h_{ij} + h (\D_N-d) h  \ri) + {1\/4}\n_i L^i \\
&+{1\/4} \le(  2 h^{ij}\n_i \n^k h_{jk} +  \n_i h^{ij} \n^k h_{jk}+ 2 \n_i h \n_j h^{ij} +  h \n_i\n_j h^{ij}\ri)+  N  \cH^\text{matter}~.
\end{split}
\ee

\subsection{Integrated constraint \label{subsecintegconst}}

We will find it very useful to also consider the integrated  second order Hamiltonian constraint
\be
H_0^{(2)} = \int_\S d^d x\sqrt{\g } \,N\cH^{(2)}~.
\ee
It is important to perform the integral with the measure that defines the canonical Hamiltonian, \ie with a factor of $N$ as shown above.   In this section, we show that the complicated expression obtained in \eqref{finalform} and \eqref{defQ} simplifies greatly upon integration. 

To show this, we will use the ADM decomposition \eqref{admdecomp} and \eqref{pidecomposition}.  We will also use the fact that, as shown above, the first order constraints will set $\htr_{i j}= \Or[\k]$ and $\pilo[ij] = \Or[\k]$. So we drop terms where $\htr_{i j}$ and $\pilo[i j]$ multiply another $\Or[1]$ quantity since this allows us to avoid writing a number of unnecessary terms that will eventually not be relevant for our analysis.

First, the integrated constraint becomes independent of $h^\r{L}_{ij} = \n_i \e_j+\n_j\e_i$.  This is trivial at first order  because $h^\r{L}_{ij}$ corresponds to an infinitesimal diffeomorphism. At second order, the cancellation is non-trivial and quite remarkable\footnote{This can be understood as follows. The Ricci scalar is constant on the background, so it is invariant under the diffeomorphism $x_i \ra x_i + \k\e_i$. As a result, the second order Hamiltonian constraint is also invariant under that diffeomorphism, as the variation of $\sqrt{\g}$ and $N$ can be ignored because we assume $\cH^{(0)} = \cH^{(1)}=0$. This diffeomorphism modifies the metric according to $h_{ij} \ra h_{ij}+\n_i\e_i+\n_j\e_i+ \k h^{(2)}_{ij} + O(\k^2)$. At linearized order, this generates an arbitrary $h^\r{L}_{ij}$ and shows that the first order constraint is independent of the longitudinal metric. At second order, we also generate a subleading term whose explicit expression is $h^{(2)}_{ij} =\n_i \e_k\n_j\e^k-\e_i\e_j + \g_{ij} \e_k\e^k +\cL_\e h_{ij}^\r{TT}$. By applying the above diffeomorphism to the constraint at $h^\r{L}=0$, we obtain
\be
\le. N\cH^{(2)}\ri|_{h^\r{L}=0} = N\cH^{(2)} - {1\/2} \n_i J^i[h^{(2)}]~,
\ee
where we used the fact that since $\cH^{(2)}$ captures terms up to $\Or[\kappa^0]$ the subleading term $h^{(2)}$ can only affect it through those terms that have an explicit factor of $\k^{-1}$. This shows the dependence on $h^\r{L}$ in $N\cH^{(2)}$ is indeed captured by a total divergence.}. It follows from the fact that we can write the constraint as
\be
\label{nch2rewrite}
\begin{split}
N \cH^{(2)} &= 2 N \le(  \Pi^{ij}\Pi_{ij} -{1\/d-1}\Pi^2\ri)  -{1\/8}N\, h^{ij}_\r{TT} (\D_N +2)h_{ij}^\r{TT}   +  N  \cH^\text{matter} \\
&+ {1\/2} \n_i M^i + {1\/4}\n_i L^i[h^\r{TT}]   - {1\/2\k} \n_i J^i + \Or[\htr],
\end{split}
\ee
where the $\Or[\htr]$ term is subleading as explained above, and where the dependence in $\e_i$ is fully captured by the divergence of the following current
\be
\label{defMi}
\begin{split}
M^i &= N \le( - \n_j M^{ij} + \htt_{jk} \n^j\n^k \e^i-2 \n^i \htt_{jk} \n^j \e^k - \htt[ij] \n_j \n_k \e^k + 2 d \htt[ij] \e_j\ri) \\
&+ \n_j N \le( M^{ij} + \htt[ij] \n_k\e^k - \htt[ik] \n_k \e^j + 2 \e_k \n^j \htt[ik]\ri),
\end{split}
\ee
where we have defined
\be
M^{ij}\equiv \e^i \e^j + \e^k \n^j \n^i \e_k + \g^{ij}\le( (d-2) \e_k\e^k - \e_k \D \e^k\ri).
\ee
Above, the symbol $L^i[h^\r{TT}]$ means that  \eqref{lidef} is evaluated only on $h^\r{TT}_{ij}$ and this evaluation reduces to
\be
L^i[h^\r{TT}]= N\, h_{jk}^\r{TT} \n^k h^{ij}_\r{TT}  -{3\/2} N \, h^{jk}_\r{TT}\n_i h_{jk}^\r{TT} + \n^i N \, h_{jk }^\r{TT} h^{jk}_\r{TT} -\n^j N \, h^{ik}_\r{TT} h_{jk}^\r{TT} ~.
\ee
The validity of this rewriting is checked in the associated Mathematica notebook  \cite{xActgithub}.

From equation \eqref{nch2rewrite} it is clear that the integration of  $N \cH^{(2)}$ over the entire Cauchy slice leads to boundary terms that involve  $M^i, L^i[\htt]$ and $J^i$.  The terms involving $M^i$ and  $ L^i[h^\r{TT}]$ are quadratic in the metric fluctuation and since we have imposed normalizable boundary conditions their decay at large $r$ is faster than the growth of the area of the sphere.  Therefore the boundary contribution from these terms vanishes.  On the other hand, the boundary term  involving $ J^i$, upon integration over the boundary, gives  the ADM energy $\Hbdy$.

Let's now consider the kinetic piece\be
 2 \int_\S d^d x \sqrt{\g} \, N \le(  \Pi^{ij}\Pi_{ij} -{1\/d-1}\Pi^2\ri) ~.
\ee
To analyze the term quadratic in  $\pitr[i j]$ it is convenient to write the decomposition of section \ref{subsecadmdecomp} as
\be\label{exprpiT}
\pitr[][ij] = {1\/d-1}\le( \g_{ij}  \pitr - \a_{ij}\ri)
\ee
where 
\be
\label{alphaijsolution}
 \a_{ij} = N \n_i\n_j \a+ \n_i N \n_j\a+ \n_j N \n_i\a
\ee
in terms of a scalar operator-valued field $\a$ that satisfies the analogue of \eqref{alphaeqn} for $\pitr$:
\be\label{alphaPiT}
 (\D-d)(N\a)  = \pitr~.
\ee
From the expression \eqref{exprpiT}, we see that the term quadratic in $\pitr[i j]$ in the kinetic piece can be written
\be
\int_\S d^d x \sqrt{\g} \, N \le(  \pitr[][ij] \pitr[i j] -{1\/d-1}\Pi^2_\r{T}\ri)  =  -{1\/d-1}\int_\S d^d x\sqrt{\g}\, N \a_{ij} \pitr[ij]~.
\ee
Using \eqref{alphaijsolution},  we can write
\be
N \a_{ij} = \n_i\a_j + \n_j\a_i ,\qq \a_i\equiv {1\/2}N^2 \n_i\a~,
\ee
and we finally obtain
\be
\label{pitvanishes}
\int_\S d^d x \sqrt{\g} \, N \le(  \pitr[ij] \pitr[][i j] -{1\/d-1}\Pi^2_\r{T}\ri)  =  -{2\/d-1}\int_\S d^d x\sqrt{\g}  \, \n_i \le( \pitr[ij]\a_j\ri),
\ee
which becomes a boundary term. Since $\pitr$ vanishes at the boundary, the boundary term vanishes.\footnote{More precisely, we only need $\pitr < \Or[r^{(d-4)/2}]$ for the boundary term to vanish.} In a similar way, we can show that the cross terms involving $\pitt[ij]$ and $\pitr[ij]$ vanish. Recall that $\pilo[ij]$ does not appear since it vanishes at $\Or[1]$ in perturbation theory by the first order  momentum constraint.  This shows that $\pitr[ij]$ disappears from the integrated constraint.

Finally, the integrated Hamiltonian constraint takes the form
\be\label{Hcan2}
H_\r{0}^{(2)} =  - \Hbdy+\int_\S d^d x\sqrt{\g} \,N \hbulk
\ee
where
\be
\label{defhbulk}
\hbulk  = 2 \, \pitt[][ij] \pitt[ij] -{1\/8}\, h^{\r{TT}ij}( \D_N+2 ) h_{ij}^\r{TT}+  \cH^\text{matter}
\ee
and the explicit expression of $\Hbdy$ is given in \eqref{defHbdy}. The constraint $H_{0}^{(2)} = 0$ can be understood as the equality of the ADM energy $\Hbdy$ with a bulk energy defined as the second term of \eqref{Hcan2}. Here we see that this relationship follows naturally from the Hamiltonian constraint. 
\section{Solving the constraints \label{sec:solving}}

We now describe how the constraints discussed in the previous section can be solved to reveal a remarkable structure of correlations in gravitational wavefunctionals.  

The analysis of section \ref{secpert} immediately yields solutions to the first order constraints. We find from the first order Hamiltonian constraint that
\be
\label{firstordersol}
\cH^{(1)} \psifull = 0 \Rightarrow \htr \psifull = 0 + \Or[\kappa].
\ee
This equation should be interpreted as telling us that the {\em support} of the wavefunctional is negligible when the value of $\htr$ is parametrically larger than $\Or[\kappa]$. The first order momentum constraint tells us that
\be
\cH^{(1)}_i \psifull = 0 \Rightarrow {\d \psifull \over \d \hlo_{i j}} = 0 + \Or[\kappa].
\ee
The interesting features in the solutions appear at the next order in perturbation theory, and this is what we will focus on.

\subsection{Integrated Hamiltonian constraint \label{subsecintham}}
We first describe how to solve the integrated Hamiltonian constraint \eqref{Hcan2}. Here we look for wavefunctionals $\psiint$ with a specified dependence on $\htt, \phi$ and $\Hbdy$. The reason it is possible to restrict to only these variables  is that, as explained in section \ref{subsecintegconst}, the other degrees of freedom all drop out of the integrated Hamiltonian constraint. Note that $\Hbdy$ corresponds to a single degree of freedom from $\htr_{i j}$ as can be seen from  \eqref{honlytdepend} and \eqref{defHbdy}.  In section \ref{sec:pointwise}, we describe how the remaining dependence on $\htr_{i j}$ and $\hlo_{i j}$ can be fixed in the full wavefunctional.

We remind the reader that a factor of  $\kappa^{-1}$ is implicit in the definition of $\Hbdy$, which can be seen in \eqref{defHbdy}. Therefore, in perturbation theory, it is natural to think in terms of $\kappa \Hbdy$. The first order solutions to $H_\r{0}^{(2)} = 0$  all have a {\em degenerate} value of $\kappa \Hbdy = 0 + \Or[\kappa]$ by  virtue of equation \eqref{firstordersol}.  To work out the solution at $\Or[\kappa]$ is a standard problem in degenerate perturbation theory. We need to look for solutions that diagonalize the ``perturbation'' in \eqref{Hcan2}, which is the bulk term. 
\be
\label{bulkisE}
\int_\S d^d x \sqrt{\g}N \, \hbulk\, \psiint = E\, \psiint~.
\ee
Here the eigenvalue of the integrated bulk term is $E$ and the superscript $\{a\}$ simply reminds us that the eigenvalues of the bulk Hamiltonian operator are degenerate and the wavefunctional is not completely specified by only a value of $E$.
Then, the integrated constraint implies that 
\be\label{ADMenergyE}
\Hbdy\, \psiint = E\, \psiint~.
\ee
Since the integral $\hbulk$ depends only on the propagating degrees of freedom $h^\r{TT}_{ij}$ and $\phi$, as explained in section \ref{subsecintegconst}, it  is useful to introduce an \emph{auxiliary} Hilbert space of wavefunctionals  that depend only on $h^\r{TT}_{ij}$ and $\phi$. We will see that these states form a Fock space. 

In this auxiliary space, equation \eqref{bulkisE} simply becomes 
\be
\int_\S d^d x \sqrt{\g}N \,\hbulk\, \psifockE = E\, \psifockE~.
\ee
The equation above is the same as \eqref{bulkisE} except that the wavefunctional has no dependence of $\Hbdy$.

We show below that this can be solved by taking a factorized basis of wavefunctionals that depend, respectively, on only the transverse-traceless metric fluctuation and the matter fluctuation.
\be
\label{factbasis}
\psifockE = \psigrav \psimat ~,
\ee
where
\begin{align}
\label{eigenHgraviton} & \int_\S d^d x\sqrt{\g} N \le[ 2  \pitt[ij] \pitt[][ij] -{1\/8}\, h^{\r{TT}ij}(    \D_N+2 ) h_{ij}^\r{TT} \ri]\psigrav = E_\r{g}\, \psigrav~,\\ 
&\int_\S d^d x\sqrt{\g} N \, \cH^\text{matter}\psimat = E_\r{m}\,\psimat~, \label{eigenHmatter}
\end{align}
with $E = E_\r{g}+ E_\r{m}$. Solutions to \eqref{eigenHgraviton} and \eqref{eigenHmatter} are also degenerate although we have suppressed additional labels on the right hand side of \eqref{factbasis} to lighten the notation. 

In the subsections below, we describe, in some detail, the solutions to  \eqref{eigenHgraviton} and \eqref{eigenHmatter}. The eigenvalues $E$ in \eqref{bulkisE} are obtained after introducing a {\em normal ordering} prescription to regulate $\hbulk$. We specify this prescription below.

Here we note that once these solutions are found, the solution to the integrated Hamiltonian constraint is simply
\be
\label{solint}
\psiint = \psifockE \otimes |\Hbdy=E \rangle.
\ee
Our choice of notation above emphasizes that the spectrum of $\Hbdy$, which is a single-degree of freedom, is discrete.  

Any linear combination of solutions of the form \eqref{solint} is also a solution. The solution \eqref{solint} will be very important for our analysis since it shows how the Hamiltonian constraint, at second order,  {\em correlates} the energy of the dynamical degrees of freedom in the wavefunctional to the value of $\Hbdy$, which is given by an integrated component of the metric.

We will see later that the constraints fully determine the physical state $\psifullE$. The full wavefunctional is obtained by dressing the Fock state $\psifockE$  with the appropriate $h^\r{T}_{ij}$ and $h^\r{L}_{ij}$ dependence, as will be explained in section \ref{sec:pointwise}. 

\subsubsection{Graviton wavefunctionals \label{subsecgravwave}}

We will describe here the solutions of \eqref{eigenHgraviton}. From the integrand appearing in equation \eqref{eigenHgraviton}, it is natural to consider a basis of transverse-traceless eigenfunctions $h^{(n)}_{ij}$ satisfying
\be\label{eigenprobWDW}
{-}N^2(\D_N +2) h_{ij}^{(n)} = \w_n^2 h_{ij}^{(n)}.
\ee
As shown in Appendix~\ref{App:EigenProblem}, this eigenvalue problem is equivalent to the standard quantization of the graviton in global AdS$_{d+1}$.  We will take this basis to be normalized with respect to the inner product
\be
\label{orthhnrel}
{1\/2}\int_\S d^d x\sqrt{\g}\, N^{-1}\,h^{(m)}_{ij} h^{(n)ij} = \d_{mn}~.
\ee
We then use the decomposition 
\be
\htt_{i j} = \sum_n c_n h_{ij}^{(n)},
\ee
and the variables $c_n$ can be written, using the orthogonality condition above as
\be
c_n = {1 \over 2} \int_{\S}  d^{d} x\sqrt{\gamma}\, N^{-1} \htt_{i j} h^{(n) i j} ~.
\ee
Using the chain rule we see that
\be
 \pitt[ij] \,\psigrav = -{i \over \sqrt{\gamma}} {\d \/\d \htt_{i j} }\psigrav = -{i \over 2 N} \sum_n {\partial \psigrav \over \partial c_n} h^{(n) i j}~,
\ee
so that we have
\be
\begin{split}
2 \int_\S d^d x\sqrt{\g}N \, \pitt[ij] \pitt[][ij]\psigrav &= -{1\/2}\int_{\S} d^d {x}\sqrt{\gamma} N^{-1} \sum_{n,m} {\partial^2 \psigrav \over \partial c_n \partial c_m} h^{(n) i j} h^{(m)}_{i j}  \\
&= -\sum_n {\partial^2 \psigrav \over \partial c_n^2}~,
\end{split}
\ee
where we have again used the orthogonality relation \eqref{orthhnrel}.

Then equation \eqref{eigenHgraviton} reduces to
\be
\sum_n\le( {-}{\p^2\/\p c_n^2} + {1\/4}c_n^2 \w_n^2 \ri) \psigrav =  E_\r{g}\,\psigrav~.
\ee
We define the raising and lowering operators
\be
A^\dg_n ={1 \over \sqrt{\omega_n}} \left( {\p\/\p c_n} - {1\/2}\w_n c_n \right),\qq A_n =-{1 \over \sqrt{\omega_n}} \left( {\p\/\p c_n} +{1\/2} \w_n c_n \right),\qq [A_n,A^\dg_m ] =\d_{mn}~.
\ee
We also assume that $\hbulk$ should be normal ordered so that all annihilation operators, $A_n$ are placed to the right of creation operators $A^{\dg}_n$. With this simplification the constraint becomes
\be
\sum_n  \omega_n A_n^\dg A_n  \,\psigrav =  E_\r{grav}\, \psigrav~.
\ee
Our normal ordering prescription ensures that the energy vanishes for the vacuum, which is defined as
\be
A_n\, \psivacg= 0~,\quad \text{for all } n~.
\ee
This is the vacuum wavefunctional for the transverse-traceless gravitons. It has the expression
\be
\psivacg=\cN  \prod_n \exp\le( -{1\/4} \w_n c_n^2 \ri)= \cN\exp\le( - {1\/8}\int_\S d^d x\sqrt{\g} \,h^{\r{TT}ij}\sqrt{- (\D_N+2)}\, h_{ij}^\r{TT} \ri)~
\ee
up to a normalization constant $\cN$ that we specify below. In the flat space limit, this reproduces the results of \cite{Kuchar:1970mu} obtained using a similar method, or in \cite{Hartle:1984ke} from a Euclidean path integral.

The space of solutions is then a Fock space spanned by states of the form
\be
\label{excitedgrav}
\psigrav = 
{1 \over \prod_{i} \sqrt{d_i!}} (A_{n_1}^\dg)^{d_1} (A_{n_2}^\dg)^{d_2} \dots \psivacg ~,
\ee
with energy 
\be
 E_\r{g} =  \sum_i d_i \,\w_{n_i}~.
\ee
We have written the wavefunctionals that appear in equation \eqref{excitedgrav} in terms of the action of operators on the vacuum wavefunctionals. But they can also be written, as usual,  in terms of Hermite polynomials. Note that the validity of perturbation theory requires that, in the Fock space, \eqref{excitedgrav} we restrict attention to states where $\omega_{n_i} \ll {1 \over \kappa}$.

The natural measure on this space of wavefunctionals is simply
\be
\label{gravmeasure}
D \htt =\prod_n d c_n~,
\ee
and we choose the normalization constant $\cN$ so that with respect to this measure the wavefunctionals that appear in \eqref{excitedgrav}  are unit normalized
\be
\innerp[\psigravnoarg, \psigravnoarg] \equiv \int D \htt \psigrav \psigrav^*  = 1 ~.
\ee
Of course, wavefunctionals that differ by even a single value of $d_i$ in equation \eqref{excitedgrav} are orthogonal.

\subsubsection{Matter wavefunctionals \label{subsecmatterwave}}

The matter part of the wavefunctional can be obtained in a similar way as for the transverse-traceless gravitons. To illustrate this, we will consider a minimally coupled massive scalar field. From \eqref{scalarConstraints}, it follows that the canonical Hamiltonian is
\be
\cH^\text{matter}={1\/2}  \int_\S d^d x\sqrt{\g} N  \le(\g^{-1}\pi^2 - \phi (\D_N -m^2)\phi \ri)~,\-
\ee
where we have imposed a normalizable boundary condition at infinity for the scalar field to remove a boundary term. The operator $\D_N$ appearing here is the same as in \eqref{DeltaN}.

We consider eigenfunctions $\phi^{(n)}$ satisfying
\be\label{eigenScalar}
{-}N^2(\D_N -m^2) \phi^{(n)} = \tilde\w_n^2 \phi^{(n)}~,
\ee
normalized so that
\be
\int_\S d^d x\sqrt{\g} N^{-1}\,\phi^{(m)} \phi^{(n)}  = \d_{mn}~.
\ee
Using that
\be
\pi(x) = - i {\d\/\d\phi(x)}~,
\ee
we can perform the same analysis as for the graviton. We obtain a Fock space constructed from the frequencies $\tilde\w_n$.

We can check that the Wheeler-DeWitt analysis reproduces the correct frequencies by considering the equation of motion in the full spacetime
\be\label{scalarfulleq}
(\hat\Box-m^2)\phi=0~,
\ee
which becomes on the Cauchy slice,
\be
{-}N^{-2} \p_t^2\phi + (\D_N-m^2)\phi = 0~.
\ee
In the same manner as for the graviton, this shows that $\tilde\w_n$ as defined in \eqref{eigenScalar} are indeed the frequencies obtained from \eqref{scalarfulleq}.

In global AdS$_{d+1}$ with normalizable boundary conditions, the resulting spectrum is  \cite{Aharony:1999ti}
\be
\tilde\w_n = \D+\l +2n,\qq n\in \bZ_{\geq 0}~.
\ee
where $\l$ labels a spherical harmonic of $S^{d-1}$ with eigenvalue $\l(\l+d-2)$ and the conformal dimension is 
\be
\label{conformaldim}
\D=  {1\/2}(d+\sqrt{d^2+4m^2})~.
\ee
In precise analogy with the analysis above,  we expand the matter field as
\be
\phi=\sum_{n} \tilde{c}_{n} \phi^{(n)}.
\ee
The equation \eqref{eigenHmatter} then reduces to
\be
{1 \over 2} \sum_n\le( {-}{\p^2\/\p \tilde{c}_n^2} + \tilde{c}_n^2 \w_n^2 \ri) \psimat =  E_\r{m}\psimat~.
\ee
We then define
\be
\tilde{A}^\dg_n = {1 \over \sqrt{2 \tilde{\omega}_n}} \left({\p\/\p \tilde{c}_n} - \tilde{\w}_n \tilde{c}_n \right),\qq \tilde{A}_n =-{1 \over \sqrt{2 \tilde{\omega}_n}}  \left({\p\/\p \tilde{c}_n} + \tilde{\w}_n \tilde{c}_n \right) ,\qq [\tilde{A}_n,\tilde{A}^\dg_m ] = \d_{mn}~,
\ee
and the vacuum wavefunctional, which is annihilated by all the $\tilde{A}_n$ operators is given by
\be
\label{excmatstates}
\psivacmat=  \wt{\cN}  \prod_n \exp\le( -{1\/2} \tilde{\w}_n \tilde{c}_n^2 \ri) = \wt{\cN} \exp\le( - {1\/2}\int_\S d^d x\sqrt{\g} \,\phi \sqrt{- (\D_N-m^2)}\, \phi \ri)~,
\ee
where $\wt\cN$ is a normalization constant. Once again,  excited states can be obtained by acting with creation operators:
\be
\psimat     = 
{1 \over \prod{\sqrt{d_i!}}} (A_{n_1}^\dg)^{d_1} (A_{n_2}^\dg)^{d_2} \dots \psivacmat ~.
\ee
As above, we normal order the matter contribution to $\hbulk$ so that the annihilation operators $A_{n}$ are placed to the right of the creation operators $A_n^{\dg}$. With this convention, the energy is given by
\be
 E_\r{m} =  \sum_i d_i \,\tilde{\w}_{n_i}~.
\ee
We remind the reader that, as in the case of graviton wavefunctionals, within perturbation theory,  we are restricted to states of the form \eqref{excmatstates} where $\tilde{\omega}_{n_i} \ll {1 \over \kappa}$.
The natural measure on this space of wavefunctionals is simply
\be
\label{matmeasure}
D \phi =\prod_n d \tilde{c}_n~,
\ee
and we choose the normalization constant so that the wavefunctionals are unit normalized
\be
\innerp[\psimatnoarg, \psimatnoarg] \equiv \int D \phi\, \psimat \psimat^*   = 1 ~.
\ee
As above, wavefunctionals that differ by even a single value of $d_i$ in the expression \eqref{excmatstates} are orthogonal. 

The analysis of matter wavefunctionals completes our analysis of the auxiliary Fock space. These wavefunctionals can be combined with the transverse-traceless graviton wavefunctionals obtained above as displayed in \eqref{factbasis}. The resulting wavefunctional enters the solution of the integrated Hamiltonian constraint displayed in \eqref{solint}.

\subsection{Pointwise constraints}\label{sec:pointwise}

In the previous section, we have described how to solve the integrated Hamiltonian constraint. However, the Hamiltonian and momentum constraints, displayed in \eqref{secondconst}, actually comprise an infinite set of constraints --- one for each spacetime point. In this section we will present the leading order solution to these pointwise constraints. We will also describe a procedure to obtain higher order solutions.

We show in section \ref{secholproof} that the main result of this paper --- which is that wavefunctionals that coincide near the boundary must also coincide in the bulk --- does {\em not} require the detailed form of the dependence of the wavefunctionals on $\hlo$ and $\htr$ in the bulk. For us, it is only important that the pointwise constraints can be used to  {\em uniquely} lift a solution of the integrated Hamiltonian constraint displayed in \eqref{solint} to a solution of the full constraints \eqref{secondconst}. So, in the bulk of this section, we focus on a  procedure that makes it evident that the pointwise constraints can be used to perturbatively fix the dependence of the wavefunctional on the pointwise values of $\htr_{i j}$ and $\hlo_{i j}$.  In the solution \eqref{solint}, it was only the dependence on $\Hbdy$ --- which is the integral of a particular component of $\htr_{i j}$ --- that was fixed. Therefore our procedure  leads to the following uplift.
\be
\label{schemeuplift}
\psiint \xrightarrow{\text{pointwise~constraints}} \psifullE~.
\ee
In section \ref{secindirectargument} we then provide an indirect argument that leads to the same conclusion: namely that the uplift \eqref{schemeuplift} can be performed uniquely.  The details, and checks of the explicit solution itself are presented in Appendix \ref{appleadingorder}.

\subsubsection{Rewriting the pointwise constraints}
We start by putting the pointwise Hamiltonian and momentum constraint in a  convenient form.   In this section, we often display the dependence of the wavefunctional on the individual components of the ADM decomposition of the metric fluctuation and momenta using notation like $\psifullexp$.

\paragraph{\bf Hamiltonian constraint.}
First, consider the second order Hamiltonian constraint. From expression \eqref{finalform} the Hamiltonian constraint is equivalent to
\be\label{HamconstDij}
\cD^{i j} h_{i j}(x) \psifullexp  = \kappa\, Q(x) \psifullexp~,
\ee
where we have defined the differential operator
\be\label{dijexplicit}
\cD^{i j}   \equiv {1 \over 2}  \left(\n^i \n^j - \gamma^{i j} \n_k \n^k + (d-1) \gamma^{i j} \right) 
\ee
and $Q$ is defined in \eqref{defQ}.

As explained near \eqref{honlytdepend}, the LHS of \eqref{HamconstDij} only depends on $\htr$ since the operator $\cD^{i j}$ annihilates the longitudinal and the transverse-traceless components. So we can also write the equation above as
\be
\label{pointwise}
\cD^{i j} \htr_{i j}(x) \psifullexp  = \kappa\, Q(x) \psifullexp~,
\ee
which can be rewritten as
\be
\label{greenfnsol}
\htr_{i j}(x) \psifullexp  = \kappa \int_{\S} d^d x' \sqrt{\gamma} '\,G_{i j}(x, x') Q(x') \psifullexp ~,
\ee
where the Green's function $G_{i j}$ satisfies $\cD^{i j} G_{i j}(x,x') = {1 \over \sqrt{\gamma}} \delta(x, x')$ with boundary conditions that it vanishes as $x'$ approaches the boundary. 
We emphasize that \eqref{greenfnsol} is just an exact rewriting of \eqref{pointwise} and not really a solution.

The equation \eqref{greenfnsol} may seem complicated.  However, we can develop a perturbative procedure to solve \eqref{greenfnsol} as follows. The idea, as indicated originally by ADM \cite{Arnowitt:1962hi} and then elaborated in \cite{York:1972sj,Kuchar:1970mu,kuchar1991problem} is to think of the momentum $\pitr[ij]$ as a notion of ``local'' time. Therefore we can think of the pointwise constraint \eqref{pointwise} as telling us how initial data on a slice ``evolve'' as we change time locally but keep the endpoints of the Cauchy slice fixed.

Thus, we must view $\pitr[ij]$ to be the ``position'' variable while $\htr_{ij}$ is the conjugate momentum. This idea can be implemented by performing a partial Fourier transform of the wavefunctional
\be
\label{psitowtpsi}
\psifullfour  = \int D h^\r{T}\, e^{-i \int_\S d^d x\sqrt{\g}\, \pitr[i j] \htr_{ij}} \psifullexp
\ee
and we will slightly abuse notation by also denoting this wavefunctional by $\Psi$. This allows us to rewrite the constraint \eqref{greenfnsol} as 
\be
\label{fteqn}
{i \over \sqrt{\gamma}} {\d \over \d \pitr[ij] (x)} \psifullfour   = \kappa \int_{\S} d^d x'\, \sqrt{\gamma} ' G_{i j}(x,x')  Q(x')  \psifullfour~.
\ee

\paragraph{Momentum constraint.}
We now rewrite the momentum constraint using a similar procedure. We do not display all intermediate steps since the procedure used is almost identical to the procedure used above.

We start with the form of the constraint as shown in equation \eqref{secondOrderMomentum}. Then we note that it can be written in the form
\be
\label{momrewriting}
{-}{i \over \sqrt{\gamma}} {\d \over \d \hlo_{i j} (x)} \psifullfour   = {\kappa \over 2} \int_{\S} d^d x'\, \sqrt{\gamma} ' \tilde{G}^{i j k} (x,x')  Q_k(x')    \psifullfour~.
\ee
Here the Green's function $\wt{G}^{i}$ is the solution to
\be
\nabla_i \wt{G}^{i j k}(x, x') = {1 \over \sqrt{\gamma} '}\delta^{d}(x,x') \gamma^{j k}
\ee
with boundary conditions so that $\wt{G}^{i j k}(x, x')$ vanishes as $x$ is taken to the boundary. 

Note that in \eqref{fteqn} the operator $Q$  still involves both $\pitr[ij]$ and also $h^\r{T}_{ij}$ which should be interpreted as ${i \over \sqrt{\g}} {\d \over \d\Pi_{T}^{ij}}$ while acting on $\Psi$. Similarly in \eqref{momrewriting}, the right hand side $Q_k$ still involves $\hlo_{i j}$.  So it may appear that we have not achieved much by recasting the pointwise constraints in the form \eqref{fteqn} and \eqref{momrewriting}. Nevertheless we can take advantage of the factor of $\kappa$ that appears in \eqref{fteqn} to develop an iterative procedure to solve this equation.

\subsubsection{Leading order solutions}

We start by considering the wavefunctionals described in section \ref{secpert} that have a specified dependence on $h^\r{TT}_{ij}$ and the matter field $\phi$ and are eigenfunctions of the energy with eigenvalue $E$. 
We then specify that for the constant function $\pitr[ij](x) = 0$ and  $\hlo_{i j}(x) = 0$ we have
\be
\label{initvalwavefunc}
\left.\psifullfourE \right|_{\pitr[i j] = 0, \hlo_{i j} = 0} = \psifockE~.
\ee
 We then truncate \eqref{fteqn} and \eqref{momrewriting} by dropping occurrences of $\htr_{i j}$ and $\pilo[i j]$
\be
\label{truncqh}
Q^{(0)}(x) \equiv Q(x) \Big|_{\htr_{i j} = 0, \pilo[i j] = 0}; \qquad  Q_i^{(0)}(x) \equiv  Q_i(x)\Big|_{\htr_{i j} = 0, \pilo[i j] = 0}~.
\ee
As $\htr_{i j}$ and $\pilo[i j]$ are $\Or[\k]$ by the first order constraints, this corresponds to restricting to the leading order.

Then the leading order wavefunctional solution satisfies
\be
\label{truncatedeqn}
\begin{split}
&{i \over \sqrt{\gamma}} {\d \over \d \pitr[ij] (x)} \psifullfour   = \kappa \int_{\S} d^d x'\sqrt{\gamma'}\, G_{i j}(x,x')  Q^{(0)}(x')  \psifullfour, \\
{-}&{i \over \sqrt{\gamma}} {\d \over \d \hlo_{i j} (x)} \psifullfour   = {\kappa \over 2} \int_{\S} d^d x'\sqrt{\gamma'} \,\tilde{G}^{i j k}(x,x')  Q_k^{(0)}(x')    \psifullfour~.
\end{split}
\ee
Note that, for consistency, we must adopt the same normal ordering prescription for $Q^{(0)}$ and $Q^{(0)}_k$ that was adopted in section \ref{subsecintham}. This normal ordering prescription leads to the subtraction of a position-dependent constant in \eqref{truncatedeqn}.

These leading order equations can be solved by performing a change of variable for $\Pi^\r{T}_{ij}$ and $h^\r{L}_{ij}$. It proves convenient to define a ``time'' variable $\bt$ by the equation
\be
\Pi_{ij}^\r{T} = D_{ij} \bt~.
\ee
This is the generalization to AdS of the time variable used for example in \cite{Arnowitt:1962hi,Kuchar:1970mu}. Note that this is related to the variable  $\a$ appearing in \eqref{alphaPiT} by $\bt = -{2\/d-1} N \a$. 

Differentiating with respect to $\bt$ instead of $\pitr$ simplifies the Hamiltonian constraint to
\be
\le[ {-} {i\/\sqrt{\g}}{\d\/\d\bt(x)}- \k \,Q^{(0)}(x)\ri] \psifullfourE=0~.
\ee
Similarly, using  $\e^i$  instead of $h^\r{L}_{ij}$ allows to write the momentum constraint as
\be
\le[-{i\/\sqrt{\g}} {\d\/\d \e_i(x)} - \k\, Q_i^{(0)}(x)\ri]\psifullfourE= 0~.
\ee
These equations are derived in Appendix \ref{appleadingorder}. 
We can look for a solution of the form
\be
\label{solpointguess}
\psifullfourE= \exp(i \k \cS)\,\psifockE+ \Or[\k^2],
\ee
where the exponent $\cS$ must satisfy
\be
{1\/\sqrt{\g}}{\d \cS\/\d \bt(x)} = Q^{(0)}(x) ,\qq
{1\/\sqrt{\g}} {\d \cS \/\d \e^i(x)}  = -Q_i^{(0)}(x)~.
\ee
Remarkably, the solution can be found as it takes the simple form
\be\label{Sexpression}
\cS = \int d^d x\sqrt{\g}\le(-{2\/3} \,{\bf t} \le( \Pi^{ij}_\r{T}\Pi_{ij}^\r{T} -{1\/d-1}\Pi^2_\r{T}\ri)+2\, {\bf t}\,\Pi^{ij}_\r{TT}\Pi_\r{T}^{ij}+Q^{(0)}_0 {\bf t} -\e^i \cH_i^\r{matter}\ri).
\ee
It is proven in Appendix \ref{appleadingorder} that this is indeed the solution. This relies on a non-trivial permutation symmetry in the terms of $\cS$ that are cubic and quadratic in $\bt$. We can confirm that the approximation used in \eqref{truncqh} is valid since we can explicitly check on the solution that $\htr\Psi, (\htr)^2\Psi, \pilo \Psi$ and $(\pilo)^2\Psi$ are all subleading in $\k$.

By inverting the Fourier transform \eqref{psitowtpsi} we can also obtain wavefunctionals in the original metric representation
\be
\psifullE = \int  D\pitr \,e^{i \int  d^d x \sqrt{\g}\,\pitr[ij] \htr_{ij} } \psifullfourE  ~.
\ee
We can see that the dependence on $\htr_{ij}$ is captured by an integral that is qualitatively similar to the Airy function. 


\subsubsection{An iterative solution algorithm}

We can obtain solutions to the pointwise constraints at higher order by using an iterative procedure. At $\Or[\kappa^2]$  one must also account for the terms that involve  $h^\r{T}_{ij} = {i \over \sqrt{\g}}{\d \over\d \pitr[i j]}$ and $\pilo[i j] = -{i\/\sqrt{\g}}{\d \over \d \hlo_{i j}}$ on the right hand side of \eqref{fteqn} and \eqref{momrewriting}.  But it is clear that to obtain the solution to $\Or[\kappa^2]$ one only needs to account for the action of these terms on the $\Or[\kappa]$ solution obtained through \eqref{truncatedeqn}. In fact if one expands the wavefunctional in a power series in $\kappa$ then this pattern continues at higher order in perturbation theory:  at each order in perturbation theory, these operators act on the lower-order terms and produce a ``source term'' on the right hand side of the first order differential equation \eqref{fteqn} and \eqref{momrewriting}.  

Note that at higher orders it is not enough to keep only the terms involving $\htr$ in $Q$ but it is also necessary to include the other higher-order terms from the expansion of the Hamiltonian constraint \eqref{HconstraintExpansion}. But provided this is done, the procedure above can be extended to higher order.

It is clear that to leading order in $\kappa$ the dependence on $\pitr[ij]$ as one approaches the boundary  continuously goes over to the dependence obtained in the solutions of  \ref{subsecintegconst}. However, the solutions obtained there were very simple because  $\pitr[ij]$ drops out from the integrated constraint as described in \eqref{pitvanishes}. At a general bulk point this does not happen and therefore \eqref{fteqn} leads in general to a complicated set of coupled differential equations.

\subsubsection{An indirect argument implying a bijection between solutions to the pointwise and integrated constraints \label{secindirectargument}}
The subsection above proposed an iterative algorithm to uniquely uplift a solution of the integrated constraint to a solution of the full pointwise constraints and an explicit solution to leading order.  However, it is possible to argue indirectly, even without the help of the explicit solution or the algorithm above, that there is a one-to-one map between solutions of the integrated constraint and solutions of the full pointwise constraints.

This is because it is  possible to obtain a description of the low-energy Hilbert space of gravity coupled to matter by other means. One common procedure adopted is simply to fix the gauge, which allows an identification of the independent degrees of freedom. As expected, these degrees of freedom correspond to the transverse-traceless graviton and the matter fields. Another equivalent procedure is to examine the set of all classical solutions of the theory and then quantize them.  Both procedures can be seen to lead to precisely the Fock space described in section \ref{subsecintham}. The solutions that we have described here are also in one-to-one correspondence with this Fock space. This implies that there are no additional solutions that we have missed, and nor does our procedure yield any spurious solutions.

\subsection{Inner product}\label{sec:innerproduct}

To complete the definition of the canonical theory, we need to give the definition of the inner product.  The inner product has been the subject of some discussion in the literature \cite{DeWitt:1967yk}. Here we will propose a specific definition of the inner product at leading order in perturbation theory and demonstrate its consistency.

Consider two solutions of the constraints that we denote by $\Psi_1$ and $\Psi_2$. We propose that the inner product between these two solutions obtained above is defined as
\be\label{definnerproduct}
\innerp[\Psi_2, \Psi_1] = \int D h^\r{TT} D\phi \,\psifullfour[1] \,\psifullfour[2]^\ast~
,
\ee
where $\ast$ refers to complex conjugation. Note that the integral is only over the propagating degrees of freedom $h^\r{TT}$ and $\phi$ and is performed at {\em fixed} values of $\pitr[i j]$ and $\hlo_{i j}$. 

To see that this definition makes sense, we must show that the inner product doesn't depend on the value of $\hlo_{ij}$ and $\pitr[i j]$ at which the wavefunctionals are evaluated. At leading order in $\kappa$ this follows directly from the ``evolution'' equations obeyed by the wavefunctionals in these variables. In particular, by conjugating equation \eqref{truncatedeqn}, we find that
\be
{-}{i \over \sqrt{\gamma}} {\d \over \d \pitr[ij] (x)} \psifullfour[2]^*   = \kappa \int_{\S} d^d x' \sqrt{\gamma} '\, G_{i j}(x,x')  Q^{(0)\ast}(x')  \psifullfour[2]^*~.
\ee
Note that in the basis used above, $Q^{(0)}(x)$ is {\em not} a real operator due to the presence of cross terms in its definition  that mix, for instance, $\pitt[i j]$ and $\pitr[i j]$.  But since $\pitr[i j]$ is realized, in the basis used above, as ${-i {\d \over \d \pitt[i j]}}$ complex conjugation of this operator introduces a negative sign. Nevertheless, by integrating by parts, and using the identities 
\be
\begin{split}
&\int D \htt D \phi\,\Psi_1\Big( {\d \over \d \htt_{i j}}\Psi_2^* \Big)= -\int D \htt D \phi\,\Big({\d \over \d \htt_{i j}} \Psi_1 \Big)\Psi_2^* ~, \\
&\int D \htt D \phi \,\Psi_1 \Big( {\d^2 \over \d \htt_{i j} \d \htt_{k l}}\Psi_2^* \Big) = \int D \htt D \phi\, \Big({\d^2 \over \d \htt_{i j} \d \htt_{k l}}\Psi_1 \Big)\Psi_2^*~, \\
\end{split}
\ee
we find that 
\be
\begin{split}
\int D \htt D \phi\, \Psi_1 Q^{(0)\ast} (x')\Psi_2^*  =  \int D \htt D \phi\, \Big(Q^{(0)}(x') \Psi_1 \Big)  \Psi_2^* ~.
\end{split}
\ee
In the sequence of equations above, we have suppressed the arguments of the wavefunctionals for clarity. 

Now, using the evolution equation we find that
\be
\begin{split}
&{i \over \sqrt{\gamma}} {\d \over \d \pitr[i j] (x)} \innerp[\Psi_2, \Psi_1] \\
&= \int_\S d^d x'  \int D \htt  D\phi \,  \left(Q^{(0)}(x') G_{i j} (x, x')\Psi_1 \Psi_2^\ast -  \Psi_1  G_{i j}(x, x') Q^{(0)\ast}(x') \Psi_2^\ast \right) = 0~.
\end{split}
\ee
Similarly, the second order momentum constraint \eqref{secondOrderMomentum}   equates $-{i\/\sqrt{\g}}\,{\d\/\d h_{ij}^\r{L}}$ with a self-adjoint operator, which ensures that
\be
-{i\/\sqrt{\g}}{\d\/\d h_{ij}^\r{L}} \innerp[\Psi_1, \Psi_2] = 0~,
\ee
and  the inner product is independent of  $h^\r{L}_{ij}$. 

This inner product reproduces the Fock space inner product if we use the natural measure 
\be
D h^\r{TT} =\prod_n d c_n,\qquad D \phi = \prod_n d \tilde{c}_n~.
\ee
Then, using the above normalization, we find the simple result
\be
\innerp[\psifullEnoarg, \psifullEpnoarg] = \delta_{E, E'} \delta_{\{a\}, \{a'\}}~.
\ee

\section{Holography of information \label{secholproof}}

In previous sections we have analyzed the form of the Hamiltonian and momentum constraints. We have shown that these constraints force a certain component of the metric fluctuation to have specific correlations with the excitations of the matter  fields and transverse-traceless gravitons. We will now show that these correlations are sufficient to completely identify a state in the bulk from boundary correlators.

More precisely, we will establish the result.
\begin{result}
\label{holresult}
If two pure or mixed states of the theory coincide at the boundary of AdS for an infinitesimal interval of time then they must coincide everywhere in the bulk.
\end{result}

An intuitive way to think of our  strategy to establish this result is as follows. At the boundary, we have available to us the boundary values of the metric and other matter fields. Let us first consider pure states. Then the correlations that we have analyzed at length in section \ref{sec:solving} allow us to determine the energy of a state from the measurement of a certain component of the metric at the  boundary.  The value of this component is suppressed by a factor of $\kappa$ but our analysis is already sufficient to reveal its nontrivial value. 

A determination of the energy is {\em not} sufficient to determine the state.  Since a pure state must be a superposition of energy eigenstates, the determination of the energy still leaves us with an ambiguity of relative phases between different energy eigenstates and also an ambiguity associated with degeneracies in energy eigenstates. 

To resolve this ambiguity, we exploit the fact that  energy eigenstates are necessarily delocalized states. This is true just by virtue of the Heisenberg uncertainty principle.  We demonstrate that the ambiguity associated with degeneracy and the ambiguity associated with the phases of eigenstates can be resolved by additional measurements of the metric and matter fields near the boundary in an infinitesimal time interval. These latter measurements are {\em not} suppressed by $\kappa$ and involve just the $\Or[1]$  fluctuations of the transverse-traceless metric component and the matter fields. The end result is that correlations of the energy and other observables near the boundary suffice to completely fix the form of the bulk state.

The extension of our result to mixed states is straightforward.  A basis of density matrices is obtained by combining a  wavefunctional corresponding to one energy eigenstate with the conjugate of a wavefunctional corresponding to another energy eigenstate. Let us denote such a basis by $\rhofullE$ where $E,E'$ are the energy eigenvalues of the wavefunctionals and $\{a\}, \{a'\}$ are additional labels necessary since energy eigenstates can be degenerate and, as usual, the density matrix has double the arguments of the wavefunctional. Any density matrix can be written as a linear combination of such elementary density matrices with certain coefficients.  Two density matrices  can only yield the same values for all moments of the energy if these coefficients satisfy certain strong constraints. As in the case of pure states, measurements of the energy are insufficient to fix these coefficients. However, we show that correlators of additional dynamical fields uniquely fix these coefficients. 

We now present a precise mathematical argument that realizes the intuition above. In preparation for this argument, we first discuss the set of boundary observables and also the set of valid mixed states in the theory before turning to the proof in section \ref{secproofmain}

\subsection{Boundary observables \label{subsecbdryobs}}
Let us briefly recapitulate the set of boundary observables. Recall that, as explained below equation \eqref{secondconst} boundary observables are automatically gauge invariant. The constraints only impose the invariance of observables under small gauge transformations, and since such transformations die off near the boundary, the constraints commute with boundary observables.

One special boundary observable that will be required is the ADM Hamiltonian, $\Hbdy$, given in equation \eqref{defHbdy}. In addition, we will require the boundary values of the metric and also the matter fields in the theory. In order to adopt a compact notation, we denote such local boundary operators collectively by 
\[
\obdry(t, \Omega); \qquad \Omega \in S^{d-1}.
\]
Note that these observables are naturally defined by a value of the boundary time, $t$, and also a position on the boundary sphere. 

For instance, consider the scalar field that we have discussed above with mass $m$. Then a gauge-invariant boundary observable is obtained through
\be
\label{boundarylimit}
\obdry(t, \Omega) = \lim_{r \rightarrow \infty} r^{\Delta} \phi(r, t, \Omega) ~,
\ee
where we are using the coordinate system in \eqref{Adsmetric} and $\Delta$ is defined in \eqref{conformaldim}.  In our notation, we assume that unlike $\Hbdy$ (defined in equation \eqref{defHbdy}), {\em no} explicit factors of ${1 \over \kappa}$ are inserted while taking the boundary limits of bulk operators. The reader should keep this distinction between $\Hbdy$ and the observables $\obdry(t, \Omega)$ in mind for the analysis below.

We pause to address a subtlety associated with the limit described in equation \eqref{boundarylimit}. In order to take the limit, the operator on the right hand side of equation \eqref{boundarylimit}, which is a bulk operator, must be first made gauge invariant in the sense of equation \eqref{opconstraint}. It can be seen that there is no unique way to dress the bulk operator in order to make it gauge invariant.  

A simple way to understand this lack of uniqueness is as follows.  In this paper, we have not invoked a specific gauge. But another way of obtaining approximately local bulk operator is simply to choose a gauge.  To every such gauge-fixed operator, there exists a gauge-invariant representation of the operator that satisfies the constraints \eqref{opconstraint}. But different choices of gauge lead to different operators. This is why the symbol $\phi(r, t, \Omega)$ does not have a unique meaning unless its precise dressing is specified. 

This lack of uniqueness changes some correlators at $\Or[\kappa]$ \cite{Donnelly:2015hta}.  Nevertheless, this issue  is not important for our analysis because we will use the operators shown in \eqref{boundarylimit} only within specific correlators. We will only need the fact that when we take the limit to the boundary, the final operator commutes with the constraints and its correlators with other local boundary operators at $\Or[1]$ are independent of how the operators was dressed in the intermediate step. The precise property used is stated precisely in equation \eqref{auxcorrelator} below  and also holds for gauge-fixed operators.

\label{metricdiscussion}
We have displayed a scalar field in \eqref{boundarylimit} but a similar limit can be taken for observables that contain the metric or other dynamical fields in the theory. In the case of observables that depend on the metric, the only element of the ADM decomposition that is relevant at $\Or[1]$ in such an observable is $\htt$. It is easiest to see this in the mixed representation of \eqref{psitowtpsi}, which can also be used for observables. Then the first order Hamiltonian and momentum constraints tell us that such an observable must be independent of $\hlo$ and $\pitr$ at $\Or[1]$. Therefore, at $\Or[1]$ such observables can only depend on $\htt$ and $\pitt$.  To lighten the notation, in the analysis below, $\op(t, \Omega)$ can stand for an insertion of either the metric or the insertion of a matter field. 


\subsection{Mixed states}
In the previous sections, we have focused on pure states in the theory. It is a short step to generalize this discussion to mixed states, and we do so now.  

In section \ref{sec:solving} we have obtained wavefunctionals that are annihilated by the constraints. A basis of density matrices is obtained by combining them:
\be
\label{rhobasis}
\rhofullE  =  \Psi_{E'}^{(a')}[h_{i j}, \phi] \Psi_{E}^{(a)}[\tilde{h}_{i j}, \tilde{\phi}]^*~,
\ee
where the wavefunctionals are normalized with respect to the inner product \eqref{definnerproduct}. Note that the density matrix depends on {\em two} metric configurations, which we have denoted above by $h_{i j}$ and $\tilde{h}_{i j}$, and two matter-field configurations, denoted above by $\phi$ and $\tilde{\phi}$.

A general density matrix is a linear combination of elements of \eqref{rhobasis}:
\[
\label{rhoconstituents}
\rhofull = \sum_{\indices} c(\indices) \rhofullE~.
\]
As usual, these density matrices satisfy the constraints that
\be
\label{validrho1}
c(\indicesrev)= c(\indices)^*~.
\ee
Moreover, the eigenvalues of the density matrix must be positive and we additionally have
\be
\label{validrho2}
\sum_{E, \{a\}} c(\indicesdiag) = 1~.
\ee
We denote expectation values of an operator $\al$ in a density matrix using the notation $\langle \al \rangle_{\rho}$. These expectation values are computed through
\be
\langle \al \rangle_{\rho} = \sum_{\indices} c(\indices) \innerp[\psifullEpnoarg, \al \,\psifullEnoarg]
\ee
where the inner product is as defined in \eqref{definnerproduct}.

\subsection{Proof of the main result \label{secproofmain}}
We are now in a position to prove the result above.

 Let $\rho_1$ and $\rho_2$ be two density matrices of the form \eqref{rhobasis} with coefficients $c_1(\indices)$ and $c_2(\indices)$ respectively. We will now show that if the we have equality  of the expectation values
\be
\label{equalcorrcond}
\langle \Hbdy^n\, \op(t_1, \Omega_1) \ldots \op(t_q, \Omega_q) \Hbdy^m \rangle_{\rho_1} = \langle \Hbdy^n\, \op(t_1, \Omega_1) \ldots \op(t_q, \Omega_q) \Hbdy^m \rangle_{\rho_2}
\ee
for arbitrary values of $n, m, q$ and and for any $t_i \in [0, \epsilon]$, we then have $\rho_1 = \rho_2$.

First note that equation \eqref{equalcorrcond} implies that
\be
\label{sumzerocond}
\sum_{\substack{E,E' \\\{a\}, \{a'\}}} \resizebox{0.87\textwidth}{!}{$ \Big[c_1(\indices) - c_2(\indices) \Big]  E^n E'^m \, \langle \op(t_1, \Omega_1) \ldots \op(t_q, \Omega_q) \rangle_{\tiny{\rhofullEnoargs}}$} =0~.
\ee
Since this is true for arbitrary values of $n,m$  it must be true that for each individual value of $E, E'$
\be
\label{indivzerocond}
\sum_{\{a\}, \{a'\}}  \Big[c_1(\indices) - c_2(\indices) \Big]   \langle \op(t_1, \Omega_1) \ldots \op(t_q, \Omega_q) \rangle_{\rhofullEnoargs} =0~,
\ee
where the  important difference with the previous equation is that \eqref{indivzerocond} does {\em not} involve any sum over $E,E'$.

We now note that
the correlators that appear in \eqref{indivzerocond} can be evaluated in the {\em auxiliary} Fock space introduced in section \ref{subsecintham}. That is,
\be\label{auxcorrelator}
\langle \op(t_1, \Omega_1) \ldots \op(t_q, \Omega_q) \rangle_{\rhofullEnoargs} = \innerp[\psifockEpnoarg,  \op(t_1, \Omega_1) \ldots \op(t_q, \Omega_q) \psifockEnoarg]+ \Or[\kappa].
\ee
Note that the correlator on the left hand side does not include $\Hbdy$ and  it only includes operators of the form \eqref{boundarylimit}. The equation above then follows from the discussion of section \ref{subsecbdryobs}. 
Computing an ordinary matter correlator with the full wavefunctional is the same at $\Or[1]$  as computing the same correlator in the Fock space.

To complete the proof, we will use the analytic properties of the  correlators that appear on the RHS of \eqref{auxcorrelator}. 
By inserting a complete set of energy eigenstates in the auxiliary Fock space, we find that
\be
\label{insertioncomplete}
\begin{split}
&\innerp[\psifockEpnoarg,  \op(t_1, \Omega_1) \ldots \op(t_q, \Omega_q) \psifockEnoarg ]     =  e^{i (E' t_1 - E t_q) } \sum_{E_j, \{a_j\}}  e^{i \sum_{i=1}^{q-1} E_i (t_{i+1} - t_i) } \\ & \times\, 
\innerp[\psifockEpnoarg, \op(0, \Omega_1) {\psifockEnoarg[1]}] \innerp[{\psifockEnoarg[1]}, \op(0, \Omega_2) {\psifockEnoarg[2]} ] \innerp[{\psifockEnoarg[2]}, \op(0, \Omega_2) {\psifockEnoarg[3]} ] \\
& \times \ldots \times \innerp[{\psifockEnoarg[q-1]}, \op(0, \Omega_{q}) \psifockEnoarg ].
\end{split}
\ee
We emphasize that the entire identity above is simply in the auxiliary Fock space, and we have used completeness and the transformation properties of the operators under time translations only in the Fock space.
This correlator is clearly analytic when the variables
\be
\label{analyticvars}
z_1 = t_1; \quad z_2 = t_2 - t_1; \quad  \ldots \quad; z_q = t_q - t_{q-1}
\ee
are continued in the upper half plane. This follows just from the positivity of energy in the auxiliary Fock space. Note that in the correlator above the term in the exponent involving $E' t_1 - E t_q$ is outside the sum over energies and when the variables $z_i$ are extended in the upper half plane each term in the exponential inside the sum picks up a factor that decays exponentially with energy.  Hence, if the correlator vanishes when $t_i \in [0, \epsilon]$ it must also vanish for  $t_i \in [0, \pi]$ by the edge of the wedge theorem \cite{streater2016pct,Haag:1992hx}. 

But, in the Fock space, the individual creation and annihilation operators can be obtained by integrating $\op(t_i)$ in a band of size $\pi$. This follows from the discrete frequencies for the excitations found in section \ref{subsecgravwave} and \ref{subsecmatterwave}. So the algebra of operators for all $t_i \in [0, \pi]$ provides a complete basis for the algebra of all operators in the Fock space. Therefore the correlator in equation \eqref{indivzerocond} vanishes for all $t_i \in [0, \pi]$ if and only if  $c_1 = c_2$. 

This proves our assertion.

\paragraph{\bf Comments on the proof \\}
We would like to comment on some subtle aspects of the proof above. 
\begin{enumerate}
\item
Note that the correlator \eqref{equalcorrcond} involves high powers of $\Hbdy$. Nevertheless, our perturbative solution can be used to reliably compute these correlators. This can be seen by rewriting the expression for the integrated constraint after Fourier transforming the wavefunctional as was done in section \ref{sec:pointwise}. The constraint then takes the form
\be
\begin{split}
&{i \over 2  \kappa}\int_{\partial \S}d^{d-1} \Omega\, n^i  (N \n^j - \n^j N) (\delta^{k}_{i} \delta^{\l}_{j} - \gamma^{k \l} \gamma_{i j}){1 \over \sqrt{\gamma}} {\delta \over \delta \pitr[k \l]} \psifullfour \\ &= \int_\S d^d x\sqrt{\g} N\, \hbulk \psifullfour + \Or[\htr_{i j}] + \Or[\kappa]
\end{split}
\ee
where we have explicitly also displayed the $\Or[\htr_{i j}]$ and higher-order terms that were dropped in the analysis of section \ref{subsecintegconst}. Now one of the key simplifications that we found in section \ref{subsecintegconst} was that $\pitr[i j]$ drops out of the integrated expression for $\hbulk$. Consequently we were able to examine wavefunctionals that satisfied 
\be
\int_\S d^d x \sqrt{\g}N \, \hbulk \psifullfourE = E \,\psifullfourE~.
\ee
We can then move to new variables 
\be
\wpitr[ij] = \kappa \,\pitr[i j], \qquad \whtr_{i j} = {i \over \sqrt{\gamma}}  {\delta \over \delta \wpitr[i j]} = {1 \over \kappa} \htr_{i j}~,
\ee
so that the equation above takes the form
\be
\begin{split}
&{i \over 2 }\int_{\partial \S} d^{d-1}  \Omega  \, n^i (N \n^j - \n^j N) (\delta^{k}_{i} \delta^{\l}_{j} - \gamma^{k \l} \gamma_{i j}){1 \over \sqrt{\gamma}} {\delta \over \delta \wpitr[k \l]} \psifullfourEtilde \\ &= E\, \psifullfourEtilde  + \kappa\, \Or[\whtr_{i j}] + \Or[\kappa].
\end{split}
\ee
Note that the factor of $\kappa$ has disappeared on the LHS above, and an additional factor of $\kappa$ has appeared in front of the functional derivatives with respect to $\wpitr[i j]$ on the second line of the RHS. This entire equation clearly has a smooth limit as $\kappa \rightarrow 0$ and this allows us to conclude that repeated applications of $\Hbdy$ produce a simple result:
\be
\begin{split}
&\Big[{i \over 2 }\int_{\partial \S} d^{d-1}  \Omega \,  n^i  (N \n^j - \n^j N) (\delta^{k}_{i} \delta^{\l}_{j} - \gamma^{k \l} \gamma_{i j}){1 \over \sqrt{\gamma}} {\delta \over \delta \wpitr[k \l]} \Big]^n \psifullfourEtilde \\ &= E^n \,\psifullfourEtilde + \Or[\kappa].
\end{split}
\ee
This is precisely what we have used above.
\item
Our perturbative analysis in section \ref{secpert} and  section \ref{sec:solving} assumes that  the states under consideration do not have energies that scale parametrically with $\Or[{1 \over \kappa}]$  so that there are no factors of $\kappa$ that we need to keep track of except for the ones that appear explicitly in perturbation theory. But the proof above requires somewhat more stringent conditions on the energies. This can be seen by  examining the passage from equation \eqref{sumzerocond} to equation \eqref{indivzerocond}. If we denote the number of energy levels below a given energy $E$ by $D(E)$ then this passage is valid provided we can take  $n,m$ in \eqref{equalcorrcond} to satisfy $n, m > D(E)$. Since we are limited to using $n, m < \Or[{1 \over \kappa}]$,  the proof above holds provided the states that enter \eqref{equalcorrcond} satisfy $D(E) < \Or[{1 \over \kappa}]$ in AdS units. We remind the reader that $\log(D(E))$ can grow no faster than $E$ on thermodynamic grounds and the linear bound is saturated by the Hagedorn behaviour of string-theory. 

This limitation should not be surprising. When $D(E) = \Or[{1 \over \kappa}]$, the expected difference in the value of {\em any} observable between two typical state is suppressed by a factor of $\Or[\sqrt{\kappa}]$. (See \cite{lloyd1988black} or section 2.4 of \cite{Raju:2020smc}) Therefore even if one were to consider all correlators, including bulk correlators, it would still be necessary to measure these correlators to highers order in $\kappa$ in order to differentiate two typical states.

We emphasize that this limitation does not mean that the result above fails to hold for high-energy states. The arguments of \cite{Laddha:2020kvp} arrive at the same result with no such limitation. So our observation simply implies that we need to refine our proof for high-energy states.
\item
We note that the proof above can also be rewritten using the projector on the vacuum as was done in \cite{Laddha:2020kvp} or by replacing powers of $\Hbdy$ with projectors onto eigenstates of $\Hbdy$. Indeed, from a physical perspective, projective measurements are more natural than correlators as was explained in \cite{Chowdhury:2020hse}. We have provided a proof using the correlators of \eqref{equalcorrcond} only to keep our argument simple and explicit.
\item
In the proof above we have utilized a small time band in order to make the assertion below \eqref{analyticvars} rigorous. We expect that it should be possible to trade this infinitesimal time band for an infinitesimal ``thickness'' in the bulk. If so, the result above can also be stated as ``if two states coincide {\em near} the boundary at a single instant of time, they must coincide everywhere in the bulk.'' However, to make this rigorous requires some delicate analysis since, in an intermediate step, it will be necessary to construct bulk operators that commute with the constraints. 
\item
The products of operators that appear in \eqref{equalcorrcond} are not necessarily Hermitian. However, the expectation value we need can always be obtained by combining the expectation values of Hermitian observables. We first write each product of operators, $A$ in \eqref{equalcorrcond} as $A = X + i Y$ where $X$ and $Y$ are Hermitian. We then have $\langle A \rangle = \langle X \rangle + i \langle Y \rangle$. For further discussion of a ``physical protocol'' that can be used to extract information about the state, by combining a boundary unitary operation with a measurement of the energy, we refer the reader to \cite{Chowdhury:2020hse}.
\item \label{infraredcomment}
The proof above takes advantage of the infrared cutoff provided by global AdS. Since the spectrum of energies is discrete, a finite set of powers of $\Hbdy$ in \eqref{equalcorrcond} are sufficient to make the passage to \eqref{indivzerocond}.  This means that the method of proof presented here must be refined before it can be applied to asymptotically flat space where there is no infrared cutoff.

We note that the result one should aim for in asymptotically flat space is clear. In \cite{Laddha:2020kvp} it was shown, using operator-theoretic techniques, that two states of massless particles that coincide in a small retarded-time band near the past boundary of $\scrip$ (or a small advanced-time band near the future boundary of $\scrim$) must be identical. We expect that a refinement of the techniques developed here, to account for the infrared subtleties of flat space, will lead to the same result.
\end{enumerate}

\section{Discussion}

\paragraph{\bf Summary of results.}

In this paper, we have explicitly shown that a careful analysis of the solutions of the gravitational constraints leads to a perturbative proof of the principle of holography of information:  any wavefunctional that satisfies the gravitational constraints in AdS is determined uniquely by its boundary values over an infinitesimal interval of time.
As we reviewed in section \ref{secsetting}, these constraints can be obtained from the straightforward canonical quantization of gravity. In the canonical formalism, states of the theory are represented as wavefunctionals of the metric and matter degrees of freedom. The requirement that these wavefunctionals yield the same amplitude for configurations that are related by diffeomorphisms of a spatial slice leads to the momentum constraint; requiring the same amplitude for configurations related by diffeomorphisms that move points in time leads to the Hamiltonian constraint, which is also called the Wheeler-DeWitt equation. The precise form of these constraints can be found in equations \eqref{HamConstraint} and \eqref{MomConstraint}. 

In section \ref{secpert}, we expanded these constraints up to second order in the metric fluctuation. An important tool introduced in section \ref{secpert} was the ADM decomposition presented in equation \eqref{admdecomp}.  This decomposition has previously been used in flat space; our results show that when suitably generalized it is also a very useful decomposition in curved space.

In section \ref{sec:solving}, we analyzed solutions to the perturbative Hamiltonian constraint. We first considered the equation obtained by integrating the Hamiltonian constraint over an entire Cauchy slice. This procedure greatly simplifies the constraint. We were able to obtain explicit solutions to the integrated constraint: these solutions are just dressed versions of wavefunctionals in an auxiliary Fock space that describe the matter excitations and the transverse-traceless metric excitations. We also showed how the pointwise Hamiltonian constraint can be solved through an iterative procedure. 

In section \ref{secholproof}, we showed that these wavefunctionals obey the remarkable property that their boundary values for an infinitesimal interval of time determine their behavior everywhere in the bulk. This result follows 
from the solutions that we obtained in sections \ref{secpert} and \ref{sec:solving}. It sheds light, in a precise and explicit setting, on how and why gravitational theories are holographic.

\paragraph{\bf Natural extensions.}
It is instructive to see what our analysis gives in the case of AdS$_3$. There are no nontrivial propagating gravitons in AdS$_3$ but it is still meaningful to define a boundary Hamiltonian that measures the total energy of the state.   
So we see that the present formalism can be applied to AdS$_3$.  It would be interesting to go further and recast the Brown-Henneaux analysis \cite{Brown:1986nw} in the language of wavefunctionals.

This work was focused on global AdS$_{d+1}$ but the analysis can also be performed for subregions of AdS. In particular, it appears straightforward to extend our analysis to the the Rindler wedge of a spherical region  \cite{Casini:2011kv} and also perhaps to more general entanglement wedges. This promises to shed light on subregion duality and entanglement wedge reconstruction and we hope to return to this problem in the near future.

\paragraph{\bf Future work.}
The analysis in this paper has been perturbative. In \cite{Laddha:2020kvp}, it was shown that with weak assumptions on the Hilbert space and the nature of boundary observables, theories of gravity must be holographic even nonperturbatively. The analysis of \cite{Laddha:2020kvp} used operator algebra techniques.  It would be very interesting if the perturbative analysis of this paper could be generalized to show that, even nonperturbatively, solutions of the WDW equation that coincide on the boundary must coincide everywhere in the bulk.  Although the nonperturbative WDW equation may seem formidable, the results of \cite{Laddha:2020kvp} suggest that obtaining such a result might be possible. 

In this paper, we have been agnostic to the matter content of the theory and its interactions. However, it is well known, from the AdS/CFT literature, that not  all low-energy effective theories can be consistently extended to obtain a UV-complete theory of quantum gravity in AdS. It would be very interesting to understand whether and how these constraints enter possible extensions of our analysis.

The results of our paper again illustrate the dramatic difference between the storage of quantum information in quantum gravity compared to quantum field theories.  In ordinary quantum field theories, it is possible to find states that differ inside a bounded region but are identical outside that region; such states localize information in the interior of some region.  The existence of such states corresponds to the ``split property'' of ordinary quantum field theories where the Hilbert space factorizes into a factor associated with the interior of the region and another factor associated with the exterior. In classical theories of gravity, configurations that differ inside a ball but coincide outside it can be constructed. For this reason, it has often been assumed that split states should also exist in quantum gravity. But our results show that this seemingly innocuous assumption is false.

It is described in \cite{Raju:2020smc} how this incorrect assumption plays a key role, both in Hawking's formulation of the information paradox and also in its various refinements (see also \cite{Chakraborty:2021rvy,Raju:2018zpn}). More interestingly, the idea that black hole radiation should obey a ``Page curve'' also relies implicitly on the same incorrect assumption of factorization. By focusing on this assumption, it was recently shown in \cite{Geng:2021hlu} that the paradigm of ``islands''  \cite{Almheiri:2020cfm} that has been used to derive this Page curve  is applicable only to theories of massive gravity and does not apply to standard theories with long-range gravity.

This paper shows how the impossibility of localizing information in a bounded region in gravity is directly related to  the structure of valid wavefunctionals in the theory.  We hope that a  study of the solutions that we have found will help to shed further light on this remarkable property of quantum gravity.

\section*{Acknowledgments}

Research at ICTS-TIFR is supported by the Government of India through the Department of Atomic Energy project RTI4001. S.R. is partially supported by a Swarnajayanti fellowship DST/SJF/PSA-02/2016-17 of the Department of Science and Technology. We are very grateful to Kyriakos Papadodimas for collaboration in the early stages of this work. We are grateful to   Tuneer Chakraborty, Joydeep Chakravarty, Hao Geng, Andreas Karch,  Alok Laddha, Ruchira Mishra, Priyadarshi Paul,   Carlos Perez-Pardavila, Siddharth Prabhu,   Lisa Randall, Marcos Riojas,  Sanjit Shashi and Pushkal Shrivastava for several useful discussions.
\appendix
\addtocontents{toc}{\protect\setcounter{tocdepth}{1}}

 \section{Split states in QED \label{appem}}
 \newcommand{\piem}{\Pi_\r{em}}
 \newcommand{\piemt}{\Pi_\r{em,T}}
 \newcommand{\pieml}{\Pi_\r{em,L}}
\newcommand{\Al}{A^{\text{L}}}
\newcommand{\At}{A^{\text{T}}}
 \newcommand{\delE}{\frac{1}{\sqrt{-g}} \pa_i\big( \sqrt{-g}\ \piem^i (x)\big)}
In this appendix, we show that ordinary gauge theories localize information much like ordinary quantum field theories, and very differently from gravity.  To illustrate this, we will solve the constraint of a $U(1)$ gauge theory coupled to matter and construct explicit wavefunctionals that are identical outside a bounded region but differ inside. Such states are called ``split states'' and the argument provided in the main text of the paper shows that split states do not exist in theories of quantum gravity.  A useful reference for the analysis of wavefunctionals in QED and ordinary quantum field theories is \cite{hatfield1991quantum}. An  analysis of the canonical quantization of QED can also be found in Appendix B of \cite{Papadodimas:2013jku}. We caution the reader that some of the conventions below differ from those of \cite{Papadodimas:2013jku} by terms involving $N$ and the determinant of the spatial metric.

\subsection{Action and constraints}
We work about the fixed global AdS background 
\be
\label{qedbackmet}
ds^2 = - N^2 dt^2 + N^{-2} dr^2 + r^2 d\Om_{d-1}^2~,
\ee
where $N$ is the same as \eqref{backgroundn}. We emphasize that in this Appendix, we are {\em not} considering a theory with dynamical gravity and so the metric  \eqref{qedbackmet} is exact. We continue to use the $d+1$ notation of the main text for covariant derivatives.
 
 The action of QED takes the form,   
 \be
\label{qedaction}
 S = -{1\/4} \int dt d^{d} x \sqrt{\gamma}N\, \hat{F}_{\mu\nu} \hat{F}^{\mu\nu} + S_\r{matter}~.
 \ee
Note that we have included the interactions of the gauge field and the matter in the term denoted as $S_{\text{matter}}$ above. The details of this action will not be important except for a few features that we mention below. But for the purpose of illustration, we consider a charged scalar field with the action, 
\be\label{scalarqed}
S_{\text{matter}} =  - \frac{1}{2} \int d^{d+1}x  \sqrt{\gamma}N\,  (\mathcal D_{\mu}\phi)^* \mathcal D^{\mu} \phi~,
\ee
where $\mathcal D_{\mu} = \partial_\mu + i A_\mu $ is the gauge covariant derivative  with the coupling constant  set to 1. 

As there is no kinetic term for the $A_0$ field, we immediately obtain a primary constraint
 \be
 \piem^t = 0~,
 \label{primaryconstem}
 \ee
whereas the canonical momentum for the spatial part of the gauge field is 
\be
\piem^i = {1 \over \sqrt{\gamma}} {\delta S \over \delta \dot{A}_i} =  -N \hat{F}^{ti}
\ee
which is just the electric field.  This is similar to the primary constraints \eqref{primaryconst} in gravity. Imposing that this constraint is preserved under time evolution leads to a secondary constraint. This is the pointwise Gauss law
 \be
 \nabla_i  {\piem^i}  = \rho~,
 \label{secondconstem}
 \ee
 where $\rho$ is the charge density of the matter.  The left hand side of \eqref{secondconstem} is  reminiscent of the momentum constraint in gravity since it is linear in the canonical momentum. However, the momentum constraint in gravity couples the metric and its canonical momentum whereas we see that \eqref{secondconstem} has no such nonlinear terms. This will allow us to present a general solution to this constraint.

For the action \eqref{scalarqed}, the momentum conjugate to the scalar field is
\be
\Pi_{\phi} = {1 \over \sqrt{\gamma}} {\delta S \over \delta \dot\phi} = \frac{1}{2N} (\dot\phi^* - i A_t \phi^*); \qquad \Pi_{\phi^*} = {1 \over \sqrt{\gamma}} {\delta S \over \delta \dot\phi^*} = \frac{1}{2N} (\dot\phi + i A_t \phi)~,
\ee
and in terms of the canonical variables, we have  
\be
\rho= i (\phi\, \Pi_{\phi} - \phi^*\, \Pi_{\phi^*})~.
\ee
But the details of the matter sector will be unimportant in the analysis below and we will only use the fact that, in the classical canonical theory, the Poisson bracket between the charge density at two distinct points, $x$ and $x'$, on the same spatial slice vanishes:
\be
\label{classicalpoisson}
\{\rho(x), \rho(x') \}_{\text{PB}} = 0~.
\ee

In the quantum theory, the states of the theory are described by  wavefunctionals of the gauge field and matter fields. The primary constraint \eqref{primaryconstem} tells us that wavefunctionals, and observables that commute with the constraints, are independent of $A_0$.  Therefore these wavefunctionals $\psi[A, \phi]$ depend on only the spatial components of the gauge field. The momentum operator is realized as
\be
\Pi^i_\r{em} = - {i \over \sqrt{\gamma}} {\d\/\d A_i}~,
\ee
 The secondary constraint then implies that
 \be
 \left[\nabla_i \piem^i - \rho\right]  \psi[A, \phi]= 0~.
 \label{gaussem}
 \ee

Since we will exclusively consider wavefunctionals that satisfy the constraints the Poisson brackets \eqref{classicalpoisson} are directly promoted to commutators in the quantum theory. Therefore we have
\be
\label{chargecommvanish}
[\rho(x), \rho(x')] = 0~,
\ee
at any two points $x, x'$ on the same spatial slice. This property will be utilized below.

\subsection{Solution to the constraints}
Since the constraints in electromagnetism are simple, it is possible to write down an exact solution to the constraints. As in the main text, it is convenient to 
decompose the gauge field into a longitudinal and a transverse part
\be
A_i = \Al_i + \At_i
\ee
which satisfy
\be
\nabla^i \At_i = 0, \qquad \Al_i = \nabla_i \chi~.
\ee
for some $\chi$ that vanishes asymptotically. The momentum can be similarly decomposed as
\be
\piem^i = \piemt^i + \pieml^i
\ee
and by a simple extension of the argument near equation \eqref{pidecomp} we find that
\be
\piemt^i = -{i \over \sqrt{\gamma}} {\delta \over \delta \At_i}; \qquad \pieml^i = -{i \over \sqrt{\gamma}} {\delta \over \delta \Al_i}~.
\ee
The constraint \eqref{gaussem} correlates the part of the wavefunctional that depends on $\Al$ with the part that depends on the charge density, leaving the part that depends on $\At$ unconstrained. A solution to the constraints is given by
\be
\label{solconem1}
\psi[A, \phi] = \exp\le[i\int_\S d^{d}x\sqrt{\gamma}\int_\S  d^{d}x' \sqrt{\gamma'}\,\Al_i(x) \nabla^i G(x, x') \rho(x') \ri]  \psi_A[\At] \psi_{\phi}[\phi]
\ee
where $\psi_{\phi}$ and $\psi_A$ are arbitrary functionals and the Green's function $G(x,x')$ satisfies
\be
\nabla_i \nabla^i G(x,x') = {1 \over \sqrt{\gamma}} \delta^{(d)}(x,x')~.
\ee
Since the spatial slice is just Euclidean AdS$_d$, the Green's function can be written as \cite{Giddings:1999jq, DHoker:2002nbb}
\be
G(x, x')= {2^{-\tilde{\Delta}} \over \tilde{\Delta}} {\Gamma(\tilde{\Delta}) \over \pi^{\tilde{\Delta} \over 2} \Gamma({\tilde{\Delta} \over 2})} \xi^{\tilde{\Delta}} \, _2F_1\big(\tfrac{\tilde{\Delta} }{ 2}, \tfrac{\tilde{\Delta}}{ 2} + 1; \tfrac{\tilde{\Delta}}{2} + 1; \xi^2 \big)
\ee
where $\tilde{\Delta} = d - 1$ and  $\xi (x,x')=(\r{cosh}\,d(x,x'))^{-1}$ and $d(x,x')$ is the geodesic distance between $x$ and $x'$. In our coordinates, we have explicitly 
\be
\xi(x,x') = \sqrt{1 + r^2} \sqrt{1 + (r')^2} + r r'\,e\cdot e'~,
\ee
where $e,e'$ are unit vectors in $\R^d$ parameterizing $S^{d-1}$.


 Note that \eqref{solconem1} is {\em not} a factorized solution since the $\rho$ in the exponent of the right hand side acts as an operator on $\psi_{\phi}$ and this forces correlations between the matter fields and the longitudinal part of the gauge field.

\subsection{Split states in QED}
Although the solution obtained above is not factorized, it is still possible to find split states. A simple example is obtained by taking two wavefunctionals $\psi^{(1)}_{\phi}[\phi]$ and $\psi^{(2)}_{\phi}[\phi]$ that are both eigenfunctions of the charge operator $\rho$:
\be
\label{rhodiffer}
\rho \,\psi^{(1)}_{\phi}[\phi] = \rho_1 \psi^{(1)}_{\phi}[\phi], \qquad  \rho \,\psi^{(2)}_{\phi}[\phi] = \rho_2 \psi^{(2)}_{\phi}[\phi]~.
\ee
Consider the case where the eigenfunctions $\rho_1$ and $\rho_2$ are both spherically symmetric, vanish outside a ball of finite radius $B_R$ centered at $r = 0$ but differ inside the ball. The fact that states of the form \eqref{rhodiffer} exist relies crucially on the fact that the charge density can be specified independently at each point in space by \eqref{chargecommvanish} and also on the fact that for ordinary matter fields it is possible to construct split wavefunctionals that agree outside a bounded region but differ inside \cite{Haag:1992hx}. 

If we impose the condition that
\be
\int_{B_R} d^d x\sqrt{\gamma}\, \rho_1  = \int_{B_R} d^d x\sqrt{\gamma} \,\rho_2 ~,
\ee
then we see that the wavefunctionals
\be
\psi^{(1)}[A,\phi] = \exp\le[i\int_\S d^{d}x\sqrt{\gamma}\int_\S  d^{d}x' \sqrt{\gamma'}\,\Al_i(x) \nabla^i G(x, x') \rho(x') \ri]  \psi_A[\At] \psi_{\phi}^{(1)}[\phi]
\ee
and
\be
\psi^{(2)}[A,\phi] = \exp\le[i\int_\S d^{d}x\sqrt{\gamma}\int_\S  d^{d}x' \sqrt{\gamma'}\,\Al_i(x) \nabla^i G(x, x') \rho(x') \ri]  \psi_A[\At] \psi_{\phi}^{(2)}[\phi]
\ee
solve the constraints for an arbitrary choice of $\psi_A[\At]$, are identical outside $B_R$ but differ inside. Note that we have used the fact that the electric field produced by $\rho_1$ and $\rho_2$, which enters in the exponents above, agrees outside $B_R$ by spherical symmetry and equality of the total charge but differs inside.

Another example of a split state is obtained  by simply taking two wavefunctionals $\psi^{(1)}[\Al, \At]$ and $\psi^{(2)}[\Al, \At]$ that are eigenstates of $\pieml$ with different eigenvalues
\be
\nabla_i \pieml^i \,\psi^{(1)}[\Al, \At] = \rho_{1}  \,\psi^{(1)}[\Al, \At]; \qquad \nabla_i \pieml^i \,\psi^{(2)}[\Al, \At] = \rho_{2}  \,\psi^{(2)}[\Al, \At]~.
\ee
Unlike the example above, $\rho_{1}$ and $\rho_{2}$ do not need to be spherically symmetric in this case but we again demand that they differ inside a ball $B_R$ but agree outside. We can then simply choose two matter wavefunctionals that satisfy \eqref{rhodiffer} and we see that the wavefunctionals
\be
\psi^{(1)}[\Al, \At] \psi^{(1)}[\phi] \qquad \text{and} \qquad \psi^{(2)}[\Al, \At] \psi^{(2)}[\phi]
\ee
differ inside $B_R$ but agree outside.

\subsection{Difference between QED and gravity}
From a technical perspective what allows us to construct split states in QED is the relation \eqref{chargecommvanish}. Unlike the charge density, the energy density {\em cannot} be independently specified at each spacetime point. This is because the commutator of the stress tensor with itself leads to the so-called Schwinger terms \cite{Deser:1967zzf}. For example, in a lattice regularization, the stress tensor at one lattice point does not commute with the stress tensor at adjacent lattice points. 

The significance of this difference can be seen by considering the global AdS vacuum. Here the specification of the total energy completely fixes the state in the bulk and so it is clear that once the integral of the stress tensor has been specified and set to vanish, there is no freedom to specify it arbitrarily in different parts of space.  In contrast, specifying the integral of the charge density leaves an infinite ambiguity in the local charge density.

There is a more physical way to understand the difference between gravity and nongravitational gauge theories. In gravity, the ``charge'' is the energy but, by the Heisenberg uncertainty principle, an excitation with a fixed total energy must be delocalized.  There is no similar principle for excitations of the electric charges or other gauge charges. This is why it is possible to find split states in ordinary gauge theories, which localize information much like other local quantum field theories, but impossible  to find split states in gravity.

\section{Graviton modes in global AdS}\label{App:GravitonModes}

We verify here that the eigenvalue problem \eqref{eigenprobWDW} coming from the Wheeler-DeWitt equation corresponds to graviton modes in AdS$_{d+1}$. We then provide an explicit solution and compute the frequencies $\w_n$ in global AdS$_4$.

\subsection{Graviton eigenvalue problem}\label{App:EigenProblem}

To relate graviton modes to the analysis of section \ref{subsecgravwave}, we should write the linearized Einstein equation in global AdS$_{d+1}$ in terms of $d$-dimensional quantities on the slice $\S$. We use hats for spacetime quantities to distinguish them from slice quantities. The background metric is taken to be
\be
ds^2 = \hga_\mn dx^\mu dx^\nu =  -N^2 dt^2 +\g_{ij}dx^i dx^j 
\ee
and the perturbation is
\be
\hh_\mn dx^\mu dx^\nu = \hh_{tt} dt^2 + 2 \hh_{ti}d t dx^i + h_{ij}dx^i dx^j~,
\ee
which we take to be transverse and traceless 
\be
\hat\n_\mu \hh^\mn  =0~,\qq \hh_\mu^{~\mu } = 0~.
\ee
This is known as the generalized de Donder gauge.

The  linearized equation of motion can be obtained by expanding the Einstein-Hilbert action to quadratic order \cite{Christensen:1979iy}
\be
S = {1\/16\pi G} \int dt d^d  x \sqrt{-\hat{g}}\, (\hat{R} - 2 \L)~.
\ee
This leads to the linearized equation
\be
(\hat\Box+2) \hh_\mn =0~.
\ee
To write this in terms of slice quantities, we use that  the non-zero Christoffel symbols of the background are
\be
\hG^i_{tt} = N\p_i N ,\qq \hG_{it}^t = N^{-1}\p_i N ,\qq \hG_{ij}^k = \G_{ij}^k~,
\ee
and a tedious but straightforward computation gives
\be\label{linearizedEinsteinslice}
(\hat\Box+2) \hh_{ij}  = - N^{-2}\p_t^2 h_{ij} + (\D_N+2) h_{ij}~,
\ee
where the Laplace-type operator $\D_N$ defined in \eqref{DeltaN} appears. The equations $(\hat\Box+2) \hh_{ti}=0$ can be used to fix the components $\hh_{ti}$ and one can check that $(\hat\Box+2) \hh_{tt}=0$ is then automatically satisfied.

The frequencies $\w_n$ of the graviton modes can be defined by the eigenvalue equation $i \p_t h_{ij}^{(n)}  = \w_n h^{(n)}_{ij}$, and we see that \eqref{linearizedEinsteinslice} indeed reduces to \eqref{eigenprobWDW}.

\subsection{Graviton spectrum in AdS$_4$}\label{App:Teukolsky}

For completeness, we give here a derivation of the graviton frequencies $\w_n$ in the case of global AdS$_4$. The background metric is 
\be\label{AdS_m}
ds^2= g_\mn dx^\mu dx^\nu= - (1+r^2)dt^2+ {dr^2\/1+r^2}  + r^2(d\t^2+\r{sin}^2\t \,d\phi^2)~.
\ee
An efficient method to obtain the graviton spectrum in AdS$_4$ is to make use of the Teukolsky equation \cite{Teukolsky:1973ha}. We start by defining a Newman-Penrose tetrad \cite{Newman:1961qr} which here takes the form
\begin{align}
l & =  {1\/1+r^2} \p_t+ \p_r~, & n & = {1\/2 } (\p_t - (1+r^2)\p_r)~,\\
m & = {1\/\sqrt{2} r}\le(\p_\t+ {i \/\r{sin}\,\t}\p_\phi\ri)~, & \bar{m}  & = {1\/\sqrt{2} r}\le(\p_\t-{i \/\r{sin}\,\t}\p_\phi\ri)~,
\end{align}
and satisfies 
\be
  g_\mn =-l_\mu n_\nu -l_\nu n_\mu +m_\mu \bar{m}_\nu + \bar{m}_\mu m_\nu~.
\ee
It consists of null vectors which are all orthogonal to each other except for $l \cdot n  = -1$ and $m\cdot \bar{m}  = 1 $.

 The Teukolsky equation can be written for any type D spacetime using the Newman-Penrose formalism. For global AdS$_4$, it takes the form
\be
\begin{split}
0 &= {r^2\/1+r^2}\p_t^2 \Psi_\eta  - r^2(1+r^2) \p_r^2\Psi_\eta  - 2(1+ 2\eta ) \p_r\Psi_\eta+  \eta{4r\/1+r^2} \p_t\Psi_\eta- {1\/\r{sin}^2\t}\p_\phi^2\Psi_\eta \\
&- {1\/\r{sin}\,\t}\p_\t(\r{sin}\,\t\, \p_\t\Psi_\eta) - 4 i \eta {\r{cos}\,\t\/\r{sin}^2\t} \p_\phi\Psi_\eta - 2 \le( 3r^2 (3+2\eta) + 2 +\eta  - {2\/\r{sin}^2\t} \ri)\Psi_\eta
\end{split}
\ee
where $\eta = \pm 1$ corresponds to the two polarizations.\footnote{This equation can also be obtained by taking the $M=a=0$ limit of the Kerr-AdS analysis of \cite{Dias:2012pp}.} 

We can consider a separated ansatz
\be
\Psi_\eta (t,r,\t,\phi) = e^{-i\w t+ i m\phi}R_\eta(r) S_\eta(\t)~,
\ee
and the master equation reduces to two coupled ODEs. The equation for $S(\t)$ can be written using the variable $x=\r{cos}\,\t$ as
\be
\p_x((1-x^2)\p_x S) + \le(\la +s - {(m + s x)^2\/1-x^2} \ri)S  = 0
\ee
and corresponds to spin-weighted spherical harmonics of spin $s$ \cite{Newman:1966ub,Goldberg:1966uu}. It is well-known that the corresponding eigenvalues are
\be
\la = \l(\l+1) - s(s+1)~,\qq \l=|s|,|s|+1,|s|+2,\dots
\ee
with azimuthal number degeneracies $m= -\l,-\l+1,\dots,\l-1,\l$. The eigenvalue $\la$ enters in the radial equation which takes the form
\bea
0\= R''(r) + {2(1+2\eta) (1+2r^2)\/r(1+r^2)} R'(r) \-
&&\hspace{1cm}+ {1\/r^2(1+r^2)^2}\le( (1+r^2)(4+ 2\eta+6r^2(3+2\eta) - \l(\l+1)) +\w^2 r^2 + 4i \eta\w r \ri)R(r)~.
\eea
For each polarization $\eta=\pm1$, the solutions are given in terms of hypergeometric functions. Imposing regularity at the origin $r=0$ selects one of the two solutions. Imposing normalizability at $r=\infty$ makes the spectrum discrete, with frequencies 
\be
\w_{\l,n} = \l+ n+1~, \qq n \in\Z_{\geq0}~,
\ee
for each polarization. For fixed $\l$ and $n$, the  degeneracy of $\w_{\l,n}$ is $2(2\l+1)$ coming from the two polarizations and the $2\l+1$ values of the azimuthal quantum number $m$.

\newcommand{\blue}[1]{{\color{blue} #1}}
 \section{Gravitational energy in AdS\label{appEnergy}}

 In this appendix, we compare the boundary Hamiltonian \eqref{defHbdy} with various expressions for the gravitational energy in AdS.

\subsection{Hawking-Horowitz prescription}

A formula of the gravitational energy in AdS was obtained in \cite{Hawking:1995fd}. For linearized perturbation, this takes the form
\be\label{energyHH}
H_\partial^\r{HH} = \frac{1}{2\kappa} \int_{\partial \Sigma}d^{d-1}\Omega  \,N n^i \nabla^j(h_{ij} - h \gamma_{ij})~.
\ee
This was derived assuming in the gauge $h_{ij}|_{\p\S} = 0$. It is easy to see that  \eqref{defHbdy} reduces to \eqref{energyHH} under this gauge condition.

\subsection{Holographic energy}

In the context of AdS/CFT, a notion of holographic energy was defined in \cite{deHaro:2000vlm}. To compare, we will write \eqref{defHbdy} in Fefferman-Graham (FG) gauge. We define a new radial coordinate $\rho$ in which the global AdS$_{d+1}$ metric \eqref{AdS_m} takes the form 
\be
\hat{\gamma}_{\mu\nu}dx^{\mu}dx^{\nu}= -\frac{(4\rho^2+1)^2}{16\rho^2} dt^2 +\frac{d\rho^2}{\rho^2}+\frac{(4\rho^{2}-1)^2}{16\rho^2}d\Omega_{d-1}
\ee 
and we assume that the perturbation satisfies $\hh_{\rho\mu}=0$. The prescription of \cite{deHaro:2000vlm} was written in terms of $\hat{h}_{00}$ but \cite{deHaro:2000vlm} also showed that the trace of the perturbation $\hat{\gamma}^{\mu \nu} \hat{h}_{\mu \nu}$ was fixed in terms of a dimension-dependent number. Therefore, in order to compare our expression with \cite{deHaro:2000vlm} it is permissible to replace $\gamma^{i j} h_{i j}$ with $\hat{h}_{00}$ up to a constant that only shifts the zero-point of the energy.   

With this substitution, the expression \eqref{defHbdy} for the energy gives
\begin{align}
\Hbdy &= \frac{1}{2\kappa} \lim_{\rho \to \infty}\rho^{d-2}\int d^{d-1} \Omega   \, (-\rho\, \p_\rho \hat{h}_{00} + 2 \hat{h}_{00}) .
 \label{E2final1}
\end{align}
In a large $\rho$ expansion, a normalizable perturbation behaves as
\be\label{D24}
\hat{h}_{00}(t, \rho, \Omega) = \rho^{2-d} \hat{h}_{00}^{(d-2)}(t, \Omega) + \dots
\ee
up to subleading terms. This gives
\be
\Hbdy =  \frac{d}{2\kappa}  \int d^{d-1}  \Omega  \,h_{00}^{(d-2)}  ,
\label{E2final2}
\ee
which matches the holographic energy of \cite{deHaro:2000vlm}.

\subsection{Iyer-Wald energy}

In the covariant phase space formalism \cite{Iyer:1994ys}, the energy is given by the boundary integral of the $(d-1)$-form
\be
{\bm\chi}_\xi (\hh) ={1\/2\k}{\bm\ve}_\mn \le(\hh^{\mu\rho} \hn_\rho\xi^\nu -{1\/2} \hh_\rho^{~\rho} \hn^\mu \xi^\nu + \xi^\rho \hn^\nu \hh^\mu_{~\rho} - \xi^\nu \hn_\rho \hh^{\mu\rho} + \xi^\nu \hn^\mu \hh_\rho^{~\rho}\ri),
\ee
where $\xi = \p_t$ and using the notation of \cite{Faulkner:2013ica}. Evaluating this on the slice $\S$ gives
\be
{\bm\chi}_\xi (\hh) =  {1\/2\k} {\bm\ve}_{it} \le(\hh^{i\rho} \hn_\rho\xi^t - \hh^{t\rho} \hn_\rho\xi^i  -{1\/2} \hh_\rho^{~\rho} (\hn^i \xi^t-\hn^t \xi^i ) + \hn^t \hh^i_{~t} -  \hn^i \hh^t_{~t}  -  \hn_\rho \hh^{i\rho} +  \hn^i \hh_\rho^{~\rho}\ri),
\ee
This can be simplified using 
\begin{align}
\hn_t \xi^t & = \hG_{tt}^t = 0 , & \hn_i \xi^i & = \hG_{it}^i = 0 , &  \\
\hn_t \xi^i & = \hG_{tt}^i = {1\/2} \g^{ij}\p_j(N^2), & 
\hn_i \xi^t & = \hG_{it}^t ={1\/2N^2} \p_i(N^2)~,
\end{align}
and we obtain
\be
{\bm\chi}_\xi (\hh) =  {1\/2\k} {\bm\ve}_i \le( \n_j N( h^{ij} - h \g^{ij}) - N \n_j( h^{ij} - h \g^{ij} )\ri),
\ee
using the relation between the volume forms $\hat{\bm\ve}_{it} =N {\bm\ve}_i$. This shows that the Iyer-Wald energy matches \eqref{defHbdy}:
\be
\Hbdy = \int_{\p\S} {\bm\chi}_\xi (\hh) ~.
\ee
In fact, the integrated Hamiltonian constraint \eqref{Hcan2} can be viewed as the quantization of a classical equation which can be expressed in this formalism. For linearized AdS spacetimes, this was detailed in \cite{Faulkner:2013ica, Faulkner:2017tkh}. The result is that the $tt$ component of Einstein's equation gives an identity
\be
\int_{\p\S}{\bm\chi}_\xi(h)=\int_\S \le( {\bm\w}_\text{grav}(h,\cL_\xi h) +{\bm\w}_\phi(\phi,\cL_\xi \phi) \ri),
\ee
for linearized on-shell perturbations. The LHS is the the energy $\Hbdy$ as shown above. The RHS involves the symplectic forms ${\bm\w}_\text{grav} $ and ${\bm\w}_\phi$ associated to gravity and matter and corresponds to a ``bulk energy'' known as the Hollands-Wald canonical energy \cite{Hollands:2012sf}.

\section{Leading order solutions \label{appleadingorder}}

We derive here the leading order solutions to the pointwise Hamiltonian and momentum constraints presented in equation \eqref{solpointguess} and equation \eqref{Sexpression}.

\ss{Hamiltonian constraint}

As explained in section \ref{sec:pointwise}, the second order Hamiltonian constraint takes the form
\be
\cD^{i j} h_{i j}^\r{T}  = \kappa \,Q^{(0)} ,
\ee
where $\cD_{ij}$ is defined in \eqref{dijexplicit} and $Q^{(0)}$ is the truncation to leading order of $Q$ which can be written
\be\label{exprQ0}
Q^{(0)} = 2  \le(  \Pi^{ij}_\r{T}\Pi_{ij}^\r{T} -{1\/d-1}\Pi_\r{T}^2\ri) + 4  \Pi_{ij}^\r{TT}\Pi_\r{T}^{ij} + Q_0^{(0)}
\ee
where we have isolated the part with no $\Pi^\r{T}$:
\be
\label{defQ00app}
\begin{split}
Q_0^{(0)} &= 2 \Pi^{ij}_\r{TT}\Pi_{ij}^\r{TT}-{1\/8}   h^{ij} (\D_N +2)h_{ij} +{1\/4}\le(  2 h^{ij}\n_i \n^k h_{jk} +  \n_i h^{ij} \n^k h_{jk}\ri) \\
&+ {1\/4}N^{-1}\n_i L^i +  \cH^\text{matter}~.
\end{split}
\ee
It is convenient to define the time variable $\bt$ by the equation
\be
\label{Piijtime}
\Pi_{ij}^\r{T} = \cD_{ij} \bt~,
\ee
which is explicitly
\be
\Pi_{ij}^\r{T} ={1 \over 2} \left(\n^i \n^j \bt- \gamma^{i j} \n_k \n^k \bt+ (d-1) \gamma^{i j} \bt\right)~.
\ee
This is the generalization to AdS of the time variable used in \cite{Arnowitt:1962hi,Kuchar:1970mu}.  Taking the trace we see that
\be
\g^{ij}\Pi_{ij}^\r{T} = {d-1\/2}(-\D+d)\bt~,
\ee
so the relation with $\a$ in \eqref{alphaPiT} is:
\be
\bt = -{2\/d-1} N \a~.
\ee
An identity that will prove important is
\be\label{Nalpha}
N\a_{ij} = \n_i\a_j+\n_j\a_i 
\ee
where we have
\be
\a_i = {1\/2}N^2\n_i\a = -{1\/(d-1)} (N \n_i \bt - \n_i N \bt)~.
\ee
This allows us to solve the Hamiltonian constraint at leading order. We have
\be
\begin{split}
\d\Psi &=\int d^d x\,{\d \Psi \/\d\Pi_\r{T}^{ij}(x)}\,\d \Pi_\r{T}^{ij}(x) \\
 &= i\int d^d x \sqrt{\g}\,  \cD^{ij} h^\r{T}_{ij} \Psi \,\d\bt~
\end{split}
\ee
using \eqref{Piijtime} and integration by parts. Hence,
\be
{-}{i\/\sqrt{\g}}{\d \Psi\/\d\bt} =  \cD^{ij}h_{ij}^\r{T} ~,
\ee
and the constraint can be written
\be
\le( {-} {i\/\sqrt{\g}}{\d\/\d\bt}- \k\, Q^{(0)}\ri) \Psi=0~.
\ee
We can write the solution in the form
\be\label{solexpr}
\Psi[{\bf t},h^\r{TT},h^\r{L},\phi]= \exp\le(i \k \cP\ri)\Psi_0[h^\r{TT},h^\r{L},\phi] + \Or[\k^2],
\ee
where $\cP$ needs to satisfy 
\be
{1\/\sqrt{\g}}{\d \cP\/\d \bt(x)} =  Q^{(0)}~.
\ee
The solution can be found and takes a remarkably simple form:
\be\label{PexprApp}
\cP = \int d^d x\sqrt{\g}\le(-{2\/3} \,{\bf t} \le( \Pi^{ij}_\r{T}\Pi_{ij}^\r{T} -{1\/d-1}\Pi^2_\r{T}\ri)+2\, {\bf t}\,\Pi^{ij}_\r{TT}\Pi_\r{T}^{ij}+{\bf t} \,Q_0^{(0)}\ri).
\ee
The first term is cubic in $\bt$ and the second term is quadratic in $\bt$. We will now check that differentiating these terms with respect to $\bt$ gives \eqref{exprQ0}.

\pg{Cubic term.} Let's consider the cubic term, allowing each entries to be different:
\be
\cP^{(3)}[{\bf t}_1,{\bf t}_2,{\bf t}_3] ={2\/3}\int d^d x
\sqrt{\g}\,{\bf t}_1 \le( \Pi^{ij}_\r{T}[{\bf t_2}]\Pi_{ij}^\r{T}[{\bf t_3}] -{1\/d-1}\Pi_\r{T}[{\bf t_2}]\Pi_\r{T}[{\bf t_3}]\ri),
\ee
We want to show that 
\be\label{P3identity}
{1\/\sqrt{\g}} {\d\/\d{\bf t}}\cP^{(3)}[{\bf t},{\bf t},{\bf t}] = 2  \le(  \Pi^{ij}_\r{T}\Pi_{ij}^\r{T} -{1\/d-1}\Pi_\r{T}^2\ri) .
\ee
Since the derivative with respect to $\bt_1$ gives one third of the RHS, we just need $\cP^{(3)}[{\bf t}_1,{\bf t}_2,{\bf t}_3]$ to be invariant under permutation of its arguments. As $\cP^{(3)}[{\bf t}_1,{\bf t}_2,{\bf t}_3]$ is manifestly invariant under ${\bf t_2}\leftrightarrow {\bf t_3}$, we just need to show that it's also invariant under ${\bf t_1}\leftrightarrow {\bf t_3}$.

First, we note that the combination gives
\be
\Pi^{ij}_\r{T}\Pi_{ij}^\r{T} -{1\/d-1}\Pi^2_\r{T}= -{1\/d-1}\Pi_{ij}^\r{T}\a^{ij}~,
\ee
so that we have
\be
\cP^{(3)}[{\bf t}_1,{\bf t}_2,{\bf t}_3] =-{2\/3(d-1)}\int d^d x
\sqrt{\g}\,{\bf t}_1 \, \Pi^{ij}_\r{T}[{\bf t_2}] \a_{ij}[{\bf t_3}]~.
\ee
We now use the identity \eqref{Nalpha} and integration by parts:
\be
\begin{split}
\cP^{(3)}[{\bf t}_1,{\bf t}_2,{\bf t}_3] &=-{4\/3(d-1)}\int d^d x\sqrt{\g}\,N^{-1}\,{\bf t}_1 \, \Pi^{ij}_\r{T}[{\bf t_2}] \n_i\a_j[\bt_3] \\
&={4\/3(d-1)^2}\int d^d x\sqrt{\g}\,N^{-1}\,{\bf t}_1 \, \Pi^{ij}_\r{T}[{\bf t_2}] \n_i(N\n_j\bt_3 -\n_j N \bt_3) \\
 &={4\/3(d-1)^2}\int d^d x 
\sqrt{\g}\,\Pi^{ij}_\r{T}[{\bf t_2}] \le( -\n_i\bt_1 \n_j{\bf t_3}-\g_{ij}\bt_1\bt_3 \ri).
\end{split}
\ee
This is manifestly symmetric under ${\bf t_1}\leftrightarrow {\bf t_3}$. Hence, $\cP^{(3)}[{\bf t}_1,{\bf t}_2,{\bf t}_3]$ is invariant under permutation of its arguments and \eqref{P3identity} is satisfied.

\pg{Quadratic term.}

Similarly, we introduce the quantity
\be
\cP^{(2)}[\bt_1,\bt_2] = 2\int d^d x \sqrt{\g} \, {\bf t}_1\,\Pi^{ij}_\r{TT}\Pi_\r{T}^{ij}[\bt_2]~.
\ee
We want to show that 
\be
{1\/\sqrt{\g}}{\d\/\d{\bf t}}\cP^{(2)}[{\bf t},{\bf t}] = 4   \,\Pi^{ij}_\r{TT}\Pi_{ij}^\r{T} ~.
\ee
This would follow if $\cP^{(2)}[\bt_1,\bt_2]$ is invariant under $\bt_1\leftrightarrow\bt_2$. We can write
\be
\begin{split}
\cP^{(2)}[\bt_1,\bt_2] &= 2\int d^d x \sqrt{\g} \, {\bf t}_1\,\Pi^{ij}_\r{TT}\cD_{ij}\bt_2, \\
&=-\int d^d x \sqrt{\g}  \,\Pi^{ij}_\r{TT}(\n^i\,{\bf t}_1 \n^j\bt_2 - \gamma^{i j} \n_k \,{\bf t}_1\n^k \bt_2 )~,
\end{split}
\ee
which is manifestly invariant under $\bt_1\leftrightarrow\bt_2$.

This shows that \eqref{PexprApp} is indeed the solution.

\ss{Momentum constraint}

The leading order constraint is
\be
{2\/\k}\g_{ij}\n_k\Pi^{jk}_\r{L} = Q_i^{(0)}~.
\ee
What plays the role of $\bt$ here is the vector $\e_i$ defined as
\be
h^\r{L}_{ij} = \n_i \e_j+\n_j \e_i~.
\ee
We then have
\be
\begin{split}
\d \Psi &= \int d^d x {\d\Psi\/\d h_{ij}^\r{L}(x)}\d h_{ij}^\r{L}(x) \\
&= -2 i \int d^d x \sqrt{\g}\,\n_i \Pi^{ij}_\r{L}\Psi \,\d\e_j~,
\end{split}
\ee
which allows to write the momentum constraint as
\be
\le(-{i\/\sqrt\g} {\d\/\d \e_i} - \k\,Q_i^{(0)}\ri)\Psi[h,\phi]= 0~,
\ee
where $Q^{(0)}_i$ is the leading order truncation of $Q_i$ which takes the form
\be
Q^{(0)}_i=  ( \n_i h_{jk}-2 \n_k h_{ij})(\Pi^{jk}_\r{T}+\Pi^{jk}_\r{TT})+  \cH_i^\r{matter}~.
\ee
This can be written in terms of $\e^i$ as
\be
Q^{(0)}_i =\le( 2(R_{k\l ij}\e^\l - \n_k\n_j\e_i )  +  \n_i h_{jk}^\r{TT}-2 \n_k h_{ij}^\r{TT}\ri)(\Pi^{jk}_\r{T}+\Pi^{jk}_\r{TT})+  \cH_i^\r{matter}~.
\ee
To properly define this operator, we should adopt the same normal ordering prescription as in section \ref{subsecintham}.

As above, we can write the solution as
\be
\Psi[h^\r{TT},h^\r{T},h^\r{L},\phi] =  \exp\le( {i\k} \cR\ri) \Psi_0[h^\r{TT},h^\r{T},\phi]+\Or[\k^2],
\ee
where $\Psi_0$ is an arbitrary functional. We need to have
\be
{1\/\sqrt{\g}}{\d \/\d \e^i}  \cR = -\,Q^{(0)}_i~.
\ee
The solution can be found explicitly to be
\be
\cR =-\int d^d x \sqrt{\g}\le( R_{ijk\l}\e^i \e^\l -\e^i \n_k\n_j\e_i+\e^i\n_i h_{jk}^\r{TT}-2\e^i \n_k h_{ij}^\r{TT}) (\Pi^{jk}_\r{T}+\Pi^{jk}_\r{TT}) + \e^i \cH_i^\r{matter}\ri).
\ee
As above, we can prove that this is a solution by showing that  the term quadratic in $\e^i$ is  symmetric in its two entries. This follows from integration by parts.

\ss{Solutions to both constraints}

We have presented above general perturbative solutions of the Hamiltonian and momentum constraint independently. Here, we will give solutions to both constraints. 

The solutions found above must be compatible with each other. This implies that the ``interaction'' part of $\cP$ and $\cR$, involving products of $\bt$ and $\e_i$ must be exactly the same. We will use below the subscript ``int'' to denote this part. It is a rather non-trivial consistency check to verify this. 

From the solution of the Hamiltonian constraint, we have
\be
\cP_\r{int} = {1\/2}\int d^d x\sqrt{\g}\,\n_iM^i\bt
\ee
using the results of section \ref{subsecintegconst}. In particular, $M^i$ is defined in \eqref{defMi}.  From the momentum constraint, this term is
\be
\begin{split}
\cR_\r{int} &= -\int d^d x \sqrt{\g}\le( 2(R_{k\l ij}\e^\l - \n_k\n_j\e_i )  +  \n_i h_{jk}^\r{TT}-2 \n_k h_{ij}^\r{TT}\ri)\Pi^{jk}_\r{T}  \\
&=\int d^d x \sqrt{\g} \,D^{jk}\le( 2(R_{k\l ij}\e^\l - \n_k\n_j\e_i )  +  \n_i h_{jk}^\r{TT}-2 \n_k h_{ij}^\r{TT}\ri) \bt
\end{split}
\ee
using integration by parts. Consistency of our solutions then requires that $\cP_\r{int}=\cR_\r{int}$ which is explicitly
\be
D^{jk}\le( R_{ijk\l}\e^i \e^\l +\n_k \e^i \n_j\e_i +\e^i\n_i h_{jk}^\r{TT}-2\e^i \n_k h_{ij}^\r{TT}\ri)  = {1\/2}\n_iM^i~.
\ee
This is a rather non-trivial identity since the LHS comes from the expansion of the momentum constraint while the RHS comes from the expansion of the Hamiltonian constraint. We have checked that this identity indeed holds, see the associated Mathematica notebook \cite{xActgithub}.

Finally, we can write the leading order solution to both constraints as
\be
\Psi [\Pi^\r{T},h^\r{TT},h^\r{L},\phi] = \exp(i \k \cS)\psi[h^\r{TT},\phi]+ \Or[\k^2],
\ee
where $\psi$ is an arbitrary functional and
\be
\cS = \int d^d x\sqrt{\g}\le(-{2\/3} \,{\bf t} \le( \Pi^{ij}_\r{T}\Pi_{ij}^\r{T} -{1\/d-1}\Pi^2_\r{T}\ri)+2\, {\bf t}\,\Pi^{ij}_\r{TT}\Pi_\r{T}^{ij}+Q^{(0)}_0{\bf t} -\e^i \cH_i^\r{matter}\ri),
\ee
with $Q^{(0)}_0$  given in \eqref{defQ00app}.

\bibliography{references}

\providecommand{\href}[2]{#2}\begingroup\raggedright\begin{thebibliography}{10}

\bibitem{Laddha:2020kvp}
A.~Laddha, S.~G. Prabhu, S.~Raju, and P.~Shrivastava, ``{The Holographic Nature
  of Null Infinity},''
  \href{http://dx.doi.org/10.21468/SciPostPhys.10.2.041}{{\em SciPost Phys.}
  {\bfseries 10} (2021) 041}, \href{http://arxiv.org/abs/2002.02448}{{\ttfamily
  arXiv:2002.02448 [hep-th]}}.

\bibitem{Raju:2020smc}
S.~Raju, ``{Lessons from the Information Paradox},''
  \href{http://arxiv.org/abs/2012.05770}{{\ttfamily arXiv:2012.05770
  [hep-th]}}.

\bibitem{Chowdhury:2020hse}
C.~Chowdhury, O.~Papadoulaki, and S.~Raju, ``{A physical protocol for observers
  near the boundary to obtain bulk information in quantum gravity},''
  \href{http://dx.doi.org/10.21468/SciPostPhys.10.5.106}{{\em SciPost Phys.}
  {\bfseries 10} no.~5, (2021) 106},
  \href{http://arxiv.org/abs/2008.01740}{{\ttfamily arXiv:2008.01740
  [hep-th]}}.

\bibitem{Maldacena:1997re}
J.~M. Maldacena, ``{The Large N limit of superconformal field theories and
  supergravity},'' \href{http://dx.doi.org/10.1023/A:1026654312961}{{\em Adv.
  Theor. Math. Phys.} {\bfseries 2} (1998) 231--252},
  \href{http://arxiv.org/abs/hep-th/9711200}{{\ttfamily arXiv:hep-th/9711200}}.

\bibitem{Gubser:1998bc}
S.~S. Gubser, I.~R. Klebanov, and A.~M. Polyakov, ``{Gauge theory correlators
  from noncritical string theory},''
  \href{http://dx.doi.org/10.1016/S0370-2693(98)00377-3}{{\em Phys. Lett. B}
  {\bfseries 428} (1998) 105--114},
  \href{http://arxiv.org/abs/hep-th/9802109}{{\ttfamily arXiv:hep-th/9802109}}.

\bibitem{Witten:1998qj}
E.~Witten, ``{Anti-de Sitter space and holography},''
  \href{http://dx.doi.org/10.4310/ATMP.1998.v2.n2.a2}{{\em Adv. Theor. Math.
  Phys.} {\bfseries 2} (1998) 253--291},
  \href{http://arxiv.org/abs/hep-th/9802150}{{\ttfamily arXiv:hep-th/9802150}}.

\bibitem{DeWitt:1967yk}
B.~S. DeWitt, ``{Quantum Theory of Gravity. 1. The Canonical Theory},''
  \href{http://dx.doi.org/10.1103/PhysRev.160.1113}{{\em Phys. Rev.} {\bfseries
  160} (1967) 1113--1148}.

\bibitem{Kuchar:1970mu}
K.~Kuchar, ``{Ground state functional of the linearized gravitational field},''
  \href{http://dx.doi.org/10.1063/1.1665133}{{\em J. Math. Phys.} {\bfseries
  11} (1970) 3322--3334}.

\bibitem{Arnowitt:1962hi}
R.~L. Arnowitt, S.~Deser, and C.~W. Misner, ``{The Dynamics of general
  relativity},'' \href{http://dx.doi.org/10.1007/s10714-008-0661-1}{{\em Gen.
  Rel. Grav.} {\bfseries 40} (2008) 1997--2027},
  \href{http://arxiv.org/abs/gr-qc/0405109}{{\ttfamily arXiv:gr-qc/0405109}}.

\bibitem{Banerjee:2016mhh}
S.~Banerjee, J.-W. Bryan, K.~Papadodimas, and S.~Raju, ``{A toy model of black
  hole complementarity},''
  \href{http://dx.doi.org/10.1007/JHEP05(2016)004}{{\em JHEP} {\bfseries 05}
  (2016) 004}, \href{http://arxiv.org/abs/1603.02812}{{\ttfamily
  arXiv:1603.02812 [hep-th]}}.

\bibitem{Raju:2019qjq}
S.~Raju, ``{Is Holography Implicit in Canonical Gravity?},''
  \href{http://dx.doi.org/10.1142/S0218271819440115}{{\em Int. J. Mod. Phys.}
  {\bfseries D28} no.~14, (2019) 1944011},
\href{http://arxiv.org/abs/1903.11073}{{\ttfamily arXiv:1903.11073 [gr-qc]}}.

\bibitem{Marolf:2008mf}
D.~Marolf, ``{Unitarity and Holography in Gravitational Physics},''
  \href{http://dx.doi.org/10.1103/PhysRevD.79.044010}{{\em Phys. Rev. D}
  {\bfseries 79} (2009) 044010},
  \href{http://arxiv.org/abs/0808.2842}{{\ttfamily arXiv:0808.2842 [gr-qc]}}.

\bibitem{Marolf:2006bk}
D.~Marolf, ``{Asymptotic flatness, little string theory, and holography},''
  \href{http://dx.doi.org/10.1088/1126-6708/2007/03/122}{{\em JHEP} {\bfseries
  03} (2007) 122}, \href{http://arxiv.org/abs/hep-th/0612012}{{\ttfamily
  arXiv:hep-th/0612012}}.

\bibitem{Jacobson:2019gnm}
T.~Jacobson and P.~Nguyen, ``{Diffeomorphism invariance and the black hole
  information paradox},''
  \href{http://dx.doi.org/10.1103/PhysRevD.100.046002}{{\em Phys. Rev. D}
  {\bfseries 100} no.~4, (2019) 046002},
  \href{http://arxiv.org/abs/1904.04434}{{\ttfamily arXiv:1904.04434 [gr-qc]}}.

\bibitem{Jacobson:2012gh}
T.~Jacobson, ``{Boundary unitarity and the black hole information paradox},''
  \href{http://dx.doi.org/10.1142/S0218271813420029}{{\em Int. J. Mod. Phys.}
  {\bfseries D22} (2013) 1342002},
\href{http://arxiv.org/abs/1212.6944}{{\ttfamily arXiv:1212.6944 [hep-th]}}.

\bibitem{Freidel:2008sh}
L.~Freidel, ``{Reconstructing AdS/CFT},''
  \href{http://arxiv.org/abs/0804.0632}{{\ttfamily arXiv:0804.0632 [hep-th]}}.

\bibitem{Cianfrani:2013oja}
F.~Cianfrani and J.~Kowalski-Glikman, ``{Wheeler-DeWitt equation and AdS/CFT
  correspondence},''
  \href{http://dx.doi.org/10.1016/j.physletb.2013.07.034}{{\em Phys. Lett. B}
  {\bfseries 725} (2013) 463--467},
  \href{http://arxiv.org/abs/1306.0353}{{\ttfamily arXiv:1306.0353 [hep-th]}}.

\bibitem{McGough:2016lol}
L.~McGough, M.~Mezei, and H.~Verlinde, ``{Moving the CFT into the bulk with $
  T\overline{T} $},'' \href{http://dx.doi.org/10.1007/JHEP04(2018)010}{{\em
  JHEP} {\bfseries 04} (2018) 010},
  \href{http://arxiv.org/abs/1611.03470}{{\ttfamily arXiv:1611.03470
  [hep-th]}}.

\bibitem{Caputa:2020lpa}
P.~Caputa, S.~Datta, Y.~Jiang, and P.~Kraus, ``{Geometrizing $ T\overline{T}
  $},'' \href{http://dx.doi.org/10.1007/JHEP03(2021)140}{{\em JHEP} {\bfseries
  03} (2021) 140}, \href{http://arxiv.org/abs/2011.04664}{{\ttfamily
  arXiv:2011.04664 [hep-th]}}.

\bibitem{Kruthoff:2020hsi}
J.~Kruthoff and O.~Parrikar, ``{On the flow of states under $T\overline{T}$},''
  \href{http://arxiv.org/abs/2006.03054}{{\ttfamily arXiv:2006.03054
  [hep-th]}}.

\bibitem{Coleman:2019dvf}
E.~A. Coleman, J.~Aguilera-Damia, D.~Z. Freedman, and R.~M. Soni, ``{$
  T\overline{T} $ -deformed actions and (1,1) supersymmetry},''
  \href{http://dx.doi.org/10.1007/JHEP10(2019)080}{{\em JHEP} {\bfseries 10}
  (2019) 080}, \href{http://arxiv.org/abs/1906.05439}{{\ttfamily
  arXiv:1906.05439 [hep-th]}}.

\bibitem{Donnelly:2018bef}
W.~Donnelly and V.~Shyam, ``{Entanglement entropy and $T \overline{T}$
  deformation},'' \href{http://dx.doi.org/10.1103/PhysRevLett.121.131602}{{\em
  Phys. Rev. Lett.} {\bfseries 121} no.~13, (2018) 131602},
  \href{http://arxiv.org/abs/1806.07444}{{\ttfamily arXiv:1806.07444
  [hep-th]}}.

\bibitem{Tolley:2019nmm}
A.~J. Tolley, ``{$ T\overline{T} $ deformations, massive gravity and
  non-critical strings},''
  \href{http://dx.doi.org/10.1007/JHEP06(2020)050}{{\em JHEP} {\bfseries 06}
  (2020) 050}, \href{http://arxiv.org/abs/1911.06142}{{\ttfamily
  arXiv:1911.06142 [hep-th]}}.

\bibitem{Ireland:2019vvj}
A.~Ireland and V.~Shyam, ``{$ T\overline{T} $ deformed YM$_{2}$ on general
  backgrounds from an integral transformation},''
  \href{http://dx.doi.org/10.1007/JHEP07(2020)058}{{\em JHEP} {\bfseries 07}
  (2020) 058}, \href{http://arxiv.org/abs/1912.04686}{{\ttfamily
  arXiv:1912.04686 [hep-th]}}.

\bibitem{Hirano:2020nwq}
S.~Hirano and M.~Shigemori, ``{Random boundary geometry and gravity dual of $
  T\overline{T} $ deformation},''
  \href{http://dx.doi.org/10.1007/JHEP11(2020)108}{{\em JHEP} {\bfseries 11}
  (2020) 108}, \href{http://arxiv.org/abs/2003.06300}{{\ttfamily
  arXiv:2003.06300 [hep-th]}}.

\bibitem{Belin:2020oib}
A.~Belin, A.~Lewkowycz, and G.~Sarosi, ``{Gravitational path integral from the
  $T^2$ deformation},'' \href{http://dx.doi.org/10.1007/JHEP09(2020)156}{{\em
  JHEP} {\bfseries 09} (2020) 156},
  \href{http://arxiv.org/abs/2006.01835}{{\ttfamily arXiv:2006.01835
  [hep-th]}}.

\bibitem{Caputa:2019pam}
P.~Caputa, S.~Datta, and V.~Shyam, ``{Sphere partition functions
  \textbackslash{}\& cut-off AdS},''
  \href{http://dx.doi.org/10.1007/JHEP05(2019)112}{{\em JHEP} {\bfseries 05}
  (2019) 112}, \href{http://arxiv.org/abs/1902.10893}{{\ttfamily
  arXiv:1902.10893 [hep-th]}}.

\bibitem{Kenmoku:1998he}
M.~Kenmoku, H.~Kubotani, E.~Takasugi, and Y.~Yamazaki, ``{Analytic solutions of
  the Wheeler-DeWitt equation in spherically symmetric space-time},''
  \href{http://dx.doi.org/10.1103/PhysRevD.59.124004}{{\em Phys. Rev. D}
  {\bfseries 59} (1999) 124004},
  \href{http://arxiv.org/abs/gr-qc/9810042}{{\ttfamily arXiv:gr-qc/9810042}}.

\bibitem{Hajicek:1984mz}
P.~Hajicek, ``{Spherically Symmetric Systems of Fields and Black Holes. 2.
  Apparent Horizon in Canonical Formalism},''
  \href{http://dx.doi.org/10.1103/PhysRevD.30.1178}{{\em Phys. Rev. D}
  {\bfseries 30} (1984) 1178}.

\bibitem{Fischler:1989ka}
W.~Fischler, I.~R. Klebanov, J.~Polchinski, and L.~Susskind, ``{Quantum
  Mechanics of the Googolplexus},''
  \href{http://dx.doi.org/10.1016/0550-3213(89)90290-3}{{\em Nucl. Phys. B}
  {\bfseries 327} (1989) 157--177}.

\bibitem{Fischler:1989se}
W.~Fischler, D.~Morgan, and J.~Polchinski, ``{Quantum Nucleation of False
  Vacuum Bubbles},'' \href{http://dx.doi.org/10.1103/PhysRevD.41.2638}{{\em
  Phys. Rev. D} {\bfseries 41} (1990) 2638}.

\bibitem{Halliwell:1984eu}
J.~J. Halliwell and S.~W. Hawking, ``{The Origin of Structure in the
  Universe},'' \href{http://dx.doi.org/10.1103/PhysRevD.31.1777}{{\em Phys.
  Rev. D} {\bfseries 31} (1985) 1777}.

\bibitem{Hori:1993zq}
T.~Hori, ``{Exact solutions to the Wheeler-DeWitt equation of two-dimensional
  dilaton gravity},'' \href{http://dx.doi.org/10.1143/PTP.90.743}{{\em Prog.
  Theor. Phys.} {\bfseries 90} (1993) 743--745},
  \href{http://arxiv.org/abs/hep-th/9303049}{{\ttfamily arXiv:hep-th/9303049}}.

\bibitem{Louis-Martinez:1993bge}
D.~Louis-Martinez, J.~Gegenberg, and G.~Kunstatter, ``{Exact Dirac quantization
  of all 2-D dilaton gravity theories},''
  \href{http://dx.doi.org/10.1016/0370-2693(94)90463-4}{{\em Phys. Lett. B}
  {\bfseries 321} (1994) 193--198},
  \href{http://arxiv.org/abs/gr-qc/9309018}{{\ttfamily arXiv:gr-qc/9309018}}.

\bibitem{Gegenberg:1994pv}
J.~Gegenberg, G.~Kunstatter, and D.~Louis-Martinez, ``{Observables for
  two-dimensional black holes},''
  \href{http://dx.doi.org/10.1103/PhysRevD.51.1781}{{\em Phys. Rev. D}
  {\bfseries 51} (1995) 1781--1786},
  \href{http://arxiv.org/abs/gr-qc/9408015}{{\ttfamily arXiv:gr-qc/9408015}}.

\bibitem{Maldacena:2019cbz}
J.~Maldacena, G.~J. Turiaci, and Z.~Yang, ``{Two dimensional Nearly de Sitter
  gravity},'' \href{http://dx.doi.org/10.1007/JHEP01(2021)139}{{\em JHEP}
  {\bfseries 01} (2021) 139}, \href{http://arxiv.org/abs/1904.01911}{{\ttfamily
  arXiv:1904.01911 [hep-th]}}.

\bibitem{Betzios:2020nry}
P.~Betzios and O.~Papadoulaki, ``{Liouville theory and Matrix models: A Wheeler
  DeWitt perspective},'' \href{http://dx.doi.org/10.1007/JHEP09(2020)125}{{\em
  JHEP} {\bfseries 09} (2020) 125},
  \href{http://arxiv.org/abs/2004.00002}{{\ttfamily arXiv:2004.00002
  [hep-th]}}.

\bibitem{Ashtekar:1986yd}
A.~Ashtekar, ``{New Variables for Classical and Quantum Gravity},''
  \href{http://dx.doi.org/10.1103/PhysRevLett.57.2244}{{\em Phys. Rev. Lett.}
  {\bfseries 57} (1986) 2244--2247}.

\bibitem{rovelli2004quantum}
C.~Rovelli, {\em Quantum gravity}.
\newblock Cambridge university press, 2004.

\bibitem{Halliwell:1989myn}
J.~J. Halliwell, ``{Introductory Lectures on Quantum Cosmology},'' in {\em {7th
  Jerusalem Winter School for Theoretical Physics: Quantum Cosmology and Baby
  Universes}}.
\newblock 1989.
\newblock \href{http://arxiv.org/abs/0909.2566}{{\ttfamily arXiv:0909.2566
  [gr-qc]}}.

\bibitem{coleman1991quantum}
S.~Coleman, J.~B. Hartle, T.~Piran, and S.~Weinberg, {\em Quantum Cosmology And
  Baby Universes: Proceedings Of 7th Jerusalem Winter School}, vol.~7.
\newblock World Scientific, 1991.

\bibitem{Giddings:2019hjc}
S.~B. Giddings, ``{Gravitational dressing, soft charges, and perturbative
  gravitational splitting},''
  \href{http://dx.doi.org/10.1103/PhysRevD.100.126001}{{\em Phys. Rev. D}
  {\bfseries 100} no.~12, (2019) 126001},
  \href{http://arxiv.org/abs/1903.06160}{{\ttfamily arXiv:1903.06160
  [hep-th]}}.

\bibitem{Donnelly:2018nbv}
W.~Donnelly and S.~B. Giddings, ``{Gravitational splitting at first order:
  Quantum information localization in gravity},''
  \href{http://dx.doi.org/10.1103/PhysRevD.98.086006}{{\em Phys. Rev. D}
  {\bfseries 98} no.~8, (2018) 086006},
  \href{http://arxiv.org/abs/1805.11095}{{\ttfamily arXiv:1805.11095
  [hep-th]}}.

\bibitem{Donnelly:2016rvo}
W.~Donnelly and S.~B. Giddings, ``{Observables, gravitational dressing, and
  obstructions to locality and subsystems},''
  \href{http://dx.doi.org/10.1103/PhysRevD.94.104038}{{\em Phys. Rev.}
  {\bfseries D94} no.~10, (2016) 104038},
\href{http://arxiv.org/abs/1607.01025}{{\ttfamily arXiv:1607.01025 [hep-th]}}.

\bibitem{Donnelly:2017jcd}
W.~Donnelly and S.~B. Giddings, ``{How is quantum information localized in
  gravity?},'' \href{http://dx.doi.org/10.1103/PhysRevD.96.086013}{{\em Phys.
  Rev.} {\bfseries D96} no.~8, (2017) 086013},
\href{http://arxiv.org/abs/1706.03104}{{\ttfamily arXiv:1706.03104 [hep-th]}}.

\bibitem{cmp/1103942446}
M.~Henneaux and C.~Teitelboim, ``{Asymptotically anti-de Sitter spaces},''
  \href{http://dx.doi.org/cmp/1103942446}{{\em Communications in Mathematical
  Physics} {\bfseries 98} no.~3, (1985) 391 -- 424}. \url{https://doi.org/}.

\bibitem{Ashtekar:1999jx}
A.~Ashtekar and S.~Das, ``{Asymptotically Anti-de Sitter space-times: Conserved
  quantities},'' \href{http://dx.doi.org/10.1088/0264-9381/17/2/101}{{\em
  Class. Quant. Grav.} {\bfseries 17} (2000) L17--L30},
  \href{http://arxiv.org/abs/hep-th/9911230}{{\ttfamily arXiv:hep-th/9911230}}.

\bibitem{deHaro:2000vlm}
S.~de~Haro, S.~N. Solodukhin, and K.~Skenderis, ``{Holographic reconstruction
  of space-time and renormalization in the AdS / CFT correspondence},''
  \href{http://dx.doi.org/10.1007/s002200100381}{{\em Commun. Math. Phys.}
  {\bfseries 217} (2001) 595--622},
  \href{http://arxiv.org/abs/hep-th/0002230}{{\ttfamily arXiv:hep-th/0002230}}.

\bibitem{Freidel:2020xyx}
L.~Freidel, M.~Geiller, and D.~Pranzetti, ``{Edge modes of gravity. Part I.
  Corner potentials and charges},''
  \href{http://dx.doi.org/10.1007/JHEP11(2020)026}{{\em JHEP} {\bfseries 11}
  (2020) 026}, \href{http://arxiv.org/abs/2006.12527}{{\ttfamily
  arXiv:2006.12527 [hep-th]}}.

\bibitem{Freidel:2020svx}
L.~Freidel, M.~Geiller, and D.~Pranzetti, ``{Edge modes of gravity. Part II.
  Corner metric and Lorentz charges},''
  \href{http://dx.doi.org/10.1007/JHEP11(2020)027}{{\em JHEP} {\bfseries 11}
  (2020) 027}, \href{http://arxiv.org/abs/2007.03563}{{\ttfamily
  arXiv:2007.03563 [hep-th]}}.

\bibitem{Aharony:2003qf}
O.~Aharony, O.~DeWolfe, D.~Z. Freedman, and A.~Karch, ``{Defect conformal field
  theory and locally localized gravity},''
  \href{http://dx.doi.org/10.1088/1126-6708/2003/07/030}{{\em JHEP} {\bfseries
  07} (2003) 030}, \href{http://arxiv.org/abs/hep-th/0303249}{{\ttfamily
  arXiv:hep-th/0303249}}.

\bibitem{dirac2001quantum}
P.~A.~M. Dirac, {\em Lectures on Quantum mechanics}.
\newblock Dover, 2001.
\newblock Original published by Belfer Graduate School of Science in 1964.

\bibitem{Regge:1974zd}
T.~Regge and C.~Teitelboim, ``{Role of Surface Integrals in the Hamiltonian
  Formulation of General Relativity},''
  \href{http://dx.doi.org/10.1016/0003-4916(74)90404-7}{{\em Annals Phys.}
  {\bfseries 88} (1974) 286}.

\bibitem{xAct}
J.~M. Mart\'in-Garc\'ia, ``\href{http://www.xact.es}{xAct: Efficient tensor
  computer algebra for the Wolfram Language},''  (2002) .

\bibitem{Brizuela:2008ra}
D.~Brizuela, J.~M. Martin-Garcia, and G.~A. Mena~Marugan, ``{xPert: Computer
  algebra for metric perturbation theory},''
  \href{http://dx.doi.org/10.1007/s10714-009-0773-2}{{\em Gen. Rel. Grav.}
  {\bfseries 41} (2009) 2415--2431},
  \href{http://arxiv.org/abs/0807.0824}{{\ttfamily arXiv:0807.0824 [gr-qc]}}.

\bibitem{xActgithub}
``Github repository: Wheeler-dewitt holography.''
  \url{https://github.com/victorgodet/wdw-holography}, 2021.

\bibitem{york1973conformally}
J.~W. York~Jr, ``Conformally invariant orthogonal decomposition of symmetric
  tensors on riemannian manifolds and the initial-value problem of general
  relativity,'' {\em Journal of Mathematical Physics} {\bfseries 14} no.~4,
  (1973) 456--464.

\bibitem{Hartle:1984ke}
J.~B. Hartle, ``{Ground State Wave Function Of Linearized Gravity},''
  \href{http://dx.doi.org/10.1103/PhysRevD.29.2730}{{\em Phys. Rev. D}
  {\bfseries 29} (1984) 2730--2737}.

\bibitem{Aharony:1999ti}
O.~Aharony, S.~S. Gubser, J.~M. Maldacena, H.~Ooguri, and Y.~Oz, ``{Large N
  field theories, string theory and gravity},''
  \href{http://dx.doi.org/10.1016/S0370-1573(99)00083-6}{{\em Phys. Rept.}
  {\bfseries 323} (2000) 183--386},
  \href{http://arxiv.org/abs/hep-th/9905111}{{\ttfamily arXiv:hep-th/9905111}}.

\bibitem{York:1972sj}
J.~W. York, Jr., ``{Role of conformal three geometry in the dynamics of
  gravitation},'' \href{http://dx.doi.org/10.1103/PhysRevLett.28.1082}{{\em
  Phys. Rev. Lett.} {\bfseries 28} (1972) 1082--1085}.

\bibitem{kuchar1991problem}
K.~Kuchar, ``The problem of time in canonical quantization,'' in {\em
  Conceptual Problems of Quantum Gravity}, {A. Ashtekar} and {J. Stachel}, eds.
\newblock Birkhauser, 1991.

\bibitem{Donnelly:2015hta}
W.~Donnelly and S.~B. Giddings, ``{Diffeomorphism-invariant observables and
  their nonlocal algebra},''
  \href{http://dx.doi.org/10.1103/PhysRevD.94.029903,
  10.1103/PhysRevD.93.024030}{{\em Phys. Rev.} {\bfseries D93} no.~2, (2016)
  024030}, \href{http://arxiv.org/abs/1507.07921}{{\ttfamily arXiv:1507.07921
  [hep-th]}}.
[Erratum: Phys. Rev.D94,no.2,029903(2016)].

\bibitem{streater2016pct}
R.~F. Streater and A.~S. Wightman, {\em PCT, spin and statistics, and all
  that}.
\newblock Princeton University Press, 2016.

\bibitem{Haag:1992hx}
R.~Haag, {\em {Local quantum physics: Fields, particles, algebras}}.
\newblock Springer-Verlag, 1992.

\bibitem{lloyd1988black}
S.~Lloyd, {\em Black Holes, Demons and the Loss of Coherence: How complex
  systems get information, and what they do with it}.
\newblock PhD thesis, Rockefeller University, 1988.

\bibitem{Brown:1986nw}
J.~D. Brown and M.~Henneaux, ``{Central Charges in the Canonical Realization of
  Asymptotic Symmetries: An Example from Three-Dimensional Gravity},''
  \href{http://dx.doi.org/10.1007/BF01211590}{{\em Commun. Math. Phys.}
  {\bfseries 104} (1986) 207--226}.

\bibitem{Casini:2011kv}
H.~Casini, M.~Huerta, and R.~C. Myers, ``{Towards a derivation of holographic
  entanglement entropy},''
  \href{http://dx.doi.org/10.1007/JHEP05(2011)036}{{\em JHEP} {\bfseries 1105}
  (2011) 036},
\href{http://arxiv.org/abs/1102.0440}{{\ttfamily arXiv:1102.0440 [hep-th]}}.

\bibitem{Chakraborty:2021rvy}
T.~Chakraborty, J.~Chakravarty, and P.~Paul, ``{Monogamy paradox: A toy model
  in flat space},'' \href{http://arxiv.org/abs/2107.06919}{{\ttfamily
  arXiv:2107.06919 [hep-th]}}.

\bibitem{Raju:2018zpn}
S.~Raju, ``{A Toy Model of the Information Paradox in Empty Space},''
  \href{http://dx.doi.org/10.21468/SciPostPhys.6.6.073}{{\em SciPost Phys.}
  {\bfseries 6} no.~6, (2019) 073},
  \href{http://arxiv.org/abs/1809.10154}{{\ttfamily arXiv:1809.10154
  [hep-th]}}.

\bibitem{Geng:2021hlu}
H.~Geng, A.~Karch, C.~Perez-Pardavila, S.~Raju, L.~Randall, M.~Riojas, and
  S.~Shashi, ``{Inconsistency of Islands in Theories with Long-Range
  Gravity},'' \href{http://arxiv.org/abs/2107.03390}{{\ttfamily
  arXiv:2107.03390 [hep-th]}}.

\bibitem{Almheiri:2020cfm}
A.~Almheiri, T.~Hartman, J.~Maldacena, E.~Shaghoulian, and A.~Tajdini, ``{The
  entropy of Hawking radiation},''
  \href{http://arxiv.org/abs/2006.06872}{{\ttfamily arXiv:2006.06872
  [hep-th]}}.

\bibitem{hatfield1991quantum}
B.~Hatfield, {\em Quantum field theory of point particles and strings}.
\newblock Perseus Books, 1991.

\bibitem{Papadodimas:2013jku}
K.~Papadodimas and S.~Raju, ``{State-Dependent Bulk-Boundary Maps and Black
  Hole Complementarity},''
  \href{http://dx.doi.org/10.1103/PhysRevD.89.086010}{{\em Phys. Rev. D}
  {\bfseries 89} no.~8, (2014) 086010},
  \href{http://arxiv.org/abs/1310.6335}{{\ttfamily arXiv:1310.6335 [hep-th]}}.

\bibitem{Giddings:1999jq}
S.~B. Giddings, ``{Flat space scattering and bulk locality in the AdS / CFT
  correspondence},'' \href{http://dx.doi.org/10.1103/PhysRevD.61.106008}{{\em
  Phys. Rev. D} {\bfseries 61} (2000) 106008},
  \href{http://arxiv.org/abs/hep-th/9907129}{{\ttfamily arXiv:hep-th/9907129}}.

\bibitem{DHoker:2002nbb}
E.~D'Hoker and D.~Z. Freedman, ``{Supersymmetric gauge theories and the AdS /
  CFT correspondence},'' in {\em {Theoretical Advanced Study Institute in
  Elementary Particle Physics (TASI 2001): Strings, Branes and EXTRA
  Dimensions}}.
\newblock 1, 2002.
\newblock \href{http://arxiv.org/abs/hep-th/0201253}{{\ttfamily
  arXiv:hep-th/0201253}}.

\bibitem{Deser:1967zzf}
S.~Deser and D.~Boulware, ``{Stress-Tensor Commutators and Schwinger Terms},''
  \href{http://dx.doi.org/10.1063/1.1705368}{{\em J. Math. Phys.} {\bfseries 8}
  (1967) 1468}.

\bibitem{Christensen:1979iy}
S.~Christensen and M.~Duff, ``{Quantizing Gravity with a Cosmological
  Constant},'' \href{http://dx.doi.org/10.1016/0550-3213(80)90423-X}{{\em Nucl.
  Phys. B} {\bfseries 170} (1980) 480--506}.

\bibitem{Teukolsky:1973ha}
S.~A. Teukolsky, ``{Perturbations of a rotating black hole. 1. Fundamental
  equations for gravitational electromagnetic and neutrino field
  perturbations},'' \href{http://dx.doi.org/10.1086/152444}{{\em Astrophys. J.}
  {\bfseries 185} (1973) 635--647}.

\bibitem{Newman:1961qr}
E.~Newman and R.~Penrose, ``{An Approach to gravitational radiation by a method
  of spin coefficients},'' \href{http://dx.doi.org/10.1063/1.1724257}{{\em J.
  Math. Phys.} {\bfseries 3} (1962) 566--578}.

\bibitem{Dias:2012pp}
O.~J.~C. Dias, J.~E. Santos, and M.~Stein, ``{Kerr-AdS and its Near-horizon
  Geometry: Perturbations and the Kerr/CFT Correspondence},''
  \href{http://dx.doi.org/10.1007/JHEP10(2012)182}{{\em JHEP} {\bfseries 10}
  (2012) 182}, \href{http://arxiv.org/abs/1208.3322}{{\ttfamily arXiv:1208.3322
  [hep-th]}}.

\bibitem{Newman:1966ub}
E.~T. Newman and R.~Penrose, ``{Note on the Bondi-Metzner-Sachs group},''
  \href{http://dx.doi.org/10.1063/1.1931221}{{\em J. Math. Phys.} {\bfseries 7}
  (1966) 863--870}.

\bibitem{Goldberg:1966uu}
J.~N. Goldberg, A.~J. MacFarlane, E.~T. Newman, F.~Rohrlich, and E.~C.~G.
  Sudarshan, ``{Spin s spherical harmonics and edth},''
  \href{http://dx.doi.org/10.1063/1.1705135}{{\em J. Math. Phys.} {\bfseries 8}
  (1967) 2155}.

\bibitem{Hawking:1995fd}
S.~W. Hawking and G.~T. Horowitz, ``{The Gravitational Hamiltonian, action,
  entropy and surface terms},''
  \href{http://dx.doi.org/10.1088/0264-9381/13/6/017}{{\em Class. Quant. Grav.}
  {\bfseries 13} (1996) 1487--1498},
  \href{http://arxiv.org/abs/gr-qc/9501014}{{\ttfamily arXiv:gr-qc/9501014}}.

\bibitem{Iyer:1994ys}
V.~Iyer and R.~M. Wald, ``{Some properties of Noether charge and a proposal for
  dynamical black hole entropy},''
  \href{http://dx.doi.org/10.1103/PhysRevD.50.846}{{\em Phys. Rev. D}
  {\bfseries 50} (1994) 846--864},
  \href{http://arxiv.org/abs/gr-qc/9403028}{{\ttfamily arXiv:gr-qc/9403028}}.

\bibitem{Faulkner:2013ica}
T.~Faulkner, M.~Guica, T.~Hartman, R.~C. Myers, and M.~Van~Raamsdonk,
  ``{Gravitation from Entanglement in Holographic CFTs},''
  \href{http://dx.doi.org/10.1007/JHEP03(2014)051}{{\em JHEP} {\bfseries 1403}
  (2014) 051},
\href{http://arxiv.org/abs/1312.7856}{{\ttfamily arXiv:1312.7856 [hep-th]}}.

\bibitem{Faulkner:2017tkh}
T.~Faulkner, F.~M. Haehl, E.~Hijano, O.~Parrikar, C.~Rabideau, and
  M.~Van~Raamsdonk, ``{Nonlinear Gravity from Entanglement in Conformal Field
  Theories},'' \href{http://dx.doi.org/10.1007/JHEP08(2017)057}{{\em JHEP}
  {\bfseries 08} (2017) 057}, \href{http://arxiv.org/abs/1705.03026}{{\ttfamily
  arXiv:1705.03026 [hep-th]}}.

\bibitem{Hollands:2012sf}
S.~Hollands and R.~M. Wald, ``{Stability of Black Holes and Black Branes},''
  \href{http://dx.doi.org/10.1007/s00220-012-1638-1}{{\em Commun. Math. Phys.}
  {\bfseries 321} (2013) 629--680},
  \href{http://arxiv.org/abs/1201.0463}{{\ttfamily arXiv:1201.0463 [gr-qc]}}.

\end{thebibliography}\endgroup

\end{document}